\documentclass[preprint, review,12pt] {elsarticle}

\usepackage{graphics}
\usepackage{graphicx}
\usepackage{epsfig}

\usepackage{amssymb}
\usepackage{amsthm}

\usepackage{lineno}
\usepackage{color}

\usepackage{lineno}
\usepackage{color}
\usepackage{xcolor}
\usepackage[dvipsnames]{xcolor}

\usepackage{float}

\begin{document}

\begin{frontmatter}

 

\title{
%
Numerical Validation of Lyapunov–Liouville Theory and Non-Diffusive Closures in Decaying Isotropic Fluid and Scalar Turbulence.}
 

\author{Nicola de Divitiis}

\address{"La Sapienza" University, Dipartimento di Ingegneria Meccanica e 
Aerospaziale, Via Eudossiana, 18, 00184 Rome, Italy, \\
phone: +39--0644585268, \ \ fax: +39--0644585750, \\ 
e-mail: n.dedivitiis@gmail.com, \ \  nicola.dedivitiis@uniroma1.it}

\begin{abstract} 
This work presents a numerical validation of the Lyapunov--Liouville theoretical framework and its corresponding non-diffusive turbulence closures in freely decaying homogeneous isotropic turbulence (HIT). Rather than solving the full three-dimensional Navier--Stokes and passive scalar transport equations, we numerically integrate the closed von K\'arm\'an--Howarth and Corrsin equations via an automated, adaptive numerical solver. Three distinct initial states---characterized by Saffman--Birkhoff, Loitsiansky, and Gaussian velocity correlation profiles---are analyzed across a wide range of Prandtl numbers (\(Pr = 0.1\), \(1\), \(10\), and \(1000\)). The transient development phase and the subsequent self-preserving diffusive regime are examined for both the velocity and passive scalar fields, and the resulting evolution is assessed against independent DNS and experimental benchmarks from the literature. Beyond providing closures for the resolved equations, the Lyapunov--Liouville framework yields the internal structure of turbulence in terms of velocity and temperature increments; specifically, the computation of the temperature increment PDFs is performed over the investigated Prandtl numbers and further extended down to $Pr = 10^{-3}$.

Our numerical findings indicate that the proposed closure reproduces the distinct decay paths dictated by the initial states, quantified by the monomial time-decay exponents $m$ and $n$ for the velocity and temperature fields, respectively. The Saffman--Birkhoff condition yields asymptotic exponents of $m \simeq -1.25$ and $n \simeq -1.25$, whereas the Loitsiansky condition leads to an accelerated mechanical decay ($m \simeq -1.51$) alongside a higher thermal persistence ($n \simeq -0.89$). Conversely, the Gaussian profile induces a rapid decay ($m \simeq -2.7$), where the system approaches a critical lower threshold at a non-dimensional time $t \simeq 33$ (measured in units of the initial Lyapunov time) as $R_\lambda$ drops below $10$. Furthermore, the model captures the non-equilibrium evolution of characteristic scales: at $Pr = 1000$, the thermal microscale $\lambda_\theta$ drops below the Kolmogorov scale $\eta_K$, consistent with Batchelor's scaling theory. Here, the calculated Batchelor constant settles toward $C_B \simeq 3.5$, varying between $3.4$ and $5.7$ across the investigated cases. Additionally, the evaluated Kolmogorov and Obukhov--Corrsin constants align with decaying turbulence benchmarks, settling around $C_K \simeq 1.72 - 1.75$ and $C_{OC} \simeq 1.8$, respectively. Finally, the longitudinal velocity increment PDFs describe multi-scale intermittency features and higher-order moments compatible with experimental data, while the temperature increment PDFs show a transition from quasi-Gaussian statistics at low $Pr$ ($10^{-3} - 10^{-1}$) to enhanced, scale-dependent intermittency at higher $Pr$, offering a satisfactory agreement with established DNS references.
\end{abstract}


\end{frontmatter}

\newcommand{\no}{\noindent}
\newcommand{\be}{\begin{equation}}
\newcommand{\ee}{\end{equation}}
\newcommand{\bea}{\begin{eqnarray}}
\newcommand{\eea}{\end{eqnarray}}
\newcommand{\bc}{\begin{center}}
\newcommand{\ec}{\end{center}}

\newcommand{\calr}{{\cal R}}
\newcommand{\calv}{{\cal V}}

\newcommand{\bff}{\mbox{\boldmath $f$}}
\newcommand{\bfg}{\mbox{\boldmath $g$}}
\newcommand{\bfh}{\mbox{\boldmath $h$}}
\newcommand{\bfi}{\mbox{\boldmath $i$}}
\newcommand{\bfm}{\mbox{\boldmath $m$}}
\newcommand{\bfp}{\mbox{\boldmath $p$}}
\newcommand{\bfr}{\mbox{\boldmath $r$}}
\newcommand{\bfu}{\mbox{\boldmath $u$}}
\newcommand{\bfv}{\mbox{\boldmath $v$}}
\newcommand{\bfx}{\mbox{\boldmath $x$}}
\newcommand{\bfy}{\mbox{\boldmath $y$}}
\newcommand{\bfw}{\mbox{\boldmath $w$}}
\newcommand{\bfk}{\mbox{\boldmath $\kappa$}}

\newcommand{\bfA}{\mbox{\boldmath $A$}}
\newcommand{\bfD}{\mbox{\boldmath $D$}}
\newcommand{\bfI}{\mbox{\boldmath $I$}}
\newcommand{\bfL}{\mbox{\boldmath $L$}}
\newcommand{\bfM}{\mbox{\boldmath $M$}}
\newcommand{\bfS}{\mbox{\boldmath $S$}}
\newcommand{\bfT}{\mbox{\boldmath $T$}}
\newcommand{\bfU}{\mbox{\boldmath $U$}}
\newcommand{\bfX}{\mbox{\boldmath $X$}}
\newcommand{\bfY}{\mbox{\boldmath $Y$}}
\newcommand{\bfK}{\mbox{\boldmath $K$}}

\newcommand{\bfeta}{\mbox{\boldmath $\eta$}}
\newcommand{\bfrho}{\mbox{\boldmath $\rho$}}
\newcommand{\bfchi}{\mbox{\boldmath $\chi$}}
\newcommand{\bfphi}{\mbox{\boldmath $\phi$}}
\newcommand{\bfPhi}{\mbox{\boldmath $\Phi$}}
\newcommand{\bflambda}{\mbox{\boldmath $\lambda$}}
\newcommand{\bfxi}{\mbox{\boldmath $\xi$}}
\newcommand{\bfLambda}{\mbox{\boldmath $\Lambda$}}
\newcommand{\bfPsi}{\mbox{\boldmath $\Psi$}}
\newcommand{\bfomega}{\mbox{\boldmath $\omega$}}
\newcommand{\bfOmega}{\mbox{\boldmath $\Omega$}}
\newcommand{\bfeps}{\mbox{\boldmath $\varepsilon$}}
\newcommand{\bfepsn}{\mbox{\boldmath $\epsilon$}}
\newcommand{\bfzeta}{\mbox{\boldmath $\zeta$}}
\newcommand{\bfkappa}{\mbox{\boldmath $\kappa$}}
\newcommand{\bfsigma}{\mbox{\boldmath $\sigma$}}
\newcommand{\itPsi}{\mbox{\it $\Psi$}}
\newcommand{\itPhi}{\mbox{\it $\Phi$}}

\newcommand{\bint}{\mbox{ \int{a}{b}} }
\newcommand{\ds}{\displaystyle}
\newcommand{\Sum}{\Large \sum}



\bigskip

\section{Introduction \label{intro}}

The mathematical description of homogeneous isotropic turbulence (HIT) remains one of the most enduring challenges in classical physics. Since the pioneering scaling arguments proposed by Kolmogorov \cite{kolmogorov1941local} and Obukhov \cite{obukhov1941spectrum}, the statistical behavior of velocity and passive scalar fields within the inertial-convective range has been extensively investigated. In their seminal work, von K\'arm\'an and Howarth \cite{VonKarman1938statistical} derived the exact dynamical equation governing the evolution of the two-point longitudinal velocity correlation function. Analogously, Corrsin \cite{corrsin1951spectrum, Corrsin_1} extended this formulation to incorporate the dynamics of a passive scalar trancant, such as temperature, establishing the foundation for the analysis of mixed fluid-scalar interactions. 

Despite the mathematical elegance of the von K\'arm\'an--Howarth and Corrsin equations, they are structurally unclosed due to the non-linear advection terms in the Navier--Stokes and transport equations. This nonlinearity manifests as a coupling between lower-order and higher-order correlation functions---specifically linking the two-point double correlations to the two-point triple correlations---giving rise to the classical closure problem of turbulence. Over the decades, numerous semi-empirical closures and analytical theories have been proposed to resolve this impasse. These range from the Millionshchikov quasi-normal hypothesis \cite{millionshtchikov1941theory} and related classical closures \cite{Hasselmann58, Millionshtchikov69, Oberlack93, Baev, Mellor84}, to advanced spectral theories such as the Eddy-Damped Quasi-Normal Markovian (EDQNM) framework \cite{orszag1970analytical, briard2016passive}.
However, traditional Eulerian closures frequently struggle to encapsulate the full richness of small-scale intermittency, the non-local nature of the energy cascade, and the precise non-stationary dynamics of freely decaying flows without relying on arbitrary, flow-dependent tuning parameters.

To overcome these structural limitations, an alternative paradigm has emerged by investigating the kinematics of fluid motion from a statistical mechanics and dynamical systems perspective, leveraging the Lagrangian description. In this context, interscale energy transfer and the multi-scale stretching of fluid elements can be rigorously quantified through the lens of chaos theory, specifically via Lyapunov stability and Liouville analysis. Early foundations establishing the link between chaotic advection, mixing, and Lagrangian stretching were laid by Ottino \cite{ottino1989kinematics}. Building upon these concepts, the closure of the von K\'arm\'an–Howarth and Corrsin equations was successfully achieved in  \cite{deDivitiis2011, deDivitiis2014, deDivitiis2016} by evaluating the velocity field through finite-scale Lyapunov theory applied to the relative fluid motion. This framework was recently generalized to account for the Liouville spectral gap, bifurcation-driven Lagrangian--Eulerian decoupling, and state-variable statistics via non-observable quasi-probability density functions (quasi-PDFs) \cite{deDivitiis2026, deDivitiis2026a}, providing a comprehensive theoretical basis for small-scale intermittency, structural anomalies, and forward/backward scattering phenomena \cite{deDivitiis2026b}.

While these recent theoretical advancements have provided a closed framework for fully developed turbulence, its performance in non-equilibrium, freely decaying regimes warrants systematic investigation. Freely decaying HIT represents a classic benchmark for turbulence closures; immediately following an initial state, the flow undergoes a transient development phase wherein non-linear transport constructs the energy cascade from initially uncorrelated large-scale structures. During this transitorium, classical spectral quantities, such as the Kolmogorov and Obukhov--Corrsin constants, exhibit non-equilibrium fluctuations before a self-preserving decay regime is established, as observed in high-resolution direct numerical simulations (DNS) \cite{Wang1996, kaneda2003energy} and grid turbulence experiments \cite{Comte-Bellot1966, sreenivasan1996passive, Warhaft2000}.

The principal objective of the present work is to investigate the performance of the Lyapunov--Liouville theoretical framework and its corresponding non-diffusive turbulence closures under non-stationary, decaying conditions. 
{ 
Rather than performing full three-dimensional simulations of the Navier--Stokes and passive scalar fields, 
we solve the system of partial differential equations given by the closed von K\'arm\'an--Howarth and Corrsin equations, where the advection terms are modeled via Lyapunov--Liouville analysis. By implementing an automated, adaptive numerical solver, these closed equations are integrated starting from three distinct initial states---namely Saffman--Birkhoff, Loitsiansky, and Gaussian correlation profiles. This study aims to evaluate the behavior of the multi-scale energy and scalar cascades under different unsteady conditions. Specifically, we examine the transient regime and subsequent self-preserving decay of both fields, assessing the results computed via the Lyapunov--Liouville analysis against independent DNS and experimental benchmarks available in the literature.}

{
First, we provide an analysis of the time-dependent evolution of both velocity and temperature spatial correlations alongside their corresponding energy and scalar spectra across a range of Prandtl numbers ($10^{-1} \le Pr \le 1000$). Second, we track the transient and self-preserving behavior of the mechanical and thermal dissipation coefficients. Our results suggest that these coefficients do not act as universal constants but are modulated by flow conditions, exhibiting a dependency on the initial correlation topology. Third, by evaluating the resolved spectra, we estimate the structural constants of turbulence---specifically the Kolmogorov ($C_K \in (1.71, \  1.8)$), Batchelor ($C_B \in (3.4, \  5.7)$), and Obukhov--Corrsin ($C_{OC} \simeq 1.8$) constants---under different initial states, to examine how non-equilibrium decaying regimes may affect these parameters. Fourth, we analyze the non-Gaussian probability density functions (PDFs) of velocity and temperature increments, observing how skewness, flatness, and small-scale intermittency vary as a function of separation scale, non-dimensional time, and Prandtl number. Finally, this work includes the tracking of the Lagrangian bifurcation rate $S_L$ and the Eulerian bifurcation rate $S_E$ (the latter coinciding with the total dissipation rate). Within the proposed framework, the strict inequality $S_L > S_E$ emerges as a necessary condition for sustaining a turbulent state, and we examine how and when this condition is violated during the decay process.}

The remainder of this paper is organized as follows. Section 2 summarizes the core mathematical tenets of the Lyapunov--Liouville theory and the proposed closure relations, while also reviewing the internal structure of HIT through the structure functions of velocity and temperature increments alongside their corresponding statistical moments and probability density functions (PDFs).
Section 3 provides a theoretical analysis of the proposed closures, outlining their main properties. Specifically, it examines how the evolution of the velocity and temperature autocorrelations splits into two distinct regimes: a transient development phase and a subsequent diffusive decay regime, with the development phase estimated to occur within a time interval of approximately two initial Lyapunov times.
{ Section 4 describes the numerical scheme used to integrate the von K\'arman--Howarth and Corrsin equations, while Section 5 presents the detailed results and discussion, comparing our findings with established DNS and experimental benchmarks from the literature. Finally, concluding remarks are provided in Section 6.}

\bigskip

\section{Theoretical Background: Correlation Equations, Nondiffusive Closures and Internal Structure Following Lyapunov--Liouville Analysis}

This section reviews the core results of the Liouville--Lyapunov theoretical framework previously developed by the author \cite{deDivitiis2026, deDivitiis2026a}. This approach describes the turbulent energy cascade, the skewness, and the intermittency of velocity and temperature increment statistics. Furthermore, it analytically derives the closures for the von K\'arm\'an--Howarth and Corrsin equations, yields the Kolmogorov scaling law for velocity, and uncovers the internal structure of both velocity and temperature fluctuations.

A key finding of this analysis is the sharp distinction established between Lagrangian and Eulerian fluctuation regimes in fully developed turbulence \cite{deDivitiis2026}. Specifically, the following hierarchy holds:
\bea
\begin{array}{l@{\hspace{-0.cm}}l}
\underbrace{\ds S_L >> 
\sup\left\lbrace \Lambda_L \right\rbrace >>
\left\langle \Lambda_L \right\rangle_L }\gtrsim 
\underbrace{\sup\left\lbrace \Lambda_E \right\rbrace >>
\left\langle \Lambda_E \right\rangle_E >>
S_E} \\\
 \mbox{Lagrangian \ parameters} \ \ \ \  \  \ \ \mbox{Eulerian \ parameters}
\end{array}
\label{S>>>>S}
\eea
where $\langle\cdot\rangle_L$ and $\langle\cdot\rangle_E$ denote averages over the Lagrangian and Eulerian ensembles, respectively. Here, $\Lambda_L$ represents the Lyapunov exponent associated with Lagrangian particle trajectories, whereas $\Lambda_E$ is the Lyapunov exponent associated with phase-space trajectories of the Navier--Stokes equations. The terms $S_L$ and $S_E$ denote the bifurcation rates corresponding to the Lagrangian trajectories and the Navier--Stokes equations, respectively. 

The bifurcation rate defines the mean frequency of bifurcations within chaotic regimes, matching the rate at which trajectories intersect the hypersurface $\Sigma_D$ in phase space, where the Jacobian of the dynamical system becomes singular. Consequently, this quantity represents the mean crossing rate of the Jacobian-degeneracy manifold. If trajectories do not intersect $\Sigma_D$, the system -though nonlinear- remains non-chaotic. Conversely, when trajectories repeatedly cross $\Sigma_D$, chaotic dynamics emerge, causing the state variables to fluctuate on timescales dictated by the bifurcation rate. The present framework distinguishes between two such rates: Eulerian bifurcations, associated with the Navier--Stokes equations, and Lagrangian bifurcations, associated with the velocity gradient. 
Specifically, while the Eulerian bifurcation rate defines the frequency at which a trajectory in the Navier--Stokes phase space intersects the hypersurface where the associated Jacobian matrix becomes singular, the Lagrangian bifurcation rate denotes the frequency at which a Lagrangian trajectory crosses a surface in physical space where the determinant of the velocity gradient tensor vanishes.
For a comprehensive mathematical treatment, the reader is referred to \cite{deDivitiis2026}. The analytical results derived in \cite{deDivitiis2026, deDivitiis2026a} are strictly valid provided that Eq. (\ref{S>>>>S}) is satisfied.

Equation (\ref{S>>>>S}) demonstrates that fluctuations along fluid particle trajectories are significantly faster than those within the Eulerian field. Consequently, two qualitatively distinct regimes emerge:
(i) an Eulerian regime dominated by stretching, where Lyapunov exponents fluctuate on timescales slower than the growth of perturbations;
(ii) a Lagrangian regime dominated by rapid folding mechanisms, where Lyapunov exponents fluctuate much faster than the separation of trajectories.
These bifurcation rates are here calculated, according to Ref. \cite{deDivitiis2026} as
\bea
\ds S_E \backsim \left\vert\frac{d \ln u}{dt}  \right\vert, \ \ \ S_L \backsim \sqrt{R_\lambda \ \left\langle \Lambda_L^2 \right\rangle_L}
\label{b_rates}
\eea
This theoretical formulation serves as the foundation for the analysis developed in the present work. Specifically, our study is strictly built upon the closure schemes for the von K\'arm\'an--Howarth and Corrsin equations established by the author in \cite{deDivitiis2026}. In that work, the emergence of a Liouville spectral gap and the phenomenon of bifurcation-driven Lagrangian--Eulerian decoupling provided the physical-mathematical justification for nondiffusive turbulence closures. To ensure self-consistency, the governing equations and the specific functional forms of the closures utilized herein are recalled below.

In homogeneous isotropic turbulence, the evolution of the longitudinal velocity correlation function $f=\langle u_r u_r' \rangle_E/u^2$ and the temperature correlation function $f_\theta = \langle \vartheta \vartheta' \rangle_E/\theta^2$ are governed by the von K\'arm\'an--Howarth and Corrsin equations:
\bea
\begin{array}{l@{\hspace{-0.cm}}l}
\ds \frac{\partial f}{\partial t} = 
\ds  \frac{K}{u^2} +
\ds 2 \nu  \left(  \frac{\partial^2 f} {\partial r^2} +
\ds \frac{4}{r} \frac{\partial f}{\partial r}  \right) +\frac{10 \nu}{\lambda_T^2} f, \\\\
\ds \frac{\partial f_\theta}{\partial t} = 
\ds  \frac{G}{\theta^2} +
\ds 2 \kappa  \left(  \frac{\partial^2 f_\theta} {\partial r^2} +
\ds \frac{2}{r} \frac{\partial f_\theta}{\partial r}  \right) +\frac{12 \kappa}{\lambda_\theta^2} f_\theta,
\end{array} 
\label{vk-h}
\eea
subject to the following boundary conditions:
\bea
\begin{array}{l@{\hspace{-0.cm}}l}
\ds f(t, 0)=1, \ \ \ \lim_{r \rightarrow \infty} f(t, r)=0, \\\\
\ds f_\theta(t, 0)=1, \ \ \ \lim_{r \rightarrow \infty} f_\theta(t, r)=0.
\end{array}
\label{bc}
\eea
where kinematic viscosity and thermal diffusivity are assumed to be both constant and independent of temperature.
Here, $u \equiv \sqrt{\langle u_r^2 \rangle_E}$ and $\theta \equiv \sqrt{\langle \vartheta^2 \rangle_E}$. The quantities $\lambda_T \equiv \sqrt{-1/f''(0)}$ and $\lambda_\theta \equiv \sqrt{-2/f_\theta''(0)}$ represent the Taylor and Corrsin microscales, respectively.

The functions $K$ and $G$, which account for the turbulent energy cascade, are expressed in terms of the longitudinal triple velocity correlation $k(r)$ and the mixed triple correlation $k_\theta(r)$ between $u_r$ and $\vartheta$:
\bea
\begin{array}{l@{\hspace{+0.0cm}}l}
\ds K(r) = u^3 \left( \frac{\partial }{\partial r} + \frac{4}{r} \right) 
k(r), 
\ \ \mbox{where} \ \ 
\ds k(r) = \frac{\langle u_r^2 u_r' \rangle_{E}}{u^3}, \\\\
\ds G(r) = 2 u \theta^2 \left( \frac{\partial }{\partial r} + \frac{2}{r} 
\right) k_\theta(r), 
\ \ \mbox{where} \ \ 
\ds k_\theta(r) = \frac{\langle u_r \vartheta \vartheta' \rangle_{E}}{\theta^2 u}.
\end{array}
\eea
Equations (\ref{vk-h}) achieve closure once $K$ and $G$ are expressed exclusively as functions of $f$ and $f_\theta$. Based on the Liouville--Lyapunov analysis of Ref. \cite{deDivitiis2026}, these closures are given by:
\bea
\begin{array}{l@{\hspace{+0.0cm}}l}
\ds K(r) =
 u^3 \sqrt{\frac{1-f}{2}} \ \frac{\partial f}{\partial r}, \\\\
\ds G(r) = 
 u \theta^2 \sqrt{\frac{1-f}{2}} \ \frac{\partial f_\theta}{\partial r}.
\end{array}
\label{K closure}
\eea
These expressions, determined through the link between finite scale lagrangian Lyapunov exponent and velocity correlation  \cite{deDivitiis2026, deDivitiis2026a}
\bea
\ds \langle\Lambda_L^2(r)\rangle_L r^2= 2 u^2 (1-f(r)),
\label{link0}
\eea
represent non-diffusive closures, originating physically from the decoupling of Lagrangian trajectories in the presence of a spectral gap.
The closures provided in Eqs. (\ref{K closure}) yield an accurate description of the energy cascade, successfully reproducing negative values for the skewness of both velocity and temperature increments:
\bea
\begin{array}{l@{\hspace{+0.0cm}}l}
\ds H^{(3)}_u(r) \equiv 
\frac{\langle (\Delta u_r)^3 \rangle }{\langle (\Delta u_r)^2 \rangle^{3/2}} 
=
\frac{6 k(r)}{(2(1-f(r)))^{3/2}}, \\\\
\ds H^{(3)}_\theta(r) \equiv 
\frac{\langle (\Delta \vartheta)^2 \Delta u_r \rangle }
{\langle (\Delta \vartheta)^2 \rangle {\langle (\Delta u_r)^2 \rangle}^{1/2}}=
\frac{4 k_\theta (r)}{ 2(1-f_\theta(r)) (2(1-f(r)))^{1/2}}.
\end{array}
\label{H3(r)}
\eea
In the small-scale limit, these expressions satisfy:
\bea
\begin{array}{l@{\hspace{+0.0cm}}l}
\ds H^{(3)}_u(0) = \lim_{r \rightarrow 0} H^{(3)}_u(r) = - \frac{3}{7}, \\\\
\ds H^{(3)}_\theta(0) = \lim_{r \rightarrow 0} H^{(3)}_\theta(r) = - \frac{1}{5}, 
\end{array}
\label{H3(0)}
\eea
which stands in satisfactory alignment with established literature \cite{Chen92, Orszag72, Panda89, Anderson99, Carati95, Kang2003} and is consistent with the classical Kolmogorov law. Furthermore, the resulting temperature spectra appear compatible with the theoretical arguments of Kolmogorov, Obukhov--Corrsin, and Batchelor \cite{Batchelor_2, Batchelor_3, Obukhov}, as well as with experimental \cite{Gibson, Mydlarski} and numerical \cite{Rogallo, Donzis} findings available in the literature.

Although the second expression in Eqs. (\ref{K closure}) is specifically derived for the temperature correlation, this theoretical framework applies, without loss of generality, to any passive scalar governed by diffusive transport.

Moreover, in \cite{deDivitiis2026a}, a dedicated bifurcation analysis of the von K\'arm\'an--Howarth equation demonstrates that the transitional Taylor-scale Reynolds number, $R_\lambda^*$, which establishes the baseline conditions for homogeneity and isotropy, is exactly equal to $10$. Below this threshold, the homogeneous isotropic turbulence (HIT) loses its physical condition of existence, and  the proposed closures are no longer applicable.

Beyond describing the turbulent energy cascade, the theory advanced in \cite{deDivitiis2026a} uncovers the internal structure and the corresponding statistics of velocity and temperature difference fluctuations, while simultaneously recovering the Kolmogorov scaling law. Specifically, by leveraging non-observable Lyapunov bifurcation modes and their corresponding quasi-probability density functions (quasi-PDFs), this formulation reproduces the Kolmogorov scaling law, which dictates that $u/u_k \sim (u/u_k)^2 (\eta_k/\lambda_T) \sim \sqrt{R_\lambda}$, where $u$ and $u_k$ denote the velocity standard deviation and the Kolmogorov velocity scale, respectively, while $\eta_k$ and $\lambda_T$ represent the Kolmogorov length scale and the Taylor microscale.
This suggests that the non-observability of these bifurcation Lyapunov modes provides the fundamental conceptual link between the intermittency of $\Delta u_r$ and the Kolmogorov law. Indeed, without invoking such non-observability, intermittency could still be identified, but the analytical derivation of the Kolmogorov law would remain inaccessible \cite{deDivitiis2026a}.
While the Navier--Stokes non-linearity inherently implies the existence of intermittency, it is such non-observability that enforces the Kolmogorov scaling. In this perspective, intermittency that increases with $R_\lambda$ is not an \textit{ad hoc} correction, but rather the unique statistically consistent state emerging from the non-observability of bifurcation modes and the constant skewness of $\Delta u_r$. Moreover, this mechanism also induces an intermittency of the temperature difference which rises accordingly with the P\'eclet number.

The dimensionless increments $\Delta u_r$ and $\Delta \vartheta$, normalized by their respective standard deviations, can be expressed as functions of $R_\lambda$ and the P\'eclet number $Pe \equiv R_\lambda Pr$:
\bea
\begin{array}{l@{\hspace{-0.cm}}l}
\ds \frac{\Delta u_r}{\sqrt{ \langle (\Delta u_r)^2 \rangle}} = 
\frac{\zeta_u + \Psi_u (\chi(\zeta_{u+}^2-1)-(\zeta_{u-}^2-1))}{\sqrt{1+2 \Psi_u^2(1+ \chi^2) }}, 
\\\\ 
\ds  \Psi_u (r) =\Phi(r) \sqrt{R_\lambda}, \\\\
\ds \frac{\Delta \vartheta}{\sqrt{ \langle (\Delta \vartheta)^2 \rangle}} = 
\frac{\zeta_\theta + \Psi_\theta (\zeta_{\theta+}^2-\zeta_{\theta-}^2)}{\sqrt{1+4 \Psi_\theta^2 }}, \\\\
\ds \Psi_\theta(r) = \Phi(r) \sqrt{Pe} 
\end{array}
\label{struct funs}
\eea
where $\zeta_X$, $\zeta_{X +}$, and $\zeta_{X -}$ (for $X = u, \theta$) are mutually independent, zero-mean, unit-variance Gaussian random variables. Eqs. (\ref{struct funs}) define the specific structure functions governing the statistics of $\Delta u_r$ and $\Delta \vartheta$, where the term involving $\Psi_u$ accounts for the intermittency of $\Delta u_r$. Furthermore, according to the analysis of \cite{deDivitiis2026a}, $\Psi_u(0) \sim (u/u_k)^2 (\eta_k/\lambda_T)$.
Consequently, the PDFs of $\Delta u_r$ and $\Delta \vartheta$ can be formally derived via the Frobenius--Perron equation \cite{Nicolis95}:
\bea
\begin{array}{l@{\hspace{-0.cm}}l}
\ds F_{u}(\Delta u_r') = \int_\zeta \int_{\zeta_{-}} \int_{\zeta_{+}} P(\zeta, \zeta_{-}, \zeta_{+}) \
\delta (\Delta u_r'-\Delta u_r(\zeta, \zeta_{-}, \zeta_{+})) \ \ d\zeta \ d\zeta_{-} \ d\zeta_{+}, \\\\
\ds F_{\theta}(\Delta \vartheta') = \int_\zeta \int_{\zeta_{-}} \int_{\zeta_{+}} P(\zeta, \zeta_{-}, \zeta_{+}) \
\delta (\Delta \vartheta'-\Delta \vartheta(\zeta, \zeta_{-}, \zeta_{+})) \ d\zeta \ d\zeta_{-} \ d\zeta_{+},
\end{array}
\label{Frobenius-Perron}
\eea
where $\delta$ is the Dirac delta function and $P(\zeta, \zeta_{-}, \zeta_{+})$ is the trivariate Gaussian PDF:
\bea
\ds P(\zeta, \zeta_{-}, \zeta_{+}) = \frac{1}{\sqrt{(2 \pi)^3}} \exp\left(-\frac{\zeta^2+\zeta_{-}^2+\zeta_{+}^2}{2}\right),
\label{gaussian}
\eea
with $\Delta u_r(\zeta, \zeta_{-}, \zeta_{+})$ and $\Delta \vartheta(\zeta, \zeta_{-}, \zeta_{+})$ determined by Eqs. (\ref{struct funs}).

This formulation results in non-Gaussian statistics wherein the absolute values of the dimensionless moments increase monotonically with $R_\lambda$ and $Pe$. Specifically, these dimensionless statistical moments are evaluated as:
\bea
\begin{array}{l@{\hspace{+0.2cm}}l}
\ds H_u^{(n)} \equiv \frac{\left\langle (\Delta u_r )^n \right\rangle}
{\left\langle (\Delta u_r)^2 \right\rangle^{n/2} }
= \\\\
\ds \frac{1} {(1+2(1+\chi^2)  \Psi_u^2)^{n/2}} 
\ds \sum_{k=0}^n 
\left(\begin{array}{c}
n  \\
k
\end{array}\right)  \Psi_u^k
 \langle \zeta_u^{n-k} \rangle 
  \langle (\chi (\zeta_{u+}^2-1)  - (\zeta_{u-}^2-1) )^k \rangle, \\\\
\ds H_\theta^{(n)} \equiv \frac{\left\langle (\Delta \vartheta )^n \right\rangle}
{\left\langle (\Delta \vartheta)^2 \right\rangle^{n/2} }
= 
\ds \frac{1} {(1+4  \Psi_\theta^2)^{n/2}} 
\ds \sum_{k=0}^n 
\left(\begin{array}{c}
n  \\
k
\end{array}\right)  \Psi_\theta^k
 \langle \zeta_\theta^{n-k} \rangle 
  \langle (\zeta_{\theta+}^2 - \zeta_{\theta-}^2 )^k \rangle, 
\end{array}
\label{Tm1} 
\eea
where the quantities $\Phi(0)$ and $\chi = \chi(R_\lambda)$ remain to be determined. To this end, we first analyze the statistics of $\partial u_r / \partial r$, which, according to the proposed Lyapunov analysis, exhibits a constant skewness $H_u^{(3)}(0) = -3/7$. From Eqs. (\ref{Tm1}), the third-order moment is given by:
\bea
\ds H_u^{(3)}(r) = \frac{8 \Psi_u^3(\chi^3-1)}{(1+ 2 \Psi_u^2 (1+\chi^2))^{3/2}} 
\eea
Taking the limit $r \rightarrow 0$ yields:
\bea
\ds H_u^{(3)}(0) = \frac{8 \Psi_u^3(0)(\chi^3-1)}{(1+ 2 \Psi_u^2(0) (1+\chi^2))^{3/2}}.
\label{H30}
\eea
\begin{figure}[t]
\centering
\includegraphics[scale=.45, angle=0]{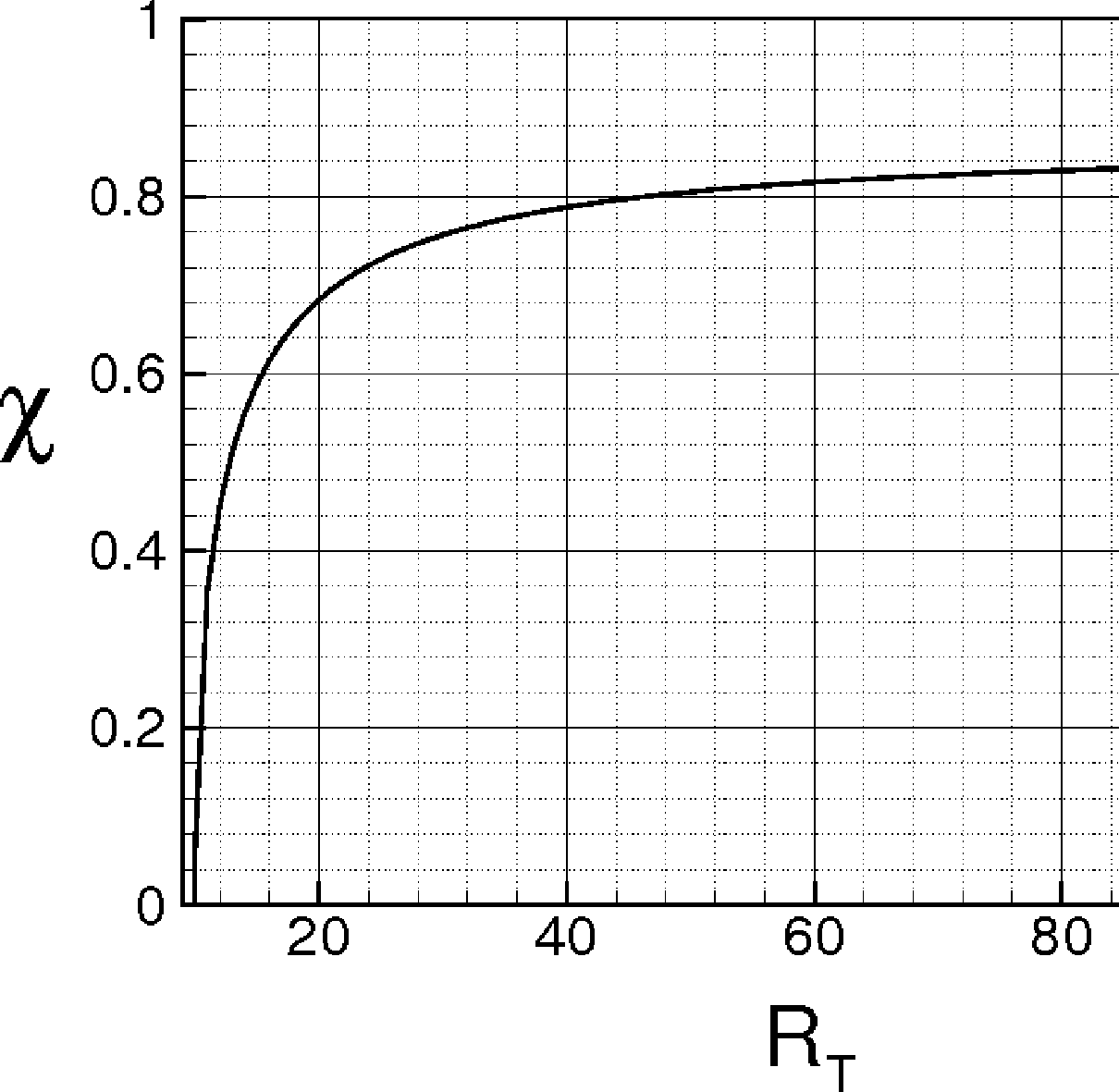}
\caption{Characteristic function $\chi$ as a function of the Taylor-scale Reynolds number, $R_\lambda$.}
\label{fig_chi_re}
\end{figure} 
Accordingly, $\chi = \chi(R_\lambda)$ is implicitly defined as a function of $\Phi(0) \sqrt{R_\lambda}$. Equation (\ref{H30}) demonstrates that $\chi(R_\lambda)$ is a monotonically increasing function of $R_\lambda$, which, for $H_u^{(3)}(0) = -3/7$, approaches the asymptotic limit $\chi_\infty = 0.8659...$ as $R_\lambda \rightarrow \infty$. For sufficiently small values of $R_\lambda$, $\chi(R_\lambda)$ can mathematically become negative. However, in fully developed turbulence, the PDF of $\partial u_r / \partial r$ must exhibit non-Gaussian tails as $\partial u_r / \partial r \rightarrow \pm \infty$, which physically dictates that $\chi$ must remain strictly positive.
Thus, the limit $\chi = 0$ is assumed to occur at $R_\lambda = R_\lambda^* = 10$, representing the threshold for homogeneous isotropic turbulence. This allows for the identification of $\Phi(0)$ via Eq. (\ref{H30}):
\bea
\ds \Phi(0) = \frac{1}{\sqrt{R_\lambda^*}} \sqrt{ \frac{{H_{u 0}^{(3)}}^{2/3}}{4-2{H_{u 0}^{(3)}}^{2/3}}} = 0.1409...
\label{phi0}
\eea 
The resulting implicit variation law $\chi = \chi(R_\lambda)$ is illustrated in Fig. \ref{fig_chi_re}.
This theoretical framework provides the basis for the analysis presented in the remainder of this work. For additional details regarding the mathematical and physical formulations of this theory, the reader is referred to \cite{deDivitiis2026, deDivitiis2026a}.

\bigskip

\section{Analysis of the Proposed Closures}

Here, we examine the fundamental properties of the proposed closures (\ref{K closure}), focusing specifically on the evolution timescales of the fully developed velocity and temperature autocorrelations. We demonstrate that these correlation functions achieve their fully developed states within finite temporal intervals that depend inherently on the initial conditions. Furthermore, beyond this transient period, the operational validity of Eq. (\ref{S>>>>S}) may no longer be sustained. 

In detail, two distinct regimes govern the evolution of the velocity and temperature correlations. The first is a "developmental" regime, during which the correlations evolve from their initial conditions, leading to a continuous reduction of the integral correlation scales until a state of local stationarity is reached, where these scales cease to decrease. The second is a purely dissipative/diffusive regime, wherein the scales tend to grow asymptotically as $\simeq \sqrt{\nu t}$. Throughout the course of this latter regime, it is critical to continuously verify whether both Eq. (\ref{S>>>>S}) and the condition $R_\lambda \ge R_\lambda^*$ remain satisfied. Once these criteria are violated, 
the homogeneous isotropic turbulence (HIT) loses its physical condition of existence, and 
the proposed non-diffusive closures (\ref{K closure}) cease to be applicable.

To analyze these dynamics, the governing equations for $u$, $\theta$, $\lambda_T$, and $\lambda_\theta$ are first derived. This is achieved by extracting the coefficients of order $r^0$ and $r^2$ from Eqs. (\ref{vk-h}) via a Taylor series expansion in even powers of $f$ and $f_\theta$ \cite{VonKarman1938statistical, corrsin1951spectrum, Corrsin_1}:
\bea
\begin{array}{l@{\hspace{+0.0cm}}l}
\ds f= 1-\frac{1}{2}\left( \frac{r}{\lambda_T(t)}\right)^2 + \frac{1}{4!} f^{IV}(t, 0) r^4+..., \\\\
\ds f_\theta= 1-\left( \frac{r}{\lambda_\theta(t)}\right)^2 + \frac{1}{4!} f_\theta^{IV}(t, 0)  r^4+ ...,
\end{array}
\label{f f_theta}
\eea
yielding the following system of differential equations:
\bea
\begin{array}{l@{\hspace{+0.2cm}}l}
\ds \frac{d u^2}{d t} = - \frac{10 \nu}{\lambda_T^2} u^2, \\\\
\ds \frac{d \theta^2}{d t} = - \frac{12 \kappa}{\lambda_\theta^2} \theta^2, 
\end{array}
\label{u2dot_thetadot}
\eea
\bea
\begin{array}{l@{\hspace{+0.0cm}}l}
\ds  \frac{d \lambda_T}{dt} = -\frac{u}{2} + \frac{\nu}{\lambda_T}
\left( \frac{7}{3} f^{IV}(t, 0) \lambda_T^4-5\right), \\\\
\ds  \frac{d \lambda_\theta}{dt} = -\frac{u}{2}\frac{\lambda_\theta}{\lambda_T} + \frac{\kappa}{\lambda_\theta}
\left( \frac{5}{6} f^{IV}_\theta(t, 0) \lambda_\theta^4-6\right).
\end{array}
\label{lambda_dot}
\eea
While Eqs. (\ref{u2dot_thetadot}) are universal and independent of the specific modeling of the triple correlations \cite{VonKarman1938statistical, Corrsin_1, corrsin1951spectrum}, Eqs. (\ref{lambda_dot}) are explicitly derived by substituting the proposed closures (\ref{K closure}).
{These equations, particularly Eq. (\ref{lambda_dot}), are reported herein solely to elucidate the analytical structure—by inspection of the right-hand sides (RHS)—of the rate of change of the various quantities. This specifically pertains to the correlation scales \(\lambda _{T}\) and \(\lambda _{\theta }\) at high Reynolds and Péclet numbers, which asymptotically converge toward a limiting solution. This asymptotic behavior serves as a useful theoretical reference to demonstrate that the transient regime is temporally bounded, lasting approximately two Lyapunov times evaluated at the initial state.
 }
\begin{figure}[H]
\centering
\includegraphics[width=9.0cm, height=13.cm]{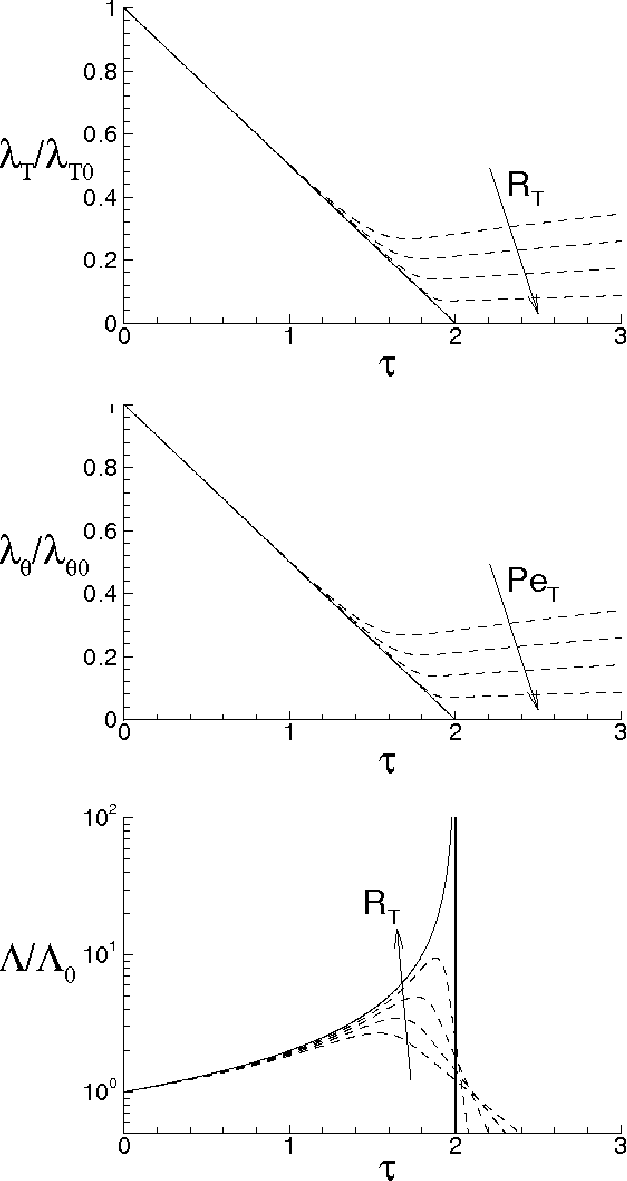}
\caption{Taylor and Corrsin microscales, along with the root-mean-square of the classical Lyapunov exponent, plotted as functions of the non-dimensional time $\tau$.}
\label{figura_3}
\end{figure}
To complement this description, it is useful to evaluate the fluctuations of the local Lagrangian Lyapunov exponent in the small-scale limit:
\bea
\begin{array}{l@{\hspace{+0.0cm}}l}
\ds \Lambda_L(t, 0) = \lim_{r \rightarrow 0} \Lambda_L(t, r).
\end{array}
\eea
By linking $f$ through Eqs. (\ref{f f_theta}) and (\ref{link0}), the root-mean-square (r.m.s.) value of $\Lambda_L(t, 0)$, denoted as $\Lambda$, is directly related to the microscales by:
\bea
\ds \Lambda(t) = \sqrt{\left\langle \Lambda_L^2(t, 0) \right\rangle_L } = \ds \frac{u}{\lambda_T}.
\eea
Based on Eqs. (\ref{lambda_dot}), we now discuss the temporal evolution of $\lambda_T$, $\lambda_\theta$, and $\Lambda$. The first terms on the right-hand side (R.H.S.) of Eqs. (\ref{lambda_dot}) represent the turbulent energy cascade mechanism, whereas the remaining terms stem directly from molecular diffusivity. While diffusion acts to increase both correlation length scales, the energy cascade continuously drives them toward smaller values. Consequently, if the cascade mechanism dominates over molecular diffusion, the conditions $d \lambda_T / dt < 0$ and $d \lambda_\theta / dt < 0$ are established.

For clarity, the idealized inviscid case ($\nu = 0$, $\kappa = 0$) is examined first. Under these conditions, $u$ and $\theta$ remain strictly constant, while $\lambda_T$, $\lambda_\theta$, and $\Lambda$ evolve with time. Specifically, the Taylor and Corrsin microscales remain proportional to each other and decrease linearly with time, governed by the dimensionless time $\tau \equiv t \Lambda(0)$:
\bea
\begin{array}{l@{\hspace{+0.0cm}}l}
\ds \frac{\lambda_T(t)}{\lambda_T(0)} \equiv 
\ds \frac{\lambda_\theta(t)}{\lambda_\theta(0)}=
1- \frac{\tau}{2}, \\\\
\ds \frac{\Lambda(t)}{\Lambda(0)}=
\frac{1}{1-\tau/2}, \\\\
\ds \tau = t \ \Lambda(0).
\end{array}
\eea
As a result, $\Lambda$ rises monotonically and diverges in a finite time. In this inviscid scenario, the energy cascade causes both microscales to systematically diminish until $\tau \rightarrow 2$, at which point the correlations are considered fully developed ($\lambda_T \rightarrow 0$, $\lambda_\theta \rightarrow 0$, and $\Lambda \rightarrow \infty$, as illustrated by the solid lines in Fig. \ref{figura_3}).
Consequently, the two correlation functions achieve their developed profiles within a finite time that scales with the initial condition $\Lambda(0)$, whereas the corresponding dimensionless time $\tau \approx 2$ exhibits a negligible dependence on the initial state. The monotonic decrease of the microscales with $\tau$ reflects the continuous spectral transfer of kinetic and thermal energy from macro- to micro-scales. As $\tau \rightarrow 2$, the divergence of $\Lambda$ implies that the velocity gradient becomes singular within a finite time dictated by $\Lambda(0)$, meaning that adjacent fluid particle trajectories separate at an exponential rate infinitely faster than the evolution of the underlying Eulerian fields.

For the viscous, diffusive case ($\nu > 0$, $\kappa > 0$), the fields undergo decay ($du/dt < 0$ and $d\theta/dt < 0$). Here, $f$ and $f_\theta$ are defined as fully developed when $d \lambda_T/dt = 0$ and $d\lambda_\theta/dt = 0$, respectively, a condition illustrated qualitatively by the dashed lines in Fig. \ref{figura_3} for varying values of $R_\lambda$ and $Pe$. 

When the initial microscales are relatively large, molecular diffusivity effects are initially overwhelmed by nonlinear convective transport. The energy cascade dominates, forcing both microscales to follow a trajectory nearly identical to the inviscid ($\nu = \kappa = 0$) case. According to Eqs. (\ref{u2dot_thetadot}) and (\ref{lambda_dot}), the evolution can be split into two sequential subregions for both $f$ and $f_\theta$. The first region corresponds to $\tau \in (0, 2)$, where $d \lambda_T/dt < 0$ and $d\lambda_\theta/dt < 0$. This domain is upper-bounded by the specific coordinates $\tau_1 \approx 2$ and $\tau_2 \approx 2$, where $d\lambda_T/dt(\tau_1) = 0$ and $d\lambda_\theta/dt(\tau_2) = 0$, respectively (with $\tau_1 \neq \tau_2$ in the general case). At these local thresholds, the kinetic and thermal energy cascades are momentarily balanced by viscosity and thermal diffusivity, marking the state where the autocorrelations are fully developed. For both fields, this transient equilibrium occurs at a finite dimensionless time $\tau \simeq 2$, scaled by the initial conditions.

Regarding the r.m.s. Lyapunov exponent $\Lambda$, it initially closely tracks the inviscid curve, reaching its maximum value at $\tau \simeq 2$ before decaying due to viscous action. The condition $d\Lambda/dt = 0$ identifies the state where chaotic dynamics, mixing, and dissipation reach their peak levels, closely matching the point where the correlations are fully developed. Beyond this peak ($d\Lambda/dt < 0$), the smallness of the microscales causes viscous dissipation to dominate over the energy cascade, leading to a progressive growth of the correlation lengths according to Eq. (\ref{lambda_dot}). This phase, which ensues immediately after the maximum of $\Lambda$, represents the classic decaying turbulence regime.

The proposed closures (\ref{K closure}) are strictly applicable only where Eq. (\ref{S>>>>S}) is satisfied, a condition under which the Navier--Stokes bifurcations generate and sustain the regime of fully developed turbulence. On the contrary, within the decaying turbulence regime characterized by $d\Lambda/dt < 0$, Eq. (\ref{S>>>>S}) will eventually cease to hold after a certain time, rendering the non-diffusive closures (\ref{K closure}) no longer valid.

In summary, the evolution of the flow field is governed by two sequential regimes. The first is a developmental regime that unfolds over a finite duration, corresponding to a dimensionless time $\tau \approx 2$ that exhibits a negligible dependence on the specific initial conditions. Regardless of the initial state, both the velocity and temperature correlation functions consistently achieve their fully developed profiles at this threshold. This phase is subsequently followed by the decay regime, during which the correlation functions exhibit a pronounced self-similarity as time progresses, with the microscales evolving asymptotically as $\lambda_T \propto \sqrt{\nu t}$ and $\lambda_\theta \propto \sqrt{\kappa t}$.

\bigskip

\section{ Numerical Scheme and Spatial-Temporal Discretization}
{

The coupled, closed von K\'arm\'an--Howarth and Corrsin equations are integrated numerically using an automated, adaptive solver compiled from standard Fortran routines. 

Spatial discretization along the $r$-coordinate is performed via a non-uniform logarithmic grid that is clustered near the origin ($r=0$) to capture the small-scale gradients characterizing the internal structure of the flow. Within the interior domain, numerical derivatives are evaluated using second-order central finite difference schemes adapted for non-uniform grid spacings. At the boundaries of the computational domain, the solver implements forward and backward finite difference formulas on the non-uniform mesh, maintaining second-order formal accuracy at the boundary treatments. 

To resolve the relevant physical scales throughout the decaying process, the minimum spatial step size $\Delta r_{min}$ at the origin is dynamically adapted to the instantaneous small-scale physics via the relation:
\bea
\Delta r_{min} = C \cdot l_m,  \  \  \mbox{with} \ C = 0.2  \  \  \mbox{and}  \ \  l_m = \min(\eta_K, \eta_B),
\eea
where $\eta_K$ and $\eta_B$ denote the Kolmogorov and Batchelor microscales, respectively. The total number of spatial grid nodes is computed based on the physical width of the spatial domain. The algorithm verifies that the required grid density remains below a maximum threshold, which corresponds to the pre-allocated dimension of the internal array structures.

Temporal marching is advanced via a standard fourth-order Runge--Kutta (RK4) integration scheme. To address mathematical stability and manage error bounds, an automated time-step control loop is implemented. The variable time-step $\Delta t$ is adjusted dynamically at each instance based on the local truncation error criterion inherent to adaptive RK4 methods, aiming to track the turbulence decay within controlled error bounds.
}

\bigskip

\section{Results and Discussion}

This section presents a numerical assessment of the Lyapunov--Liouville theoretical framework based on the integration of the closed von K\'arm\'an--Howarth and Corrsin systems, as defined by equations (\ref{vk-h}) and (\ref{u2dot_thetadot}) along with the proposed closure relations (\ref{K closure}) and the boundary conditions (\ref{bc}). To evaluate the performance of the model under non-stationary conditions, the analysis investigates the evolution of the fluid velocity and passive scalar fields starting from three distinct initial states, characterized by Saffman--Birkhoff, Loitsiansky, and Gaussian correlation profiles. 

The PDFs of the velocity and temperature increments are computed from the aforementioned numerical simulations through equations (\ref{struct funs}) and (\ref{Frobenius-Perron}). This is achieved via a dedicated computing code that generates random values of $\zeta_X$, $\zeta_{X -}$, and $\zeta_{X +}$ distributed according to a Gaussian PDF with unit standard deviation. As prescribed by the present theoretical analysis, this code requires as inputs the velocity increment skewness $H^{(3)}_u(r)$, $R_\lambda$, and $Pr$ obtained from the preceding simulations.

The numerical solutions obtained across the investigated Prandtl numbers ($Pr = 0.1, 1, 10, \mbox{ and } 1000$) are subsequently compared with independent direct numerical simulations (DNS) and experimental datasets available in the literature, showing a satisfactory alignment with established benchmarks.

Three standard cases were investigated based on different initial conditions. The first case imposes a Saffman--Birkhoff initial condition for the correlation function $f$:
\bea
\begin{array}{l@{\hspace{+0.0cm}}l}
\ds f(0, r) = \left( 1+ \frac{1}{3} \frac{r^2}{\lambda_T^2} \right)^{-3/2} \approx \frac{1}{r^3}, 
\end{array}
\label{SB}
\eea 
The second case prescribes a Loitsiansky initial condition for $f$:
\bea
\begin{array}{l@{\hspace{+0.0cm}}l}
\ds f(0, r) = \left( 1+ \frac{1}{5} \frac{r^2}{\lambda_T^2} \right)^{-5/2}\approx \frac{1}{r^5}, 
\end{array}
\label{L}
\eea 
Conversely, the third case corresponds to a Gaussian correlation function:
\bea
\begin{array}{l@{\hspace{+0.0cm}}l}
\ds f(0,r) = \exp\left( - \frac{1}{2} \frac{r^2}{\lambda_T^2}  \right)  \\\\
\end{array}
\label{G}
\eea 
For the scalar correlation $f_\theta$, the initial condition in all cases is expressed by:
\bea
\begin{array}{l@{\hspace{+0.0cm}}l}
\ds f_\theta(0, r) = \left( 1+ \frac{2}{3} \frac{r^2}{\lambda_\theta^2} \right)^{-3/2}\approx \frac{1}{r^3}, 
\end{array}
\eea 
The initial conditions for the standard deviations of velocity and temperature, the correlation microscales, and the Taylor-scale Reynolds number are identical across all cases and are given by:
\bea
\begin{array}{l@{\hspace{+0.0cm}}l}
\ds u^2(0)=1, \ \ \lambda_T(0)=1, \\\\
\ds \theta^2(0)=1, \ \ \lambda_\theta(0)=1, \\\\
\ds R_\lambda(0) = 2000
\end{array}
\eea
Subsequently, the effect of $Pr$ on the temperature spectrum and on the statistics of the temperature increment is investigated. For this purpose, we initially assume $Pr=1$ and then vary the Prandtl number to consider $Pr=0.1$, $10$, and $1000$.
\begin{figure}[t]
	\centering
	\includegraphics[width=100mm, height=100mm,]{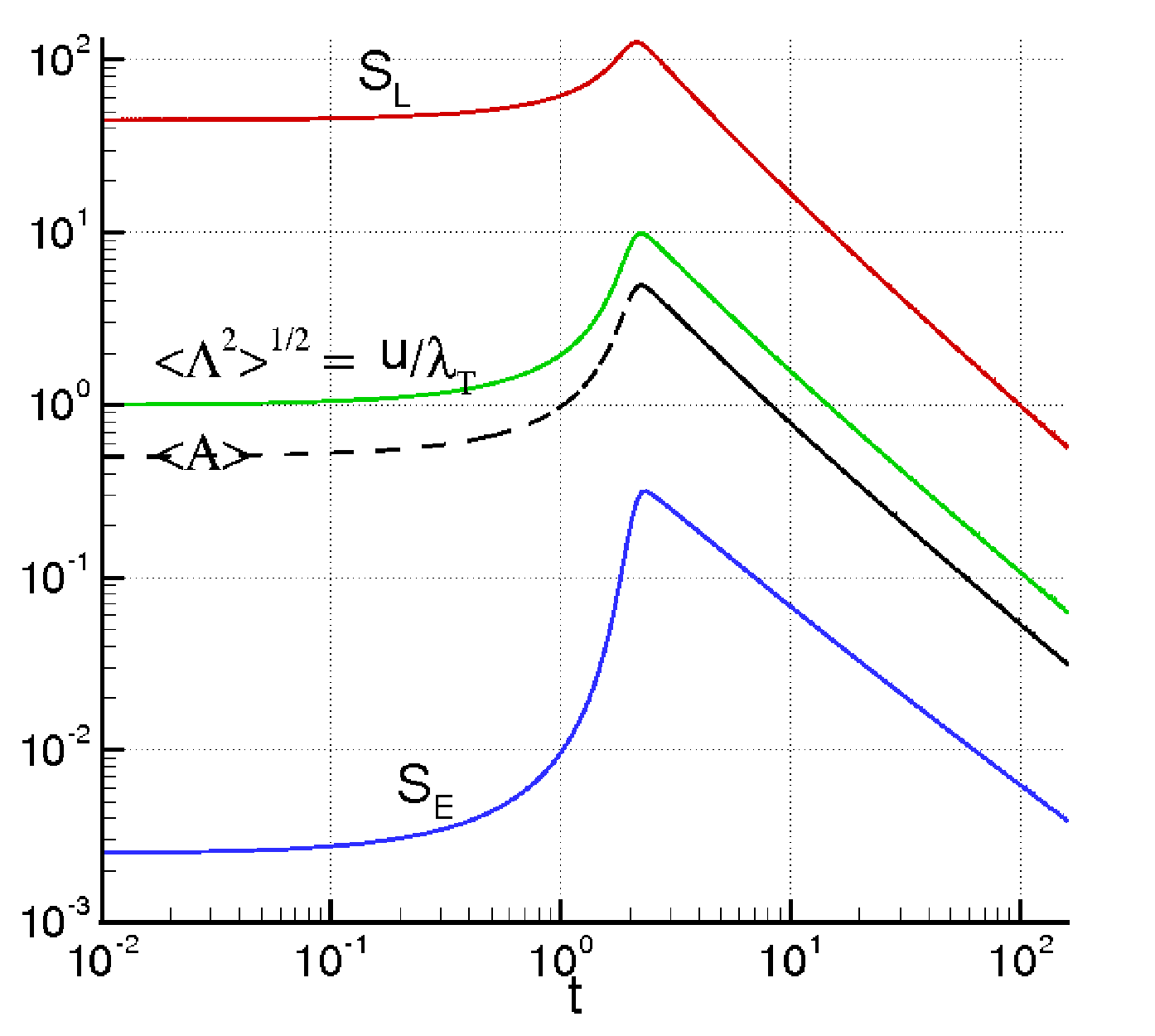}
\caption{Evolution of the Lagrangian Lyapunov exponents $\left< \Lambda_L \right>$ and $\left< \Lambda_L^2 \right>^{1/2}$ and bifurcation rates $S_E$ and $S_L$ for the Saffman--Birkhoff initial condition (Eq. \ref{SB}).}
\label{f1}
\end{figure}

{Before analyzing these numerical simulations in detail, it is instructive to establish the primary benchmark for our validation. 
The constant skewness value of the longitudinal velocity derivative, obtained with the present Lyapunov–Liouville theory, quantifies the effect of the energy cascade in isotropic turbulence. This metric is specifically selected for validation because it directly probes the primary origin of the energy cascade, to which the remaining salient quantities of the turbulent motion are physically subordinate; hence, a comparison of other homologous variables would inherently yield nearly identical alignment. This value is consistent with several results obtained through direct numerical simulations (DNS) of the Navier–Stokes equations \cite{Chen92, Orszag72, Panda89}, where it ranges from \(-0.47\) to \(-0.40\), as well as large-eddy simulations (LES) \cite{Anderson99, Carati95, Kang2003}, which yield values between \(-0.42\) and \(-0.40\). For the reader's convenience, Table 1 recalls the comparison—originally presented in \cite{deDivitiis2016}—between the skewness \(H^{(3)}_u(0)\) obtained in this analysis and the values from the aforementioned studies. The maximum absolute difference between the present value and the literature data is less than 10 \(\%\). Consequently, the hypotheses of the present analysis, and the closures (\ref{K closure}), appear to be adequate for modeling the turbulent energy cascade and spectra.
\begin{table}[h]
\centering
\caption{Comparison of the results: Skewness of $\partial u_r/ \partial r$ at various Taylor--scale Reynolds number $R_\lambda \equiv u \lambda_T/\nu$ following different authors.}
\vspace{2. mm}
\begin{tabular}{cccc} 
\hline
\hline
Reference   &  Simulation  &  $R_\lambda$  & $H^{(3)}_u(0)$  \\
\hline 
Present analysis      &  -  &  -    & -3/7 = -0.428...  \\
\cite{Chen92}       & DNS &    202 & -0.44   \\ 
\cite{Orszag72}      & DNS &    45 & -0.47    \\
\cite{Panda89}      & DNS &    64 & -0.40    \\
\cite{Anderson99}   & LES &    $<$ 71 &  -0.40          \\ 
\cite{Carati95}     & LES &   $\infty$     &  -0.40 \\  
\cite{Kang2003}     & LES &  720        &    -0.42      \\ 
\hline
\hline
 \end{tabular}
\label{table1}
\end{table} }

We now analyze the first case given by Eq.~(\ref{SB}) with $Pr=1$. Figure~\ref{f1} illustrates the variation laws of $\Lambda=u/\lambda_T$, $\langle \Lambda_L \rangle$, and the Eulerian and Lagrangian bifurcation rates as functions of time $t$, where the latter are computed via equations~(\ref{b_rates}). The time $t$ plotted in this figure and in all subsequent figures is measured in units of the initial Lyapunov time scale $1/\Lambda(0)=\lambda_T(0)/u(0)$. As observed from the plot, Eq.~(\ref{S>>>>S}) is satisfied over the entire simulation interval, thereby establishing the necessary condition for the validity of the present simulation and the proposed closures. Specifically, all quantities shown in the figure initially increase, reaching a maximum at approximately $t \approx 2$, where the levels of chaos and dissipation attain their peak values, and subsequently decrease. In particular, for this Saffman--Birkhoff initial condition, the various curves of the Lyapunov exponents and bifurcation rates appear nearly parallel to one another for $t>2$, thus providing the necessary condition for the validity of the present closures and simulation well beyond $t=2$. As previously mentioned, at $t \approx 2$, the correlation functions and the turbulent kinetic energy spectrum can be considered fully developed.

\begin{figure}[t]
	\centering
	\includegraphics[width=150mm, height=100mm,]{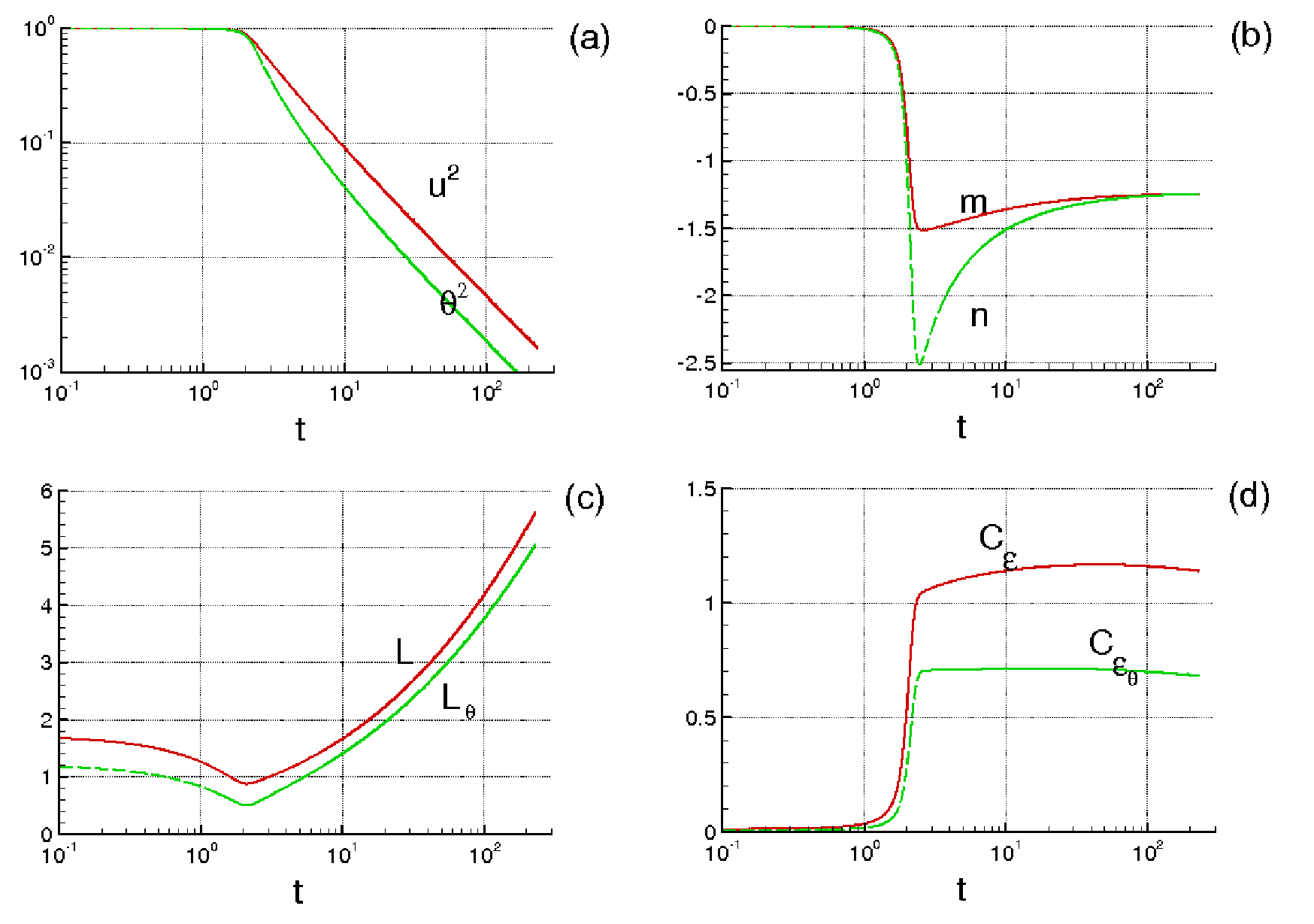}
	\caption{Evolution of physical quantities as a function of time starting from the Saffman--Birkhoff initial condition \ref{SB}. (a) Velocity and temperature standard deviations, (b) characteristic exponents of velocity and temperature, (c) integral scales, and (d) dissipation coefficients.}
\label{f2}
\end{figure}

Figure~\ref{f2}a shows the temporal variations of the velocity (red line) and temperature (green line) standard deviations, whereas Fig.~\ref{f2}b reports the characteristic exponents $m$ and $n$ associated with the power laws:
\bea
\ds u^2(t) \propto t^m, \ \ \ \theta^2(t) \propto t^n
\label{u2t2}
\eea 
Up to $t \approx 2$, marginal variations in $u^2$ and $\theta^2$ are observed due to the low dissipation rates stemming from the relatively large values of $\lambda_T$ and $\lambda_\theta$. Subsequently, a significant monotonic decay of both $u^2$ and $\theta^2$ occurs, driven by the reduction of the correlation scales and the consequent high dissipation levels. { The exponents $m$ and $n$ asymptotically tend to values close to $m \simeq -1.25$ and $n \simeq -1.25$, which are in satisfactory alignment with several literature references, including the work of Comte-Bellot \cite{Comte-Bellot1966}. }

\begin{figure}[t]
	\centering
	\includegraphics[width=150mm, height=100mm,]{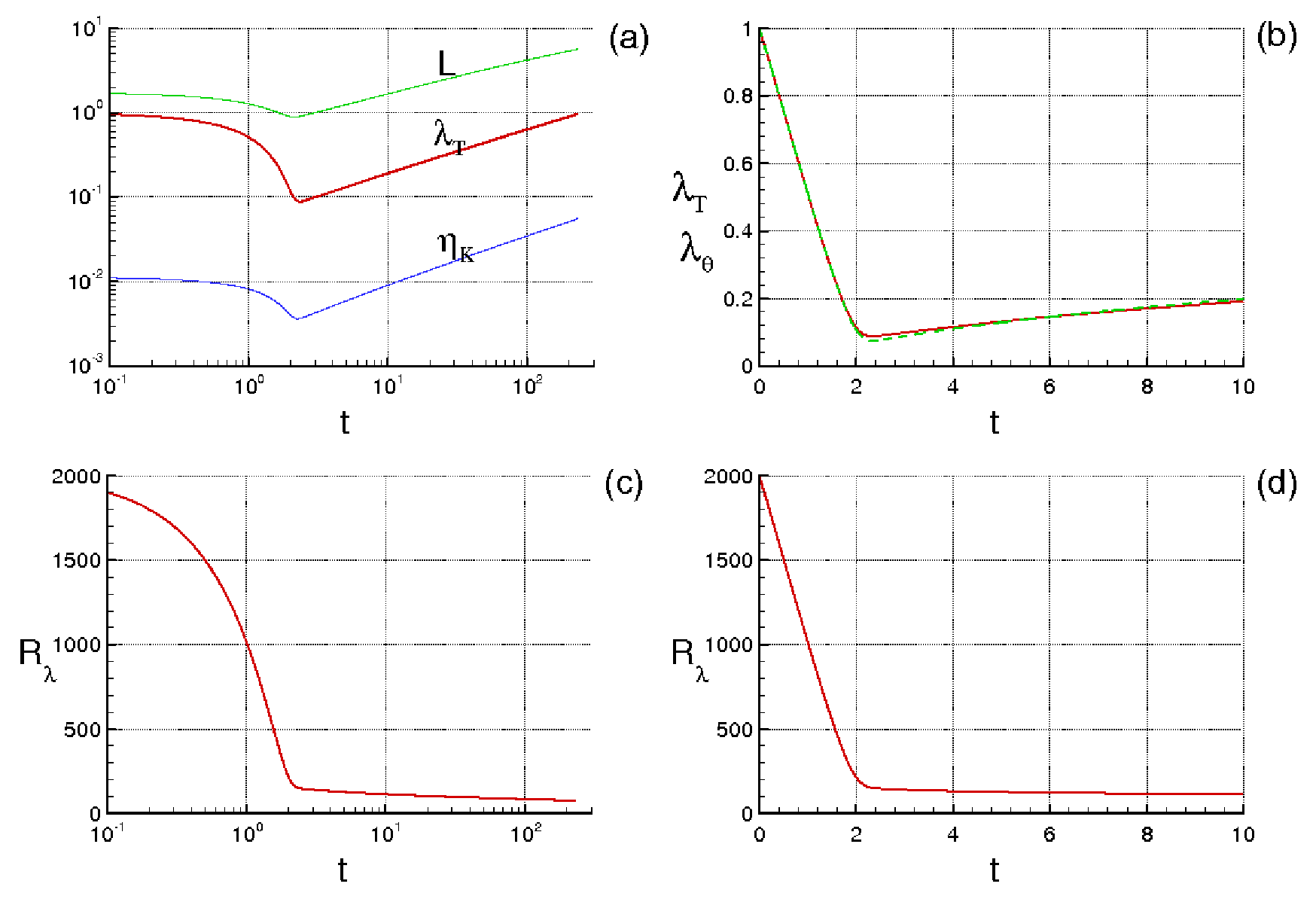}
	\caption{Evolution of physical quantities as a function of time starting from the Saffman--Birkhoff initial condition \ref{SB}. (a) Integral scale, Taylor scale, and Kolmogorov scale, (b) magnified view of the Taylor and Corrsin microscales, (c) Taylor-scale Reynolds number, and (d) magnified view of the Taylor-scale Reynolds number.}
\label{f3}
\end{figure}
The variations of the integral scales $L$ and $L_\theta$ for the respective velocity and temperature correlation functions are also reported (Fig.~\ref{f2}c). The evolution of these length scales, along with the previously discussed variations of $u^2$ and $\theta^2$, allows for the identification of the dissipation coefficients $C_\epsilon$ and $C_{\epsilon_\theta}$, defined according to the following relations \cite{Taylor1935, Obukhov, corrsin1951spectrum}:
\bea
\begin{array}{l@{\hspace{+0.0cm}}l}
\ds C_\epsilon  =  \frac{\epsilon L}{u^3}, \ \ \
\epsilon=15 \nu \frac{u^2} {\lambda_T^2} \equiv 15 \nu \Lambda^2, \\\\
\ds C_{\epsilon_\theta}  =  \frac{\epsilon_\theta L_\theta}{u \ \theta^2},  \ \ \
\epsilon_\theta=12 \kappa \frac{\theta^2}{\lambda_\theta^2},
\end{array}
\eea 
{ In the diffusive region, these coefficients exhibit an initial zone of sharp variation followed by a plateau region where $C_\epsilon \simeq 1.16$ and $C_{\epsilon_\theta} \simeq 0.72$. Given that $C_\epsilon$ and $C_{\epsilon_\theta}$ are not universal constants, the values obtained here can be considered reasonably aligned with known literature data \cite{Sreenivasan1984, Vassilicos2015}.}
\begin{figure}[h!]
	\centering
	\includegraphics[width=150mm, height=100mm,]{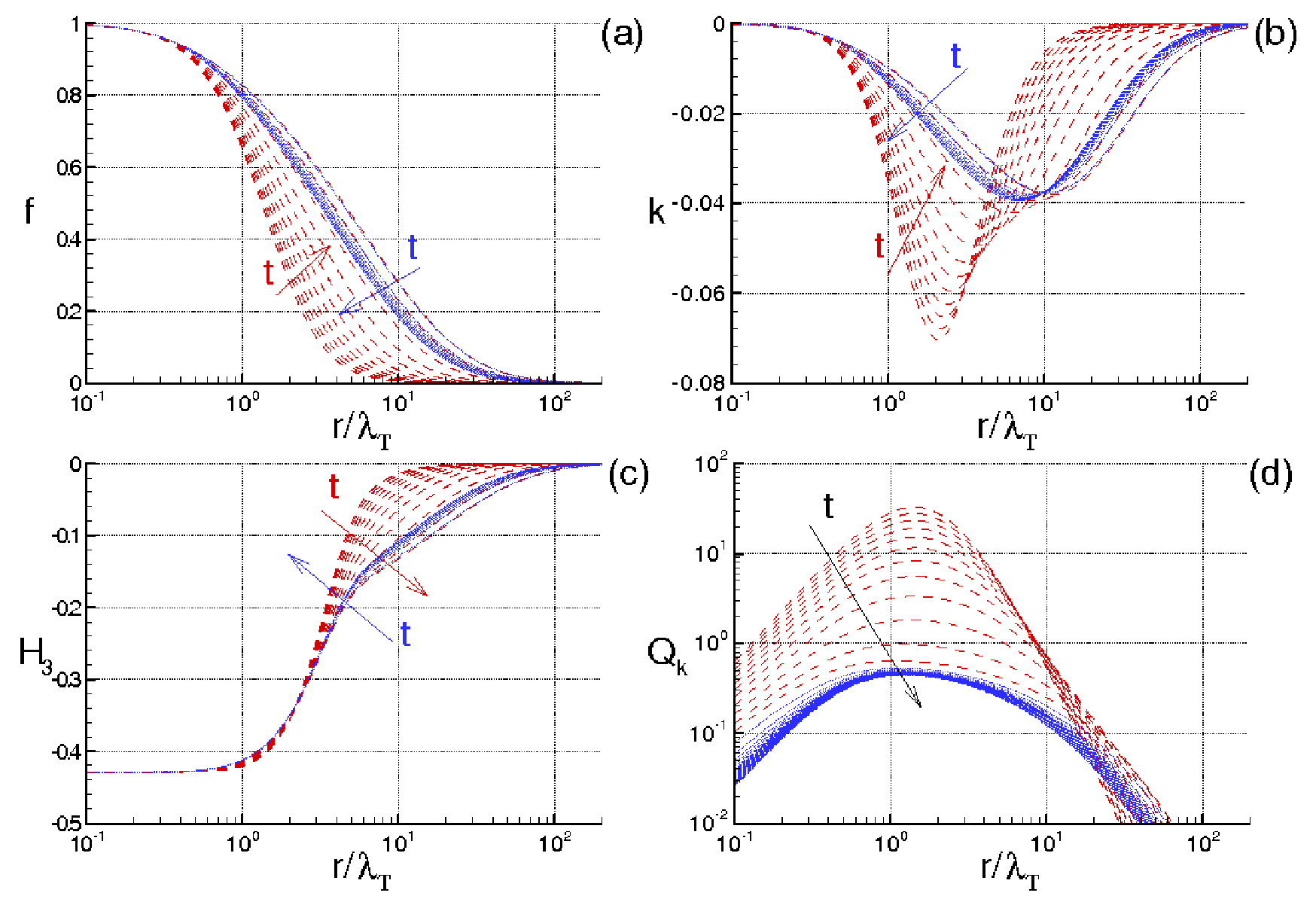}
\caption{Spatial variation laws at different time instances for the Saffman--Birkhoff initial condition \ref{SB}. (a) Longitudinal two-point velocity autocorrelation, (b) longitudinal two-point triple velocity correlation, (c) skewness of the longitudinal velocity increment, and (d) spatial Kolmogorov function. Red dashed lines represent the developing transient regime from $t=0$ to $t=2.2$ with a constant time increment of $\Delta t=0.2$ between successive curves, where the red arrow indicates the direction of increasing time. Blue dotted lines describe the diffusive regime at subsequent time instances $t > 2.2$, where the blue arrow denotes the direction of increasing time.}
\label{f4}
\end{figure}

Figure~\ref{f3} illustrates and summarizes the temporal variations of $L$, $\lambda_T$, and $\eta_k$ (Fig.~\ref{f3}a), while Fig.~\ref{f3}b shows a magnified view of the Taylor and Corrsin microscales as functions of time. Figures~\ref{f3}c and \ref{f3}d display the temporal variations of $R_\lambda$, along with a zoom of the $R_\lambda$ behavior within the time interval $(0, 10)$, respectively. These plots clearly demonstrate the presence of two distinct regimes, namely the developing and diffusive regimes, which are characterized by the aforementioned variations in the correlation scales and the corresponding Reynolds numbers. Within the developing regime, a nearly linear decrease of both $R_\lambda$ and $\lambda_T$ is observed as a function of time: $\lambda_T$, starting from $\lambda_T=1$ at $t=0$, decays to $\lambda_T \approx 0.2$ at $t \simeq 2$. Owing to $Pr=1$, $\lambda_\theta$ exhibits a very similar trend. Regarding $R_\lambda$, starting from $R_\lambda=2000$ at $t=0$, it drops to approximately $R_\lambda \simeq 165$ at $t \simeq 2$. According to the proposed model, this behavior implies that during the dimensionless time interval $(0, 2)$---corresponding to the developing regime---there is a pronounced mean flux of kinetic and thermal energy across scales, which corresponds to an energy transfer from large to small scales (energy cascade). Subsequently, for $t>2$, the diffusive regime is established, where $\lambda_T \propto \sqrt{\nu t}$ and $\lambda_\theta \propto \sqrt{\kappa t}$, while the Reynolds number tends to preserve its value. This preservation is a consequence of the Saffman--Birkhoff initial condition, which, as is well known, causes the diffusive term in the von K\'arm\'an--Howarth equation to vanish as $r \rightarrow \infty$.

Figure~\ref{f4} displays, at different time instances, the spatial variation laws as functions of $r$ for $f$, $k$, $H^{(3)}_u$, and the spatial Kolmogorov function $Q_k(r)$ defined as:
\bea
\ds Q_k(r) = - \frac{\langle \Delta u_r^3 \rangle_E}{r \epsilon}
\eea 
The dashed red curves correspond to the developing regime, with each curve representing the dimensionless times 0, 0.2, 0.4, 0.6, 0.8, 1.0, 1.2, 1.4, 1.6, 1.8, 2.0, and $2.2$, following the direction of the dashed red arrow. Conversely, the blue curves refer to the subsequent diffusive phase, where the blue arrow indicates the direction in which $t$ rises. { At the initial condition, the maximum value of $\vert k \vert$ is approximately $0.07$, which is attained near the Taylor microscale. In the subsequent instances, it decreases to approximately $0.04$ at $t \approx 2$ for separation distances on the order of $10 \lambda_T$. These values, in satisfactory alignment with known literature data \cite{Pope2000}, are connected to the proposed closure (\ref{K closure}), which yields a skewness at the origin of $H^{(3)}_u(0)=-3/7$.} This asymmetry forces the $k(r)$ curve to bend and form a hump whose extreme value settles precisely around the aforementioned $0.04 - 0.07$ threshold, while simultaneously expanding the interaction scale. 
During the same period, $H^{(3)}_u$ evolves accordingly, matching the value at the origin $H^{(3)}_u(0)=-3/7$. Specifically, both $H^{(3)}_u$ and $k$ exhibit a significantly broader correlation compared to the initial condition, a hallmark feature of the large-scale interactions characterizing turbulence. Initially, the variations of $H^{(3)}_u$ occur over a scale on the order of $\lambda_T$, whereas at $t \simeq 2$, the variations of $H^{(3)}_u$ take place over a much larger scale, on the order of $20 \div 70 \  \lambda_T$. Consequently,  $H^{(3)}_u(r)$ radically changes its shape relative to the initial condition, approaching curves belonging to the diffusive regime (in blue) that tend to overlap with each other. Regarding the Kolmogorov function, its maximum value of $4/5$ is reached at $t \simeq 2.1$ within the range $1 < r/\lambda_T < 2$, where both $d \lambda_T/dt \simeq 0$ and $dL/dt \simeq 0$. In this case, an extended plateau is not observed; instead, a relatively flat maximum appears. This behavior is likely due to the fact that the present analysis does not investigate stationary turbulence, but rather a turbulent regime in which the turbulent kinetic energy decays over time. Such nonstationarity can significantly alter the spatial Kolmogorov function. During the diffusive phase, a maximum value of $Q_k(r)$ around $0.55$ is observed, which persists at subsequent time instances.
\begin{figure}[h!]
	\centering
	\includegraphics[width=150mm, height=100mm,]{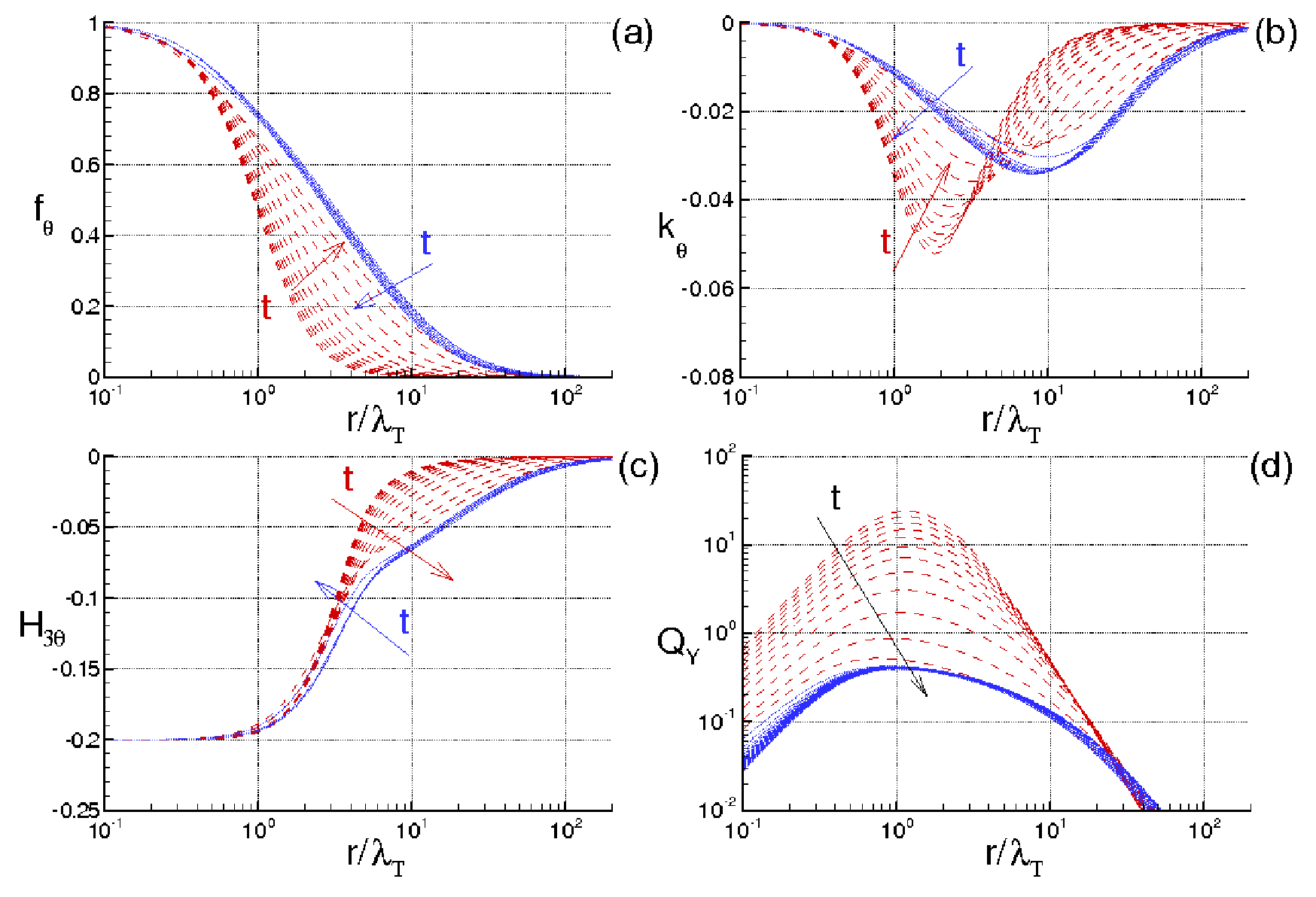}
\caption{Spatial variation laws at different time instances for the Saffman--Birkhoff initial condition \ref{SB}. (a) Two-point temperature autocorrelation, (b) mixed two-point triple temperature--velocity correlation, (c) mixed non-dimensional third-order moment of the temperature--velocity increments, and (d) Yaglom function. Red dashed lines represent the developing transient regime from $t=0$ to $t=2.2$ with a constant time increment of $\Delta t=0.2$ between successive curves, where the red arrow indicates the direction of increasing time. Blue dotted lines describe the diffusive regime at subsequent time instances $t > 2.2$, where the blue arrow denotes the direction of increasing time. }
\label{f5}
\end{figure}

Figure~\ref{f5} shows, at different time instances, the spatial variation laws as functions of $r$ for $f_\theta$, $k_\theta$, $H^{(3)}_\theta$, and the Yaglom function $Q_Y(r)$ defined as:
\bea
\ds Q_Y(r) = - \frac{\langle \Delta \vartheta^2 \Delta u_r \rangle_E}{r \epsilon_\theta}
\eea 
As before, the dashed red curves denote the developing regime, with each curve corresponding to the dimensionless times 0, 0.2, 0.4, 0.6, 0.8, 1.0, 1.2, 1.4, 1.6, 1.8, 2.0, and $2.2$, following the direction of the dashed red arrow, while the blue curves represent the subsequent diffusive phase. Because $Pr=1$, the variation laws observed are highly similar to those just described for the longitudinal velocity correlations. { At the initial condition, the maximum value of $\vert k_\theta \vert$ is approximately $0.05$, whereas at subsequent time instances, it decreases to approximately $0.033$ at $t \approx 2$. These values are in satisfactory alignment with literature data \cite{Warhaft2000}.}
During the same period, $H^{(3)}_\theta(r)$ evolves accordingly, matching in all cases the boundary condition at the origin, $H^{(3)}_\theta(0)=-1/5$. In particular, both $H^{(3)}_\theta$ and $k_\theta$ show a significantly broader correlation than the initial condition, which is typical of the large-scale interactions that characterize turbulence. Initially, the variations of $H^{(3)}_\theta$ occur on a scale of the order of $\lambda_\theta$, whereas at $t \simeq 2$, the variations of $H^{(3)}_\theta$ occur on much larger scales, on the order of $5 \div 60 \lambda_T$.
It is thus observed that $H^{(3)}_\theta(r)$ (reported as $H_{\theta 3}(r)$) changes shape radically compared to the initial condition, shifting toward the diffusive regime curves (in blue) which tend to overlap one another. Regarding the Yaglom function, its maximum value of $2/3$ is achieved at $t \simeq 2.1$ for $r/\lambda_T \approx 1$, where both $d \lambda_\theta/ dt \simeq 0$ and $dL_\theta/dt \simeq 0$. In this case, rather than an extended plateau, a relatively flat maximum is observed. This is likely due to the fact that the current analysis does not consider stationary turbulence, but a turbulent regime where the turbulent thermal energy decays over time. This nonstationarity can significantly modify the Yaglom function. During the diffusive phase, a maximum value of $Q_Y(r)$ of around $0.45$ is observed, a value that is maintained at subsequent times.
\begin{figure}[t]
	\centering
	\includegraphics[width=150mm, height=80mm,]{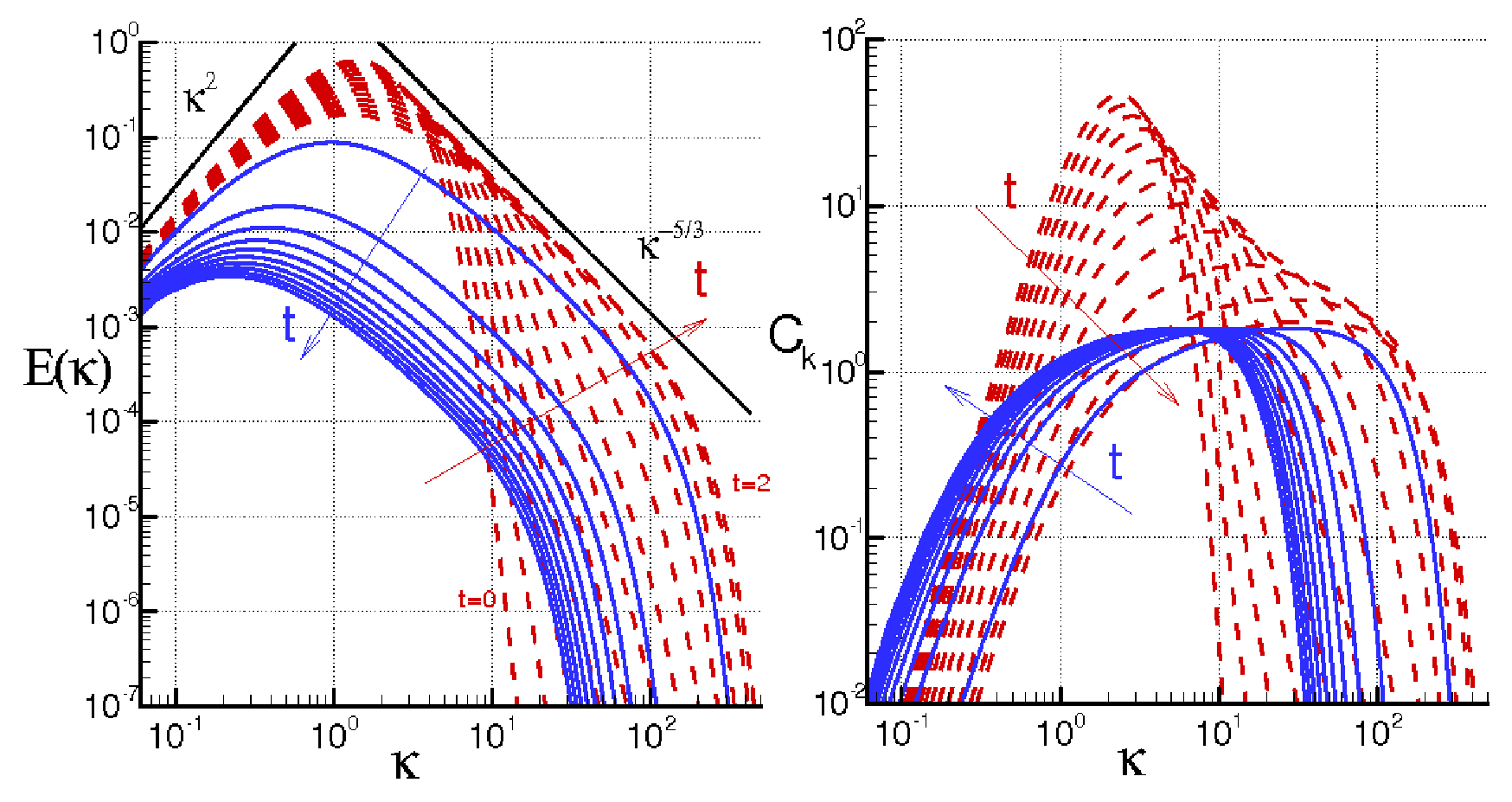}
	\caption{Turbulent kinetic energy spectra (a) and compensated Kolmogorov spectral functions (b) evaluated at different time instances for the Saffman--Birkhoff initial condition \ref{SB}.  Red dashed lines represent the developing transient regime from $t=0$ to $t=2.2$ with a constant time increment of $\Delta t=0.2$ between successive curves, where the red arrow indicates the direction of increasing time. Blue lines describe the diffusive regime at subsequent time instances $t > 2.2$, where the blue arrow denotes the direction of increasing time.  }
\label{f6}
\end{figure}

The subsequent figures show the corresponding results in terms of the energy spectra $E(t, k)$ and $E_\theta(t, k)$, which are computed from $f$ and $f_\theta$, respectively, using the following relations:
\bea
\begin{array}{l@{\hspace{+0.0cm}}l}
\ds E(t, k) 
= 
 \frac{u^2}{\pi} 
 \int_0^{\infty} f(t, r)
 k^2 r^2 
\left( \frac{\sin k r }{k r} - \cos k r  \right) d r, \\\\
\ds E_\theta(t, k) 
= 
 \frac{2 \theta^2}{\pi} 
 \int_0^{\infty} 
 \ds  f_\theta(t, r) 
k r \sin k r \ dr
\end{array}
\label{Ek}
\eea
and the compensated Kolmogorov and Obukhov--Corrsin spectral functions, defined as follows:
\bea
\begin{array}{l@{\hspace{+0.0cm}}l}
\ds\hat{C}_K(k)= \frac{E(k) k^{5/3}}{\epsilon^{2/3}}, \\\\
\ds \hat{C}_{OC}(k)= \frac{E_\theta(k) k^{5/3}}{\epsilon_\theta} \epsilon^{1/3}
\end{array}
\eea
As before, the red curves refer to the developing regime, where each individual curve is computed at the same dimensionless time instances previously defined. Regarding the kinetic energy spectrum $E(k)$, Figure~\ref{f6} illustrates a $k^2$ behavior at the origin, which stems from the Saffman--Birkhoff initial condition and is preserved at subsequent time instances. During the developing regime (red curves), kinetic energy is gradually transferred from large scales to small scales starting from the initial condition. At $t \approx 2$, the fully developed spectrum extends to the maximum wavenumber, and subsequently decays in the following diffusive regime (blue curves), where the inertial range gradually decreases in width. It can be observed that at the end of the developing regime ($t \approx 2$), the envelope of the energy spectra runs perfectly parallel to the $k^{-5/3}$ line that describes Kolmogorov's law. Both plots (a) and (b) reveal that the Kolmogorov $k^{-5/3}$ regime is limited in width. The compensated function $\hat{C}_K$ does not exhibit an extended plateau, but rather a relatively flat maximum.  { This behavior can be attributed to the fact that the present analysis does not consider stationary, forced turbulence, but rather a decaying turbulent flow where the kinetic energy continuously decreases. Nevertheless, the maximum value of the Kolmogorov function settles at $C_K \simeq 1.72$, which is in good agreement with literature results for decaying turbulence \cite{Wang1996}.}
\begin{figure}[t]
	\centering
	\includegraphics[width=150mm, height=80mm,]{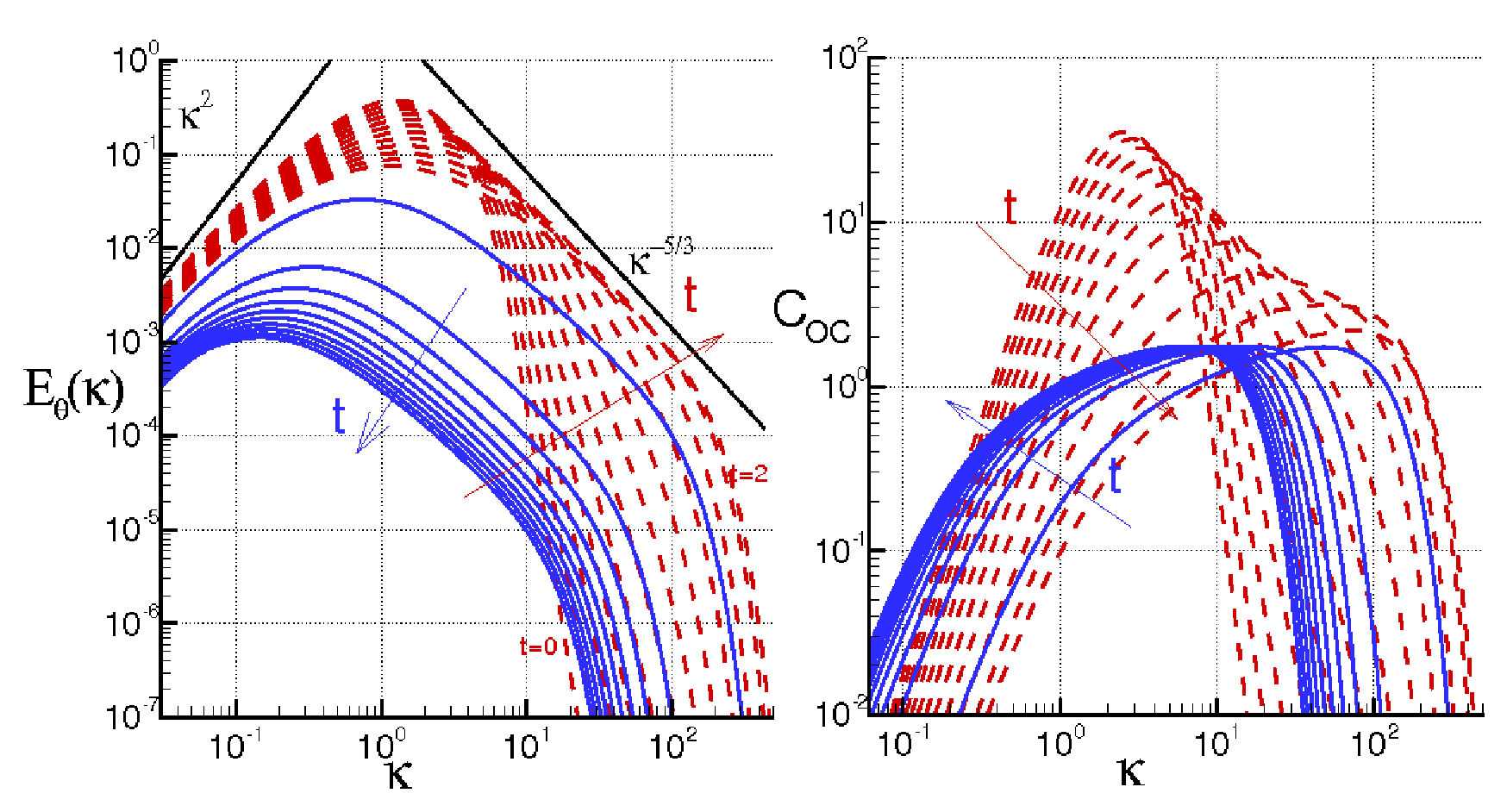}
	\caption{(a) Temperature spectra at different time instances. (b) Compensated Obukhov--Corrsin spectral functions at different time instances for the Saffman--Birkhoff initial condition \ref{SB}.  Red dashed lines represent the developing transient regime from $t=0$ to $t=2.2$ with a constant time increment of $\Delta t=0.2$ between successive curves, where the red arrow indicates the direction of increasing time. Blue lines describe the diffusive regime at subsequent time instances $t > 2.2$, where the blue arrow denotes the direction of increasing time.   }
\label{f7}
\end{figure}

Regarding the temperature spectrum $E_\theta(k)$, Fig.~\ref{f7} also shows a $k^2$ behavior near the origin due to the Saffman--Birkhoff initial condition, which tends to be preserved at subsequent time instances. During the developing regime (red curves), thermal energy is gradually transferred from large scales to small scales starting from the initial condition. At $t \approx 2$, the spectrum extends to the maximum wavenumber and subsequently decays in the following diffusive regime (blue curves), where the inertial range of the temperature spectrum gradually decreases in width. At the end of the developing regime ($t \approx 2$), the envelope of the temperature spectra runs perfectly parallel to the $k^{-5/3}$ line that describes Kolmogorov's law. Both plots (a) and (b) reveal that the Kolmogorov $k^{-5/3}$ range is nonetheless limited in width. Indeed, the compensated function $C_{OC}$ does not exhibit an extended plateau, but rather a relatively flat maximum. As noted above, this behavior can be attributed to the fact that the present analysis does not consider stationary, forced turbulence, but rather a decaying turbulent flow where the kinetic energy continuously decreases. { Nevertheless, the maximum value of the Obukhov--Corrsin function settles around $C_{OC} \simeq 1.73$, a value that closely matches the data presented in \cite{briard2016passive} concerning decaying turbulence, and is consistent with the results reported in \cite{sreenivasan1996passive}.}

\begin{figure}[t]
	\centering
	\includegraphics[width=150mm, height=50mm,]{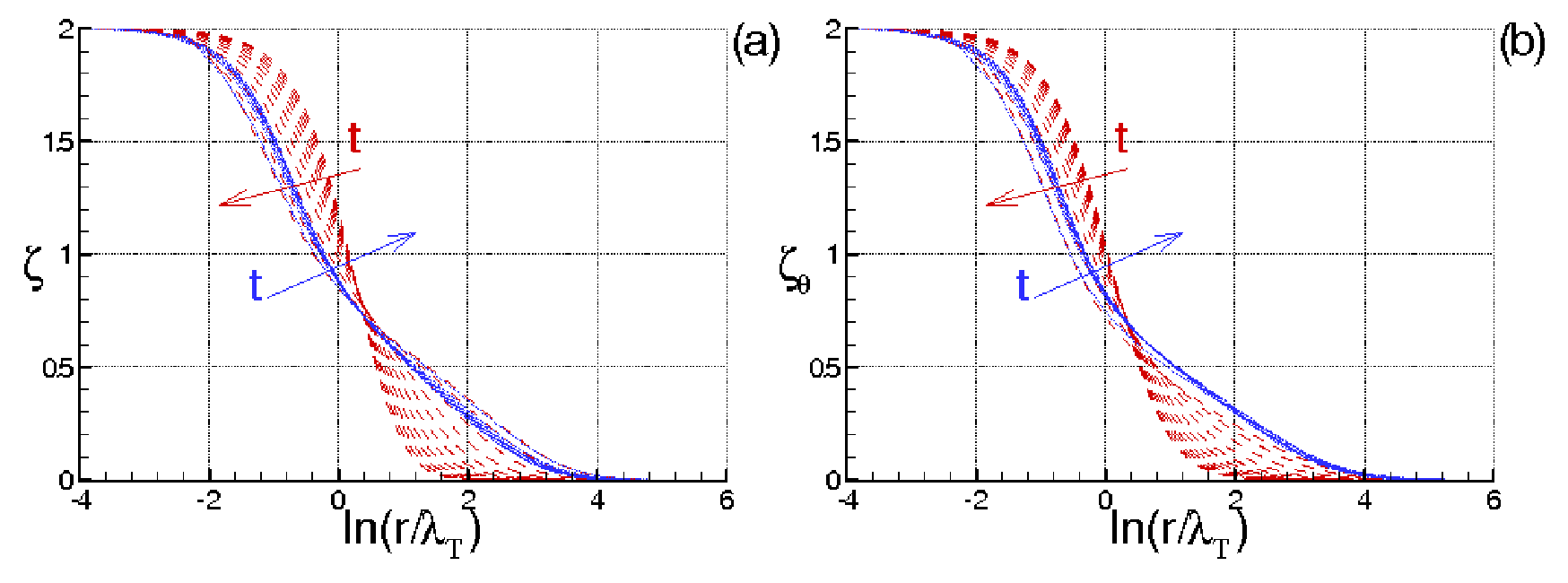}
\caption{Local scaling exponents of (a) velocity correlation and (b) temperature correlation evaluated at different time instances for the Saffman--Birkhoff initial condition \ref{SB}.  Red dashed lines represent the developing transient regime from $t=0$ to $t=2.2$ with a constant time increment of $\Delta t=0.2$ between successive curves, where the red arrow indicates the direction of increasing time. Blue dotted lines describe the diffusive regime at subsequent time instances $t > 2.2$, where the blue arrow denotes the direction of increasing time.  }
\label{f8}
\end{figure}

Figure~\ref{f8} illustrates the spatial behavior as a function of $r$, at different time instances, of the local scaling exponents for both correlation functions. These exponents are defined as:
\bea
\begin{array}{l@{\hspace{+0.0cm}}l}
\ds Z= r \frac{\partial}{\partial r} \ln \langle \Delta {\bf u} \cdot \Delta {\bf u} \rangle_E, \\\\
\ds Z_\theta = r \frac{\partial}{\partial r} \ln \langle (\Delta \vartheta)^2  \rangle_E, 
\end{array}
\eea
During the development phase of the autocorrelations, the local exponents $Z$ and $Z_\theta$ evolve with visible variations. Subsequently, during the decay phase, the different curves tend to collapse into a single one. No Kolmogorov-like plateau is observed around the value $Z=2/3$. { However, an oblique inflection point appears near the value $Z=Z_\theta=2/3$, where the absolute value of the local slope for both curves is relatively moderate. This behavior of $Z$ is entirely in line with observations in \cite{Sreenivasan2026, Migdal2026}.}
\begin{figure}[t]
	\centering 
	 \includegraphics[width=150mm, height=100mm,]{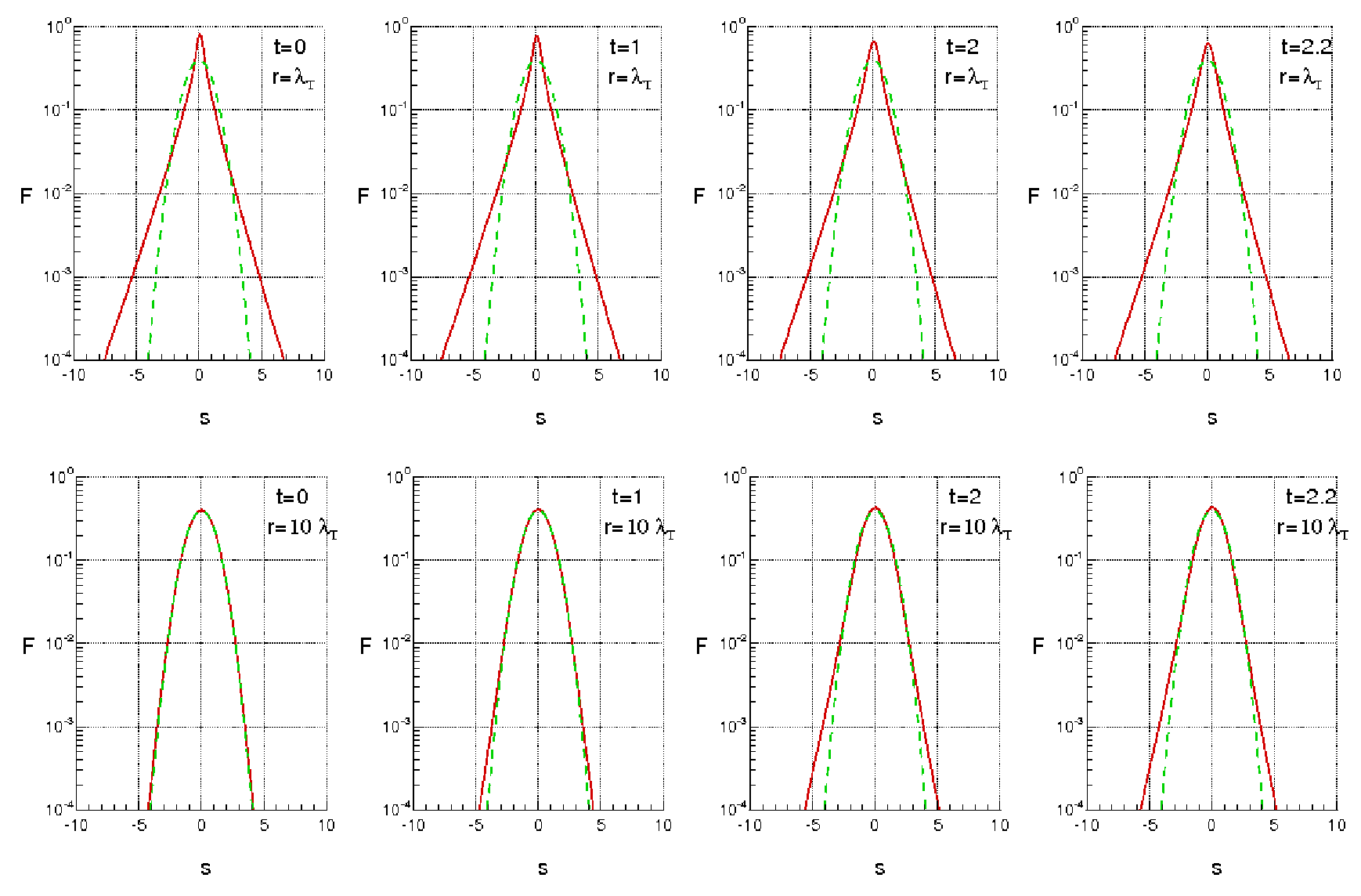}
\caption{Evolution of the PDF of the longitudinal velocity increment for $r = \lambda_T$ (top) and $r = 10\lambda_T$ (bottom) at different time instances for the Saffman--Birkhoff initial condition \ref{SB}. Green dashed lines represent the Gaussian PDF centered and with the same standard deviation value.}
\label{f9}
\end{figure}

Figure~\ref{f9} illustrates the PDF of the velocity increment for $r/\lambda_T=1$ (top) and $r/\lambda_T=10$ (bottom) during the developing regime period ($t \in (0, 2.2)$). The PDFs are expressed as a function of the dimensionless variable:
\bea
\ds s= \frac{\Delta u_r}{\sqrt{\langle (\Delta u_r)^2 \rangle_E} }
\eea
so as to obtain standardized PDFs, each with a unit standard deviation. The green dashed line represents a Gaussian distribution with a standard deviation equal to 1. Within the developing regime, we can observe that for $r/\lambda_T=1$, the intermittency decreases due to the significant reduction in the Reynolds number observed previously, while the skewness of the PDF remains nearly constant. Conversely, for $r/\lambda_T=10$, the PDF is almost Gaussian at $t=0$, whereas at subsequent instances, a slight but distinct increase in the asymmetry and intermittency of the PDF is observable, driven by the non-local nature of the energy cascade. This occurs because, at the initial condition, the non-linear cascade mechanism has not yet processed the interaction between scales; hence, the PDF at $r/\lambda = 10$ starts nearly perfectly symmetric and Gaussian. At subsequent times during the developing regime, non-linear transport operates an energy transfer that is not strictly continuous (from one scale to an adjacent one), but includes strong non-local interactions (large eddies directly stretching and distorting inertial scales). This mechanism breaks the original Gaussian symmetry. The first macroscopic intermittent events emerge (flow accelerations or decelerations over large distances), slightly lifting the tails (increasing kurtosis) of the PDF and accentuating its negative asymmetry (skewness), thereby reflecting the establishment of a direct energy cascade toward small scales.
\begin{figure}[t]
	\centering
	\includegraphics[width=100mm, height=100mm,]{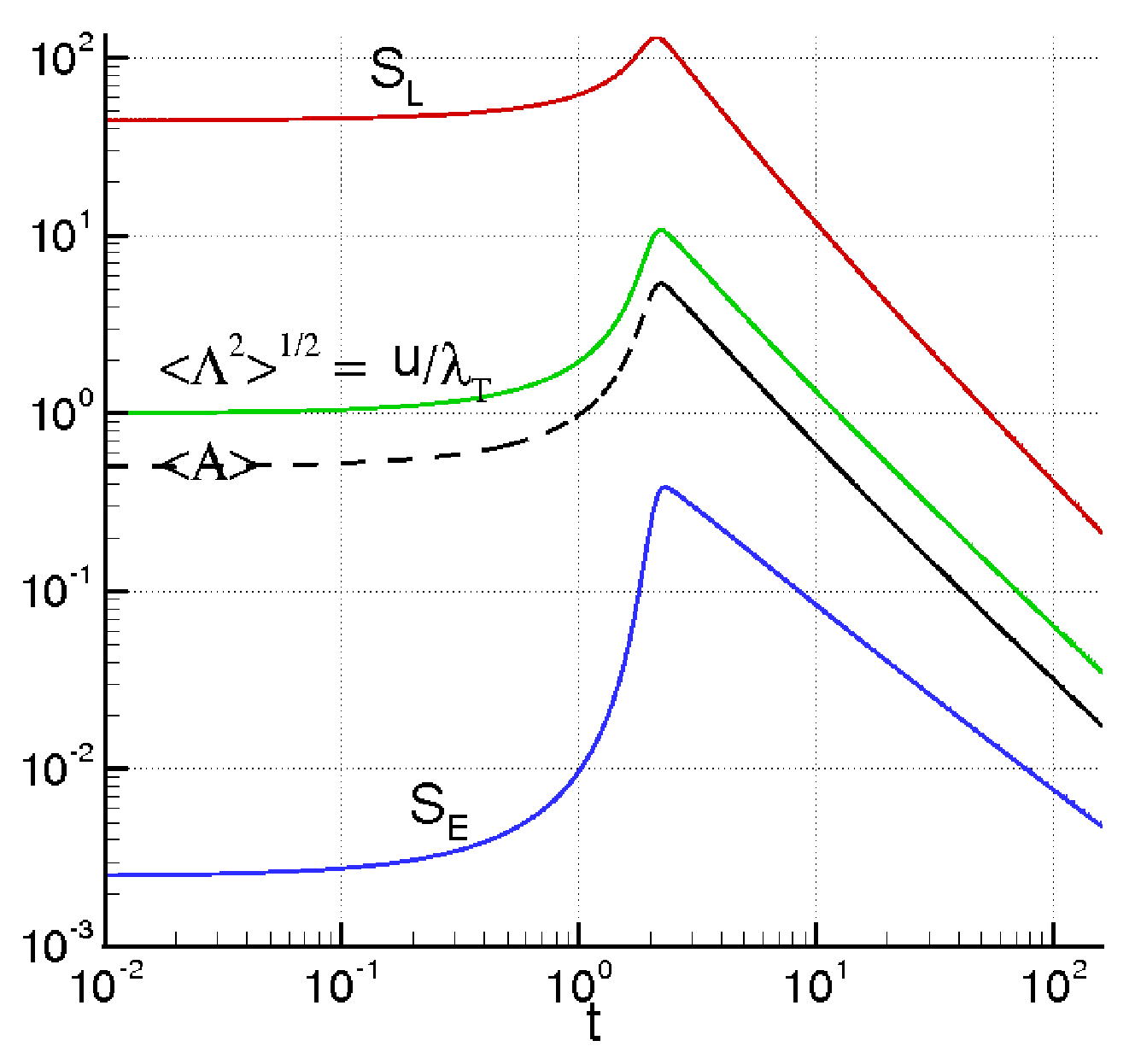}
\caption{Evolution of the Lagrangian Lyapunov exponents $\left< \Lambda_L \right>$ and $\left< \Lambda_L^2 \right>^{1/2}$ and bifurcation rates $S_E$ and $S_L$ for the Loitsiansky initial condition (Eq. \ref{L}).}
\label{f10}
\end{figure}
{This physical picture of multi-scale energy transfer connects with the asymptotic behavior evaluated at the smallest scales ($r \to 0$). In this limit, the present Lyapunov--Liouville framework yields a constant longitudinal velocity derivative skewness of $-3/7$, in compatibility with the structural requirements of isotropic turbulence. Under this formulation, the theory predicts that the corresponding velocity derivative kurtosis varies within a range of approximately $4$ to $9$ as the Taylor-scale Reynolds number ($Re_\lambda$) increases from $10$ to $2000$. These computed small-scale probability density functions and their higher-order dimensionless moments exhibit an excellent quantitative agreement with the landmark cryogenic experimental data reported by Tabeling et al.~\cite{Tabeling1996} and Belin et al.~\cite{Belin1997} for fully developed turbulent flows. }
\begin{figure}[t]
	\centering
	\includegraphics[width=150mm, height=100mm,]{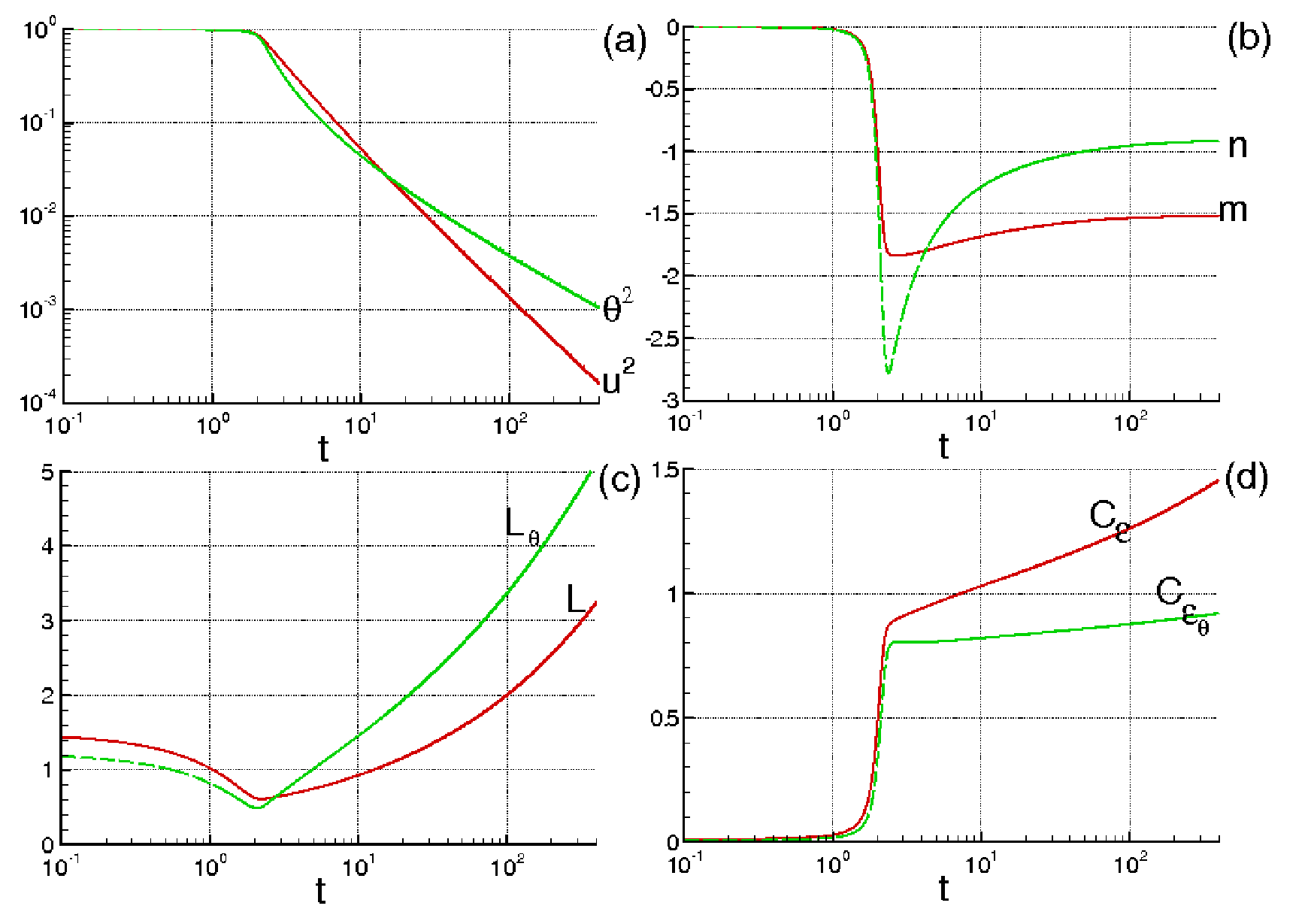}
	\caption{Evolution of physical quantities as a function of time starting from the Loitsiansky initial condition \ref{L}. (a) Velocity and temperature standard deviations, (b) characteristic exponents of velocity and temperature, (c) integral scales, and (d) dissipation coefficients.}
\label{f11}
\end{figure}

In the following figures, we analyze the evolution of the various physical quantities starting from the Loitsiansky initial condition. Figure \ref{f10} displays the Lyapunov exponents as a function of time. The development regime is closely similar to the previous case, whereas the diffusive regime exhibits a slight convergence of the curves, which do not remain strictly parallel.

\begin{figure}[t]
	\centering
	\includegraphics[width=150mm, height=100mm,]{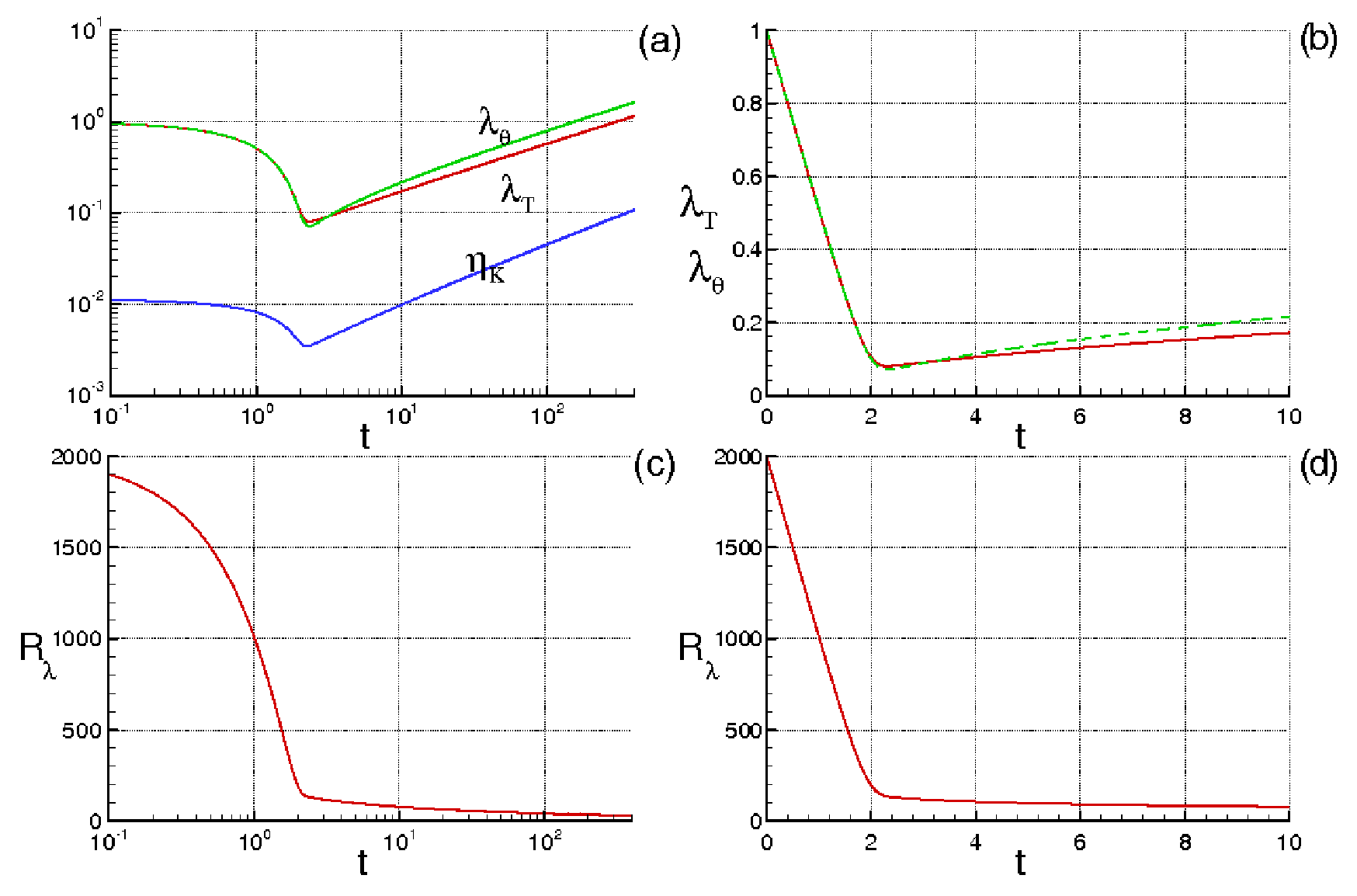}
	\caption{Evolution of physical quantities as a function of time starting from the Loitsiansky initial condition \ref{L}. (a) Integral scale, Taylor scale, and Kolmogorov scale, (b) magnified view of the Taylor and Corrsin microscales, (c) Taylor-scale Reynolds number, and (d) magnified view of the Taylor-scale Reynolds number.}
\label{f12}
\end{figure}

Consequently, a distinct behavior is observed for the velocity and temperature standard deviations, yielding different exponents $m$ and $n$ compared to the Saffman--Birkhoff case (see Fig. \ref{f11}). Specifically, $u^2$ decays faster than in the Saffman--Birkhoff case due to the higher mechanical dissipation stemming from the Loitsiansky initial velocity correlation. Conversely, $\theta^2$ exhibits higher persistence compared to both $u^2$ and the previously obtained $\theta^2$, thus deviating significantly from the current trend of $u^2$.

\begin{figure}[h!]
	\centering
	\includegraphics[width=150mm, height=100mm,]{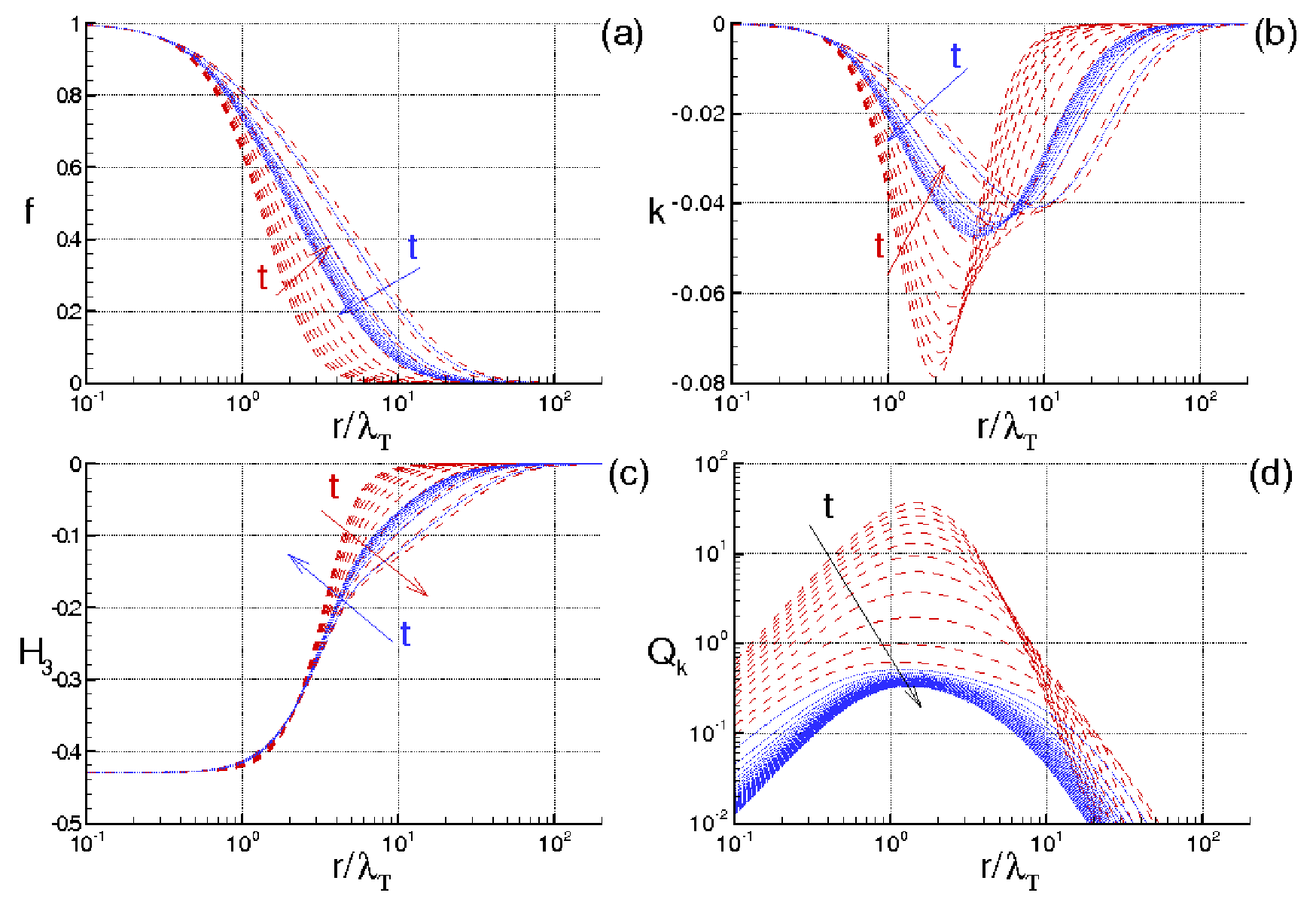}
	\caption{Spatial variation laws at different time instances for the Loitsiansky initial condition \ref{L}. (a) Longitudinal two-point velocity autocorrelation, (b) longitudinal two-point triple velocity correlation, (c) skewness of the longitudinal velocity increment, and (d) spatial Kolmogorov function.  Red dashed lines represent the developing transient regime from $t=0$ to $t=2.2$ with a constant time increment of $\Delta t=0.2$ between successive curves, where the red arrow indicates the direction of increasing time. Blue dotted lines describe the diffusive regime at subsequent time instances $t > 2.2$, where the blue arrow denotes the direction of increasing time.  }
\label{f13}
\end{figure}

This results in power-law exponents of $m \simeq -1.51$ and $n \simeq -0.89$, alongside the integral scale trends depicted in Fig. \ref{f11}c. The mechanical and thermal dissipation coefficients, $C_\epsilon$ and $C_{\epsilon_\theta}$, differ from the previous Saffman--Birkhoff case.  { $C_\epsilon$ and $C_{\epsilon_\theta}$ do not exhibit distinct plateaus---especially $C_\epsilon$---but rather maintain values of the order of unity. This confirms that both $C_\epsilon$ and $C_{\epsilon_\theta}$ are not universal constants, but are strongly dependent on the flow field and initial conditions \cite{Vassilicos2015, Warhaft2000}. } Fig. \ref{f12} illustrates the evolution of the characteristic scales and the Taylor-scale Reynolds number as a function of non-dimensional time.
\begin{figure}[h!]
	\centering
	\includegraphics[width=150mm, height=100mm,]{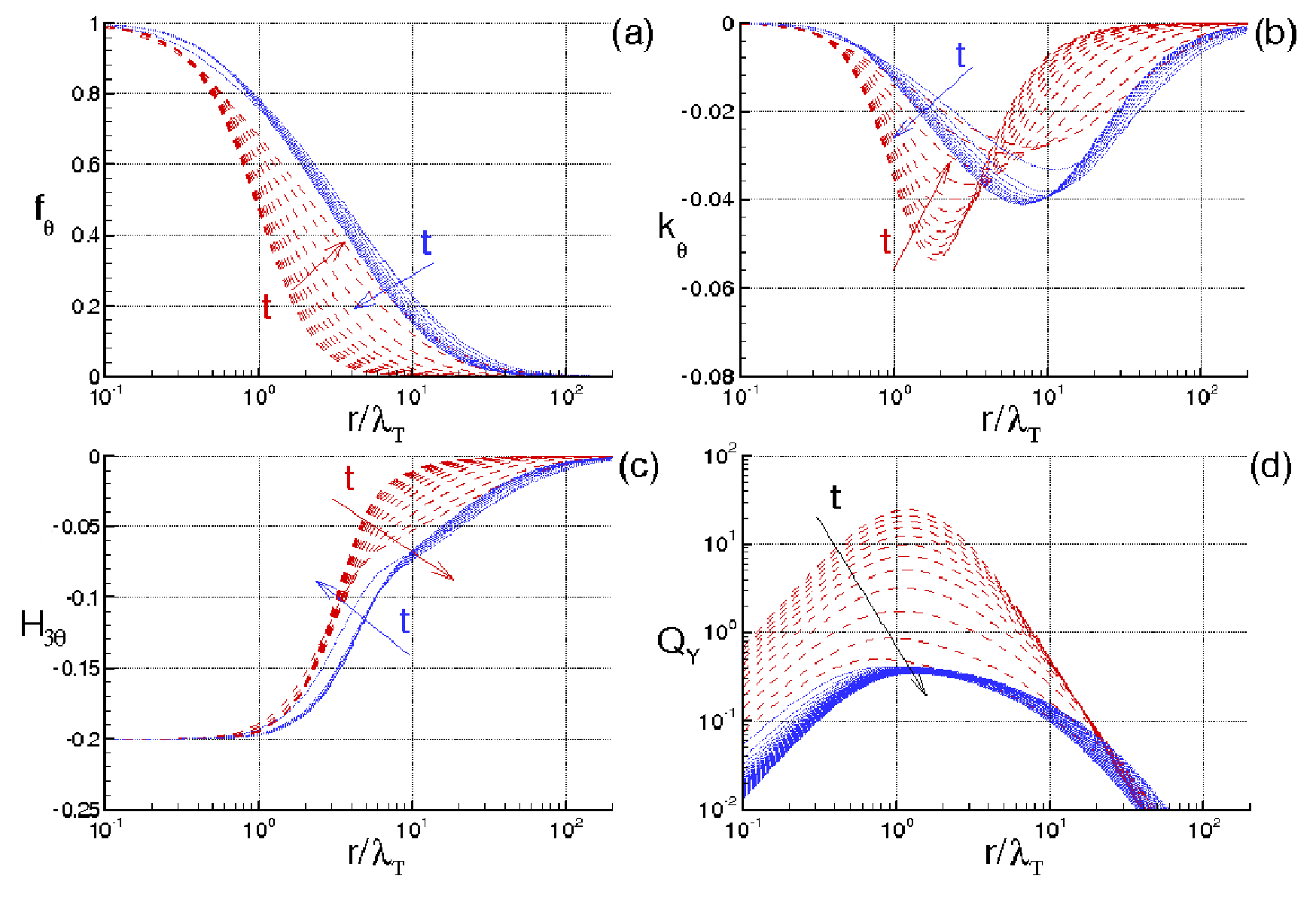}
	\caption{Spatial variation laws at different time instances for the Loitsiansky initial condition \ref{L}. (a) Two-point temperature autocorrelation, (b) mixed two-point triple temperature--velocity correlation, (c) mixed non-dimensional third-order moment of the temperature--velocity increments, and (d) Yaglom function. Red dashed lines represent the developing transient regime from $t=0$ to $t=2.2$ with a constant time increment of $\Delta t=0.2$ between successive curves, where the red arrow indicates the direction of increasing time. Blue dotted lines describe the diffusive regime at subsequent time instances $t > 2.2$, where the blue arrow denotes the direction of increasing time. }
\label{f14}
\end{figure}

In this case, it can be stated that the variations of $\lambda_T$ and $R_\lambda$ are very similar to those previously computed with the Saffman--Birkhoff initial condition within the development regime ($t \in (0, 2.2)$). Conversely, during the diffusive decay regime, visible differences emerge compared to the Saffman--Birkhoff case.
\begin{figure}[t]
	\centering
	\includegraphics[width=150mm, height=80mm,]{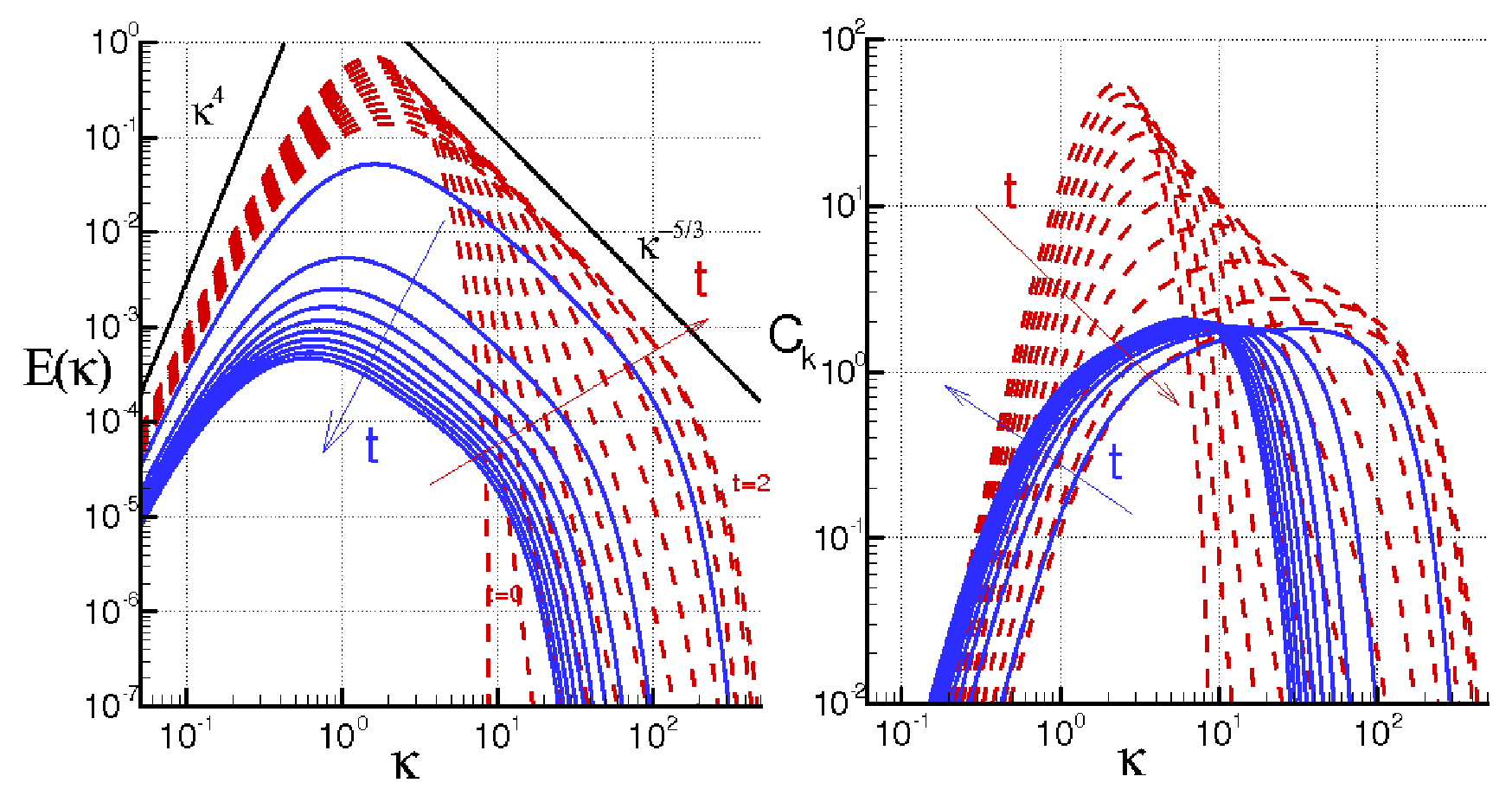}
	\caption{Turbulent kinetic energy spectra (a) and compensated Kolmogorov spectral functions (b) evaluated at different time instances for the Loitsiansky initial condition \ref{L}.  Red dashed lines represent the developing transient regime from $t=0$ to $t=2.2$ with a constant time increment of $\Delta t=0.2$ between successive curves, where the red arrow indicates the direction of increasing time. Blue lines describe the diffusive regime at subsequent time instances $t > 2.2$, where the blue arrow denotes the direction of increasing time.}
\label{f15}
\end{figure}

Specifically, $\lambda_\theta$ and $\lambda_T$ tend to diverge from each other despite $Pr=1$, while a more pronounced reduction in $R_\lambda$ is observed compared to the Saffman--Birkhoff initial condition, due to the fact that the current Loitsiansky initial condition leads to higher dissipation. Furthermore, the discrepancy between the two initial conditions results in absolute values of the two-point triple correlations $k$ and $k_\theta$ that are slightly but visibly larger than in the previous case. As before, long-range interaction effects are observed as the correlation functions evolve from the initial condition. In this case as well, at the stationary points of $\lambda_T$ and $\lambda_\theta$ ($d \lambda_T/dt \simeq 0$, $d \lambda_\theta/dt \simeq 0$ at $t \simeq 2.2$), the spatial Kolmogorov and Yaglom functions exhibit a trend characterized by a flat maximum (a strict plateau is absent here too) with values very close to $4/5$ and $2/3$, respectively (see Figs. \ref{f13} and \ref{f14}).

\begin{figure}[t]
	\centering
	\includegraphics[width=150mm, height=80mm,]{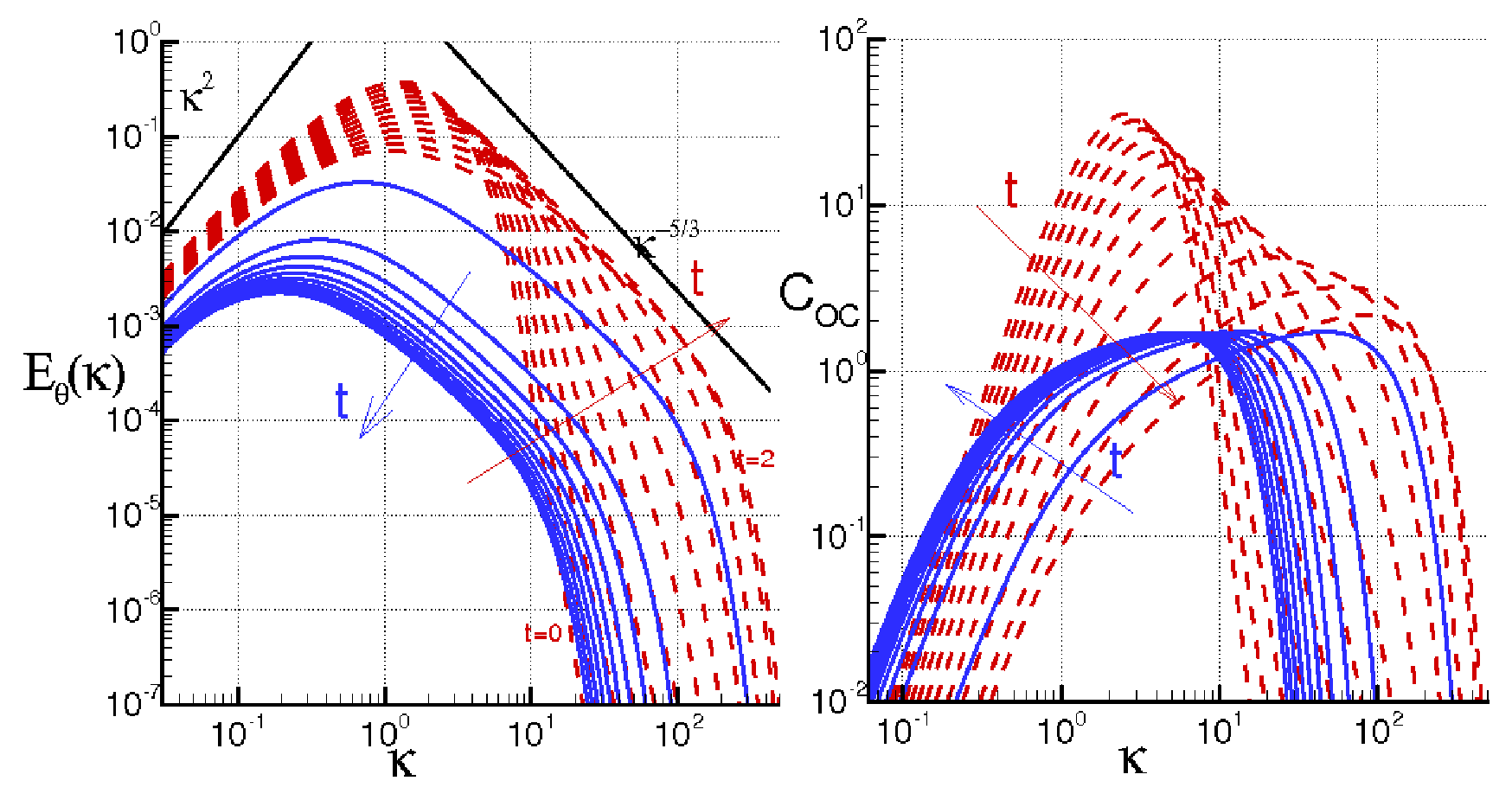}
	\caption{(a) Temperature spectra at different time instances. (b) Compensated Obukhov--Corrsin spectral functions at different time instances for the Loitsiansky initial condition \ref{L}. Red dashed lines represent the developing transient regime from $t=0$ to $t=2.2$ with a constant time increment of $\Delta t=0.2$ between successive curves, where the red arrow indicates the direction of increasing time. Blue  lines describe the diffusive regime at subsequent time instances $t > 2.2$, where the blue arrow denotes the direction of increasing time. }
\label{f16}
\end{figure}

The velocity and temperature spectra, $E(k)$ and $E_\theta(k)$, are then presented along with their corresponding compensated Kolmogorov and Obukhov--Corrsin spectral functions, respectively (see Figs. \ref{f15} and \ref{f16}).  $E(k)\approx k^4$ near the origin due to the Loitsiansky initial condition, which tends to be preserved at subsequent times after the initial state. Moreover, the envelopes of both $E(k)$ and $E_\theta(k)$ spectra in the development regime closely follow the Kolmogorov $k^{-5/3}$ power law. { In this case as well, within the diffusive regime, $\hat{C}_K$ and $\hat{C}_{O C}$ do not display a proper plateau but rather a relatively flat maximum, yielding $C_K \simeq 1.75$ and $C_{O C} \simeq 1.78$. These values are in good agreement with the literature on decaying turbulence \cite{Wang1996, briard2016passive, sreenivasan1996passive}. }
\begin{figure}[t]
	\centering
	\includegraphics[width=150mm, height=50mm,]{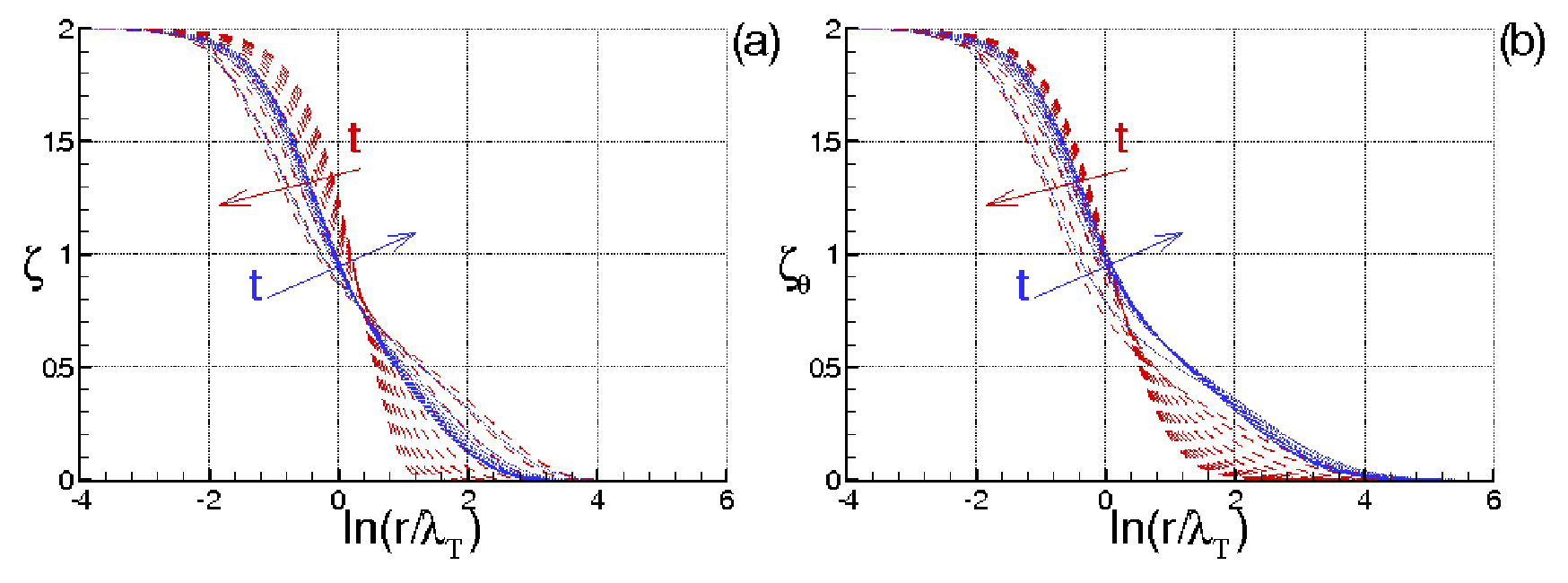}
\caption{Local scaling exponents of (a) velocity correlation and (b) temperature correlation evaluated at different time instances for the Loitsiansky initial condition \ref{L}.   Red dashed lines represent the developing transient regime from $t=0$ to $t=2.2$ with a constant time increment of $\Delta t=0.2$ between successive curves, where the red arrow indicates the direction of increasing time. Blue dotted lines describe the diffusive regime at subsequent time instances $t > 2.2$, where the blue arrow denotes the direction of increasing time.   }
\label{f17}
\end{figure}
As observed previously, the local exponents $Z$ and $Z_\theta$ exhibit significant variations during the development regime ($t \in (0, 2.2)$). Conversely, during the diffusive decay regime, the curves corresponding to different time instances for each exponent appear to collapse onto a single master curve (see Fig. \ref{f17}).

\begin{figure}[t]
	\centering
	\includegraphics[width=100mm, height=80mm,]{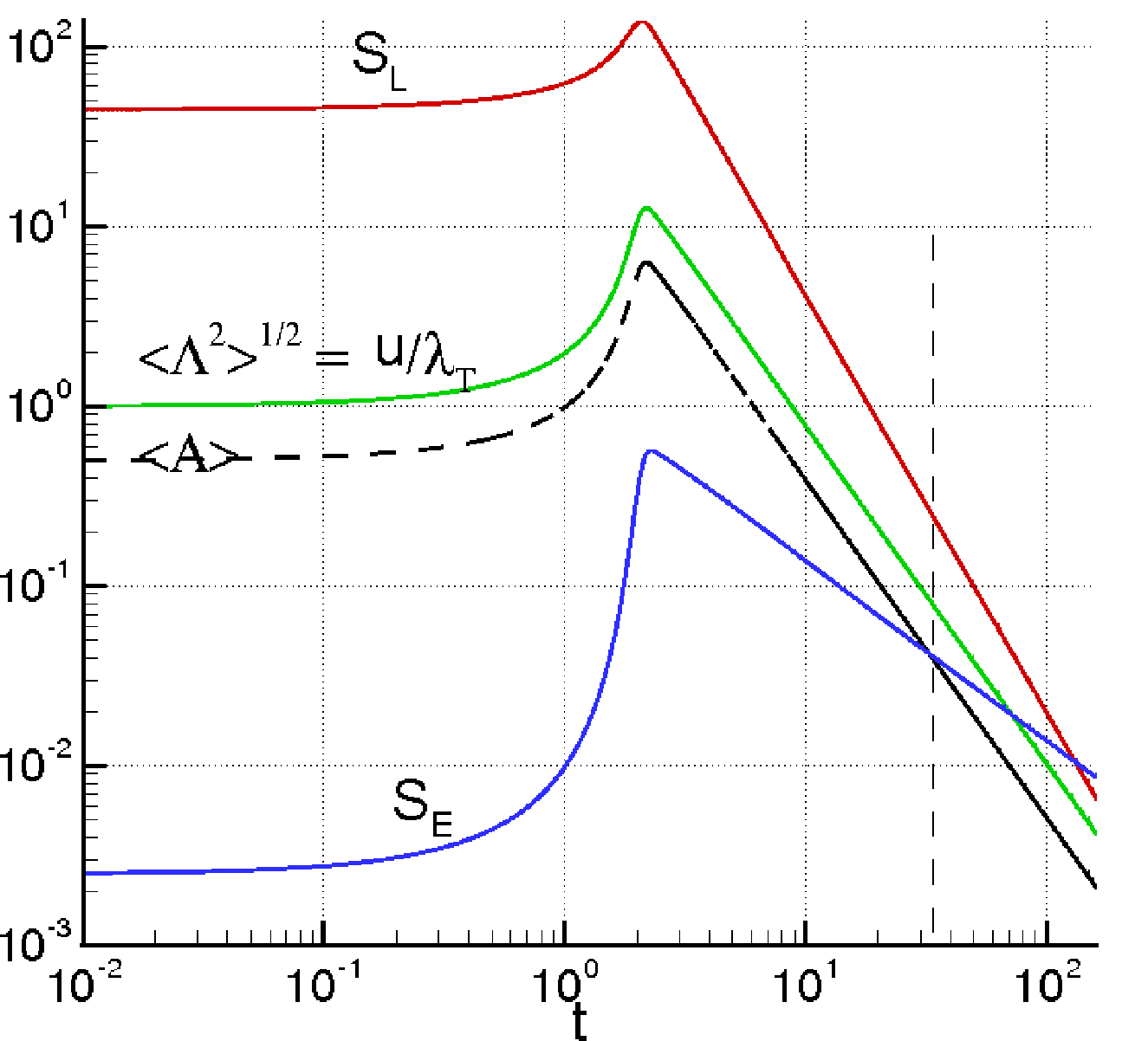}
\caption{Evolution of the Lagrangian Lyapunov exponents $\left<   \Lambda_L \right>$ and $\left<   \Lambda_L^2 \right>^{1/2}$,  and bifurcation rates $S_E$ and $S_L$ for the Gaussian initial condition (Eq. \ref{G}).}
\label{f18}
\end{figure}

Finally, we consider the case corresponding to a Gaussian initial condition for $f$ according to Eq. (\ref{G}). This configuration exhibits a distinct behavior compared to the first two, as it is characterized by an elevated dissipation rate during the diffusive decay regime. This dissipation levels induce a different evolution relative to the previously analyzed cases, limiting the preservation of kinetic energy during the diffusive decay stage and shortening the time interval over which the flow sustains a turbulent state. Beyond this period, as shown by the present analysis, the system approaches the threshold where HIT loses its physical conditions of existence.
\begin{figure}[t]
	\centering
	\includegraphics[width=150mm, height=100mm,]{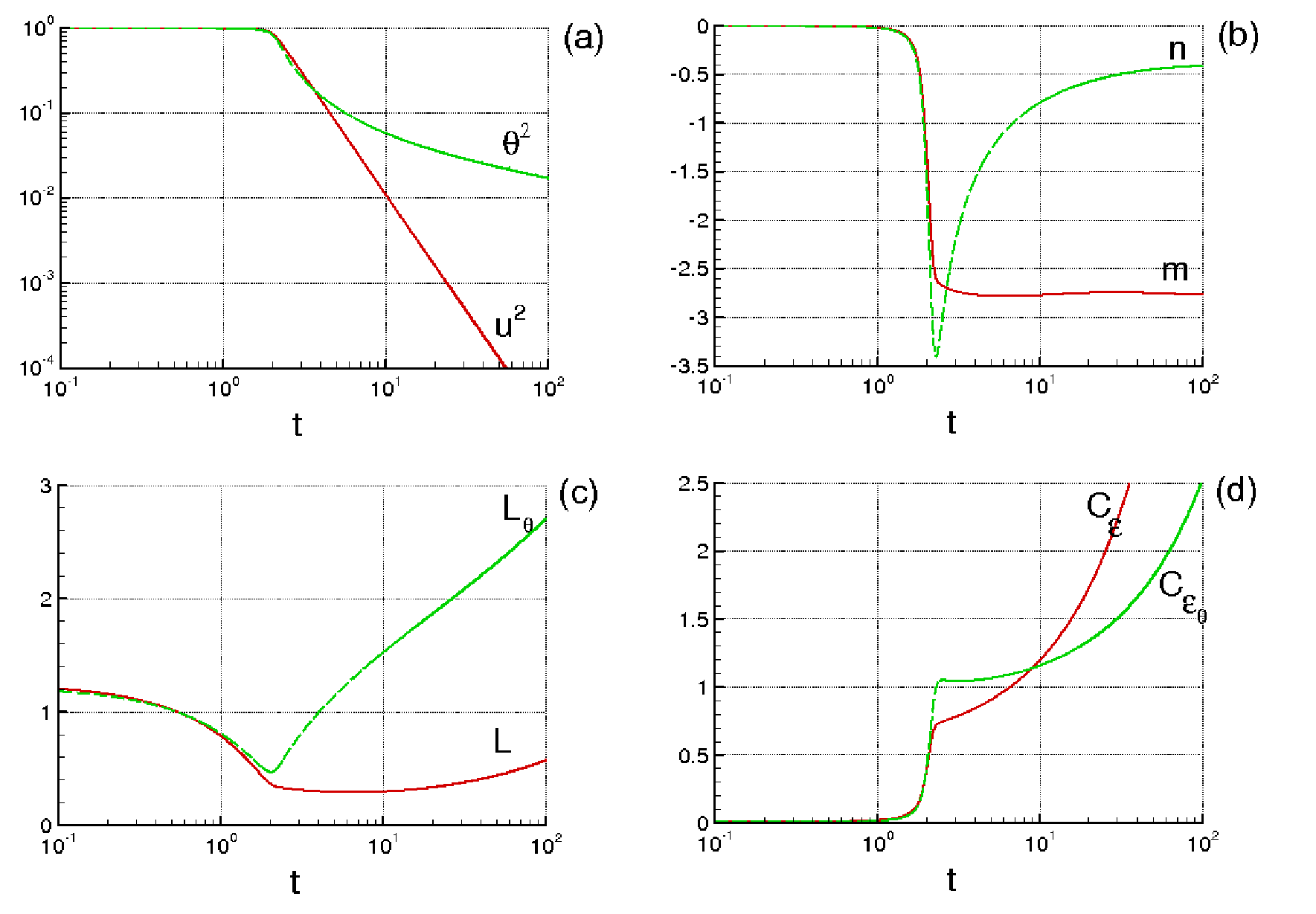}
	\caption{Evolution of physical quantities as a function of time starting from the Gaussian initial condition \ref{G}. (a) Velocity and temperature standard deviations, (b) characteristic exponents of velocity and temperature, (c) integral scales, and (d) dissipation coefficients.}
\label{f19}
\end{figure}

In detail, Fig. \ref{f18} reports the Eulerian and Lagrangian bifurcation rates, along with the mean and variance of the Lagrangian Lyapunov exponent. We observe that relation (\ref{S>>>>S}), which dictates the necessary condition for the validity of the present analysis, is satisfied over a relatively short time interval ($t \in (0, 33)$) compared to the previous cases. Indeed, at $t \simeq 33$ (indicated by the vertical dashed line), $R_\lambda = R_\lambda^* = 10$, and the dissipation rate equals the Lagrangian Lyapunov exponent.
At subsequent times, the dissipation rate exceeds the mean Lagrangian Lyapunov exponent; consequently, within the present analysis, homogeneous isotropic turbulence (HIT) loses its physical condition of existence, and the proposed closures (\ref{K closure}) are no longer applicable.
 Therefore, all subsequent diagrams for the Gaussian case—where the simulation extends beyond $t \simeq 33$—are to be considered valid only within the range $0 < t < 33$.
{
The Gaussian initial condition serves as an illustrative case that identifies the boundaries of existence for both HIT and the proposed closures within this formulation. From a physical standpoint, the specific behavior of the Gaussian profile stems from its structural configuration. Unlike the Saffman--Birkhoff and Loitsiansky conditions, where the initial velocity correlation exhibits spatial support extending beyond the Taylor microscale, the Gaussian correlation is localized. As $r \to \infty$, the Gaussian profile decays exponentially to zero; consequently, the primary spatial variations of the correlation field are concentrated in proximity to the origin and are of the order of the Taylor microscale. Because this state possesses limited large-scale energetic support, the spatial gradients near the origin appear pronounced. This induces an elevated initial dissipation rate, which translates into an increased Eulerian bifurcation rate $S_E$. As a result, the mechanical energy undergoes a rapid decay path ($m \simeq -2.7$), and the system drives the Taylor-scale Reynolds number down, crossing the threshold $R_\lambda \simeq 10$ at a non-dimensional time $t \simeq 33$. At this juncture, the condition $S_L > S_E$ is no longer satisfied, suggesting a physical limit for the sustainability of HIT and the applicability of the proposed non-diffusive closure model as the flow leaves the fully turbulent regime.
 }
Consequently, the standard deviations of velocity and temperature exhibit a distinct behavior, with their corresponding exponents \(m\) and \(n\) diverging significantly from those of the two previous cases.  Specifically, $u^2$ decays markedly faster due to the higher mechanical dissipation stemming from the initial correlation (\ref{G}). Regarding $\theta^2$, it exhibits higher persistence compared to $u^2$, thus deviating significantly from the current trend of $u^2$. In practice, after a short time, non-zero temperature fluctuations are observed within a nearly quiescent fluid. These temperature fluctuations do not have sufficient time to fully develop because the system reaches its limit of validity at $t \simeq 33$.

\begin{figure}[t]
	\centering
	\includegraphics[width=150mm, height=100mm,]{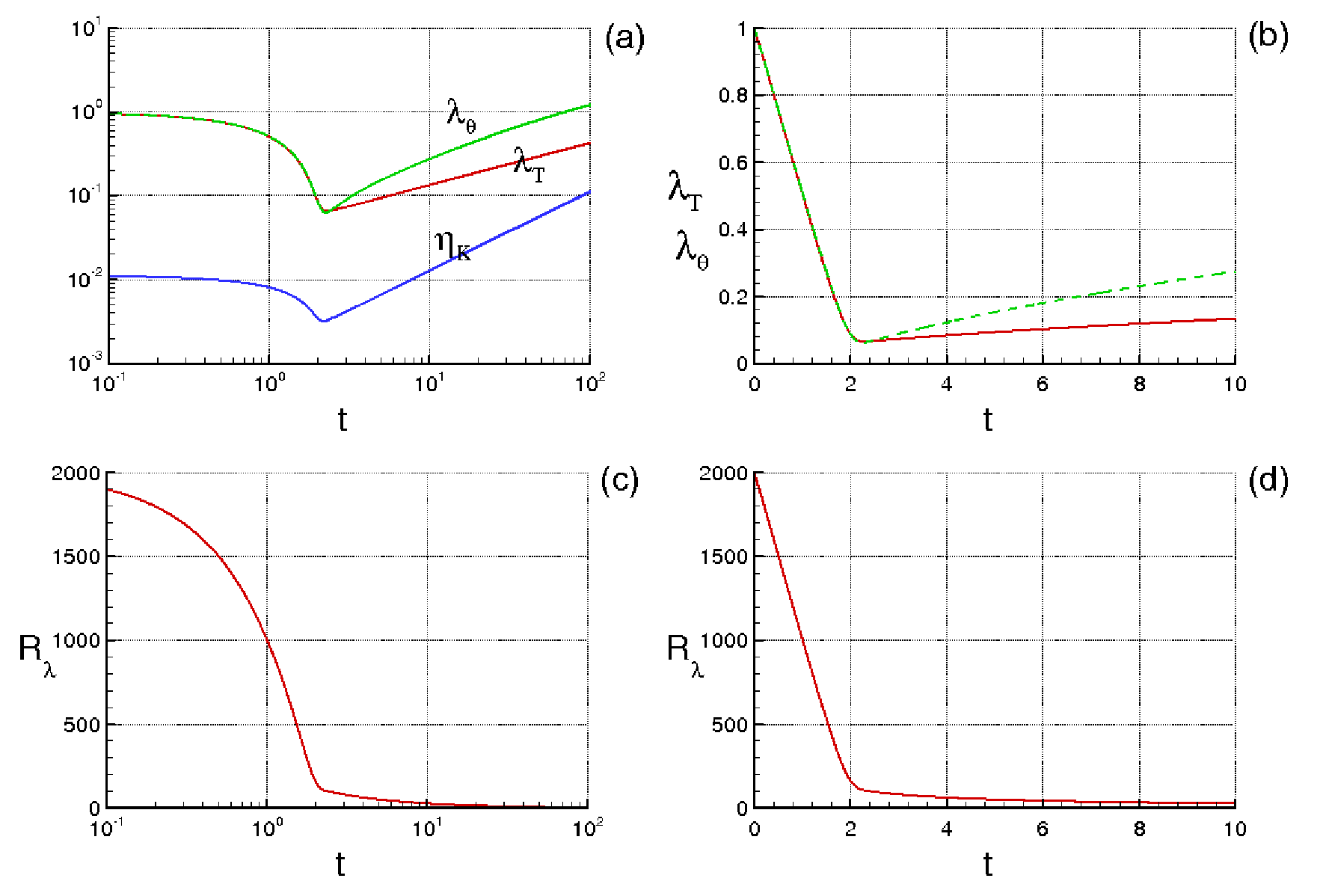}
	\caption{Evolution of physical quantities as a function of time starting from the Gaussian initial condition \ref{G}. (a) Integral scale, Taylor scale, and Kolmogorov scale, (b) magnified view of the Taylor and Corrsin microscales, (c) Taylor-scale Reynolds number, and (d) magnified view of the Taylor-scale Reynolds number.}
\label{f20}
\end{figure}

This yields a value of $m \simeq -2.7$, whereas the exponent $n$ does not reach a stationary value within the interval $(2, 33)$. The time evolution of the integral scales is depicted in Fig. \ref{f19}c. Although both the mechanical and thermal dissipation coefficients, $C_\epsilon$ and $C_{\epsilon_\theta}$, remain of the order of unity, they differ from the previous cases. Specifically, within the time interval $(2, 33)$, both $C_\epsilon$ and $C_{\epsilon_\theta}$ exhibit a monotonic trend, further confirming that they are not universal constants but rather depend on the flow field and initial conditions \cite{Vassilicos2015, Warhaft2000}.

\begin{figure}[t]
	\centering
	\includegraphics[width=150mm, height=80mm,]{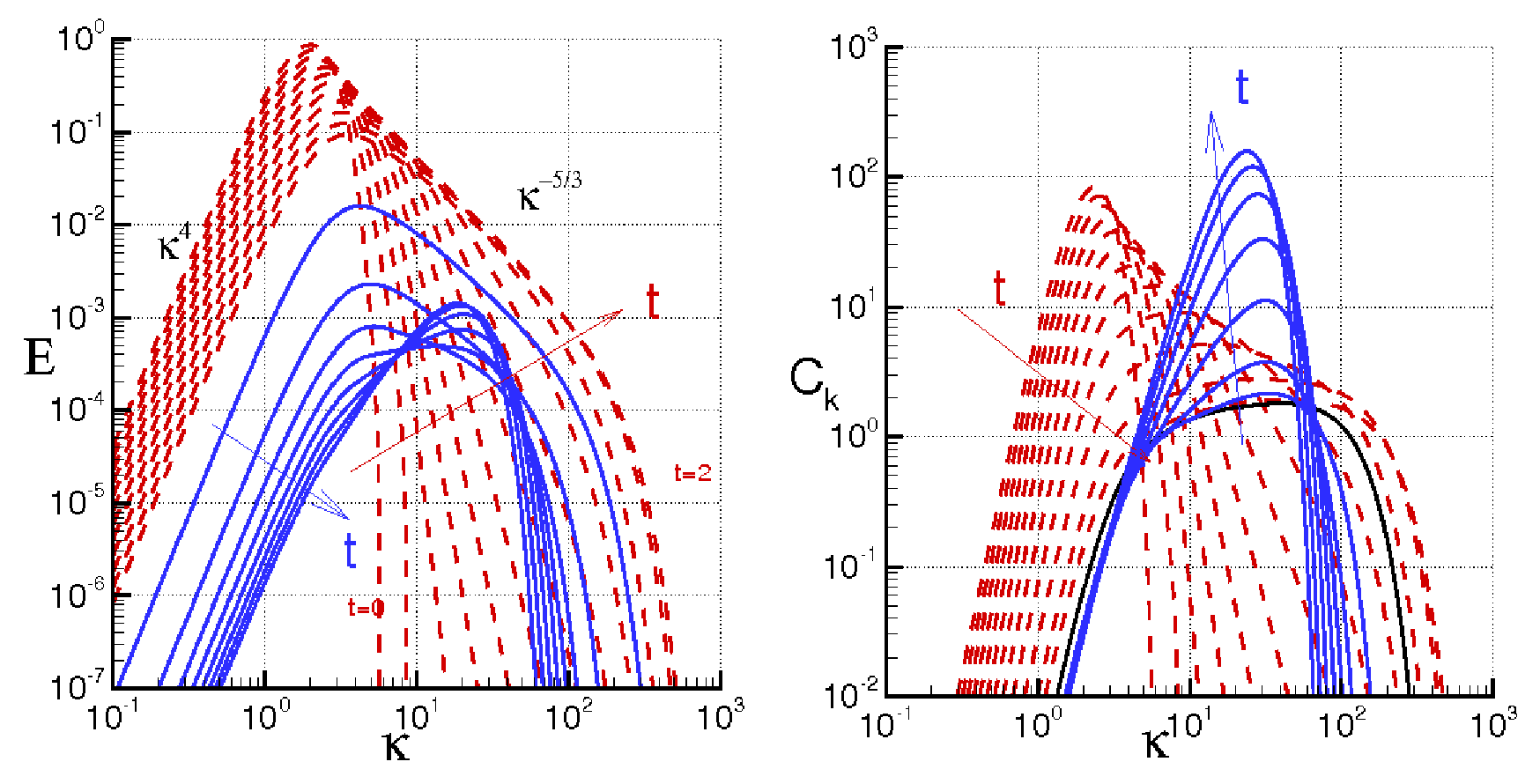}
\caption{Turbulent kinetic energy spectra (a) and compensated Kolmogorov spectral functions (b) evaluated at different time instances for the Gaussian initial condition \ref{G}.   Red dashed lines represent the developing transient regime from $t=0$ to $t=2.2$ with a constant time increment of $\Delta t=0.2$ between successive curves, where the red arrow indicates the direction of increasing time. Blue  lines describe the diffusive regime at subsequent time instances $t > 2.2$, where the blue arrow denotes the direction of increasing time.   }
\label{f21}
\end{figure}

Fig. \ref{f20} illustrates the evolution of the characteristic scales and the Taylor-scale Reynolds number as a function of non-dimensional time. It can be observed that, in terms of correlation scales and Reynolds number, the development regime is nearly identical to the previous cases. Conversely, during the diffusive decay regime, both the correlation scales and $R_\lambda$ decay much more rapidly than in the Saffman--Birkhoff and Loitsiansky cases.

\begin{figure}[t]
	\centering
	\includegraphics[width=150mm, height=80mm,]{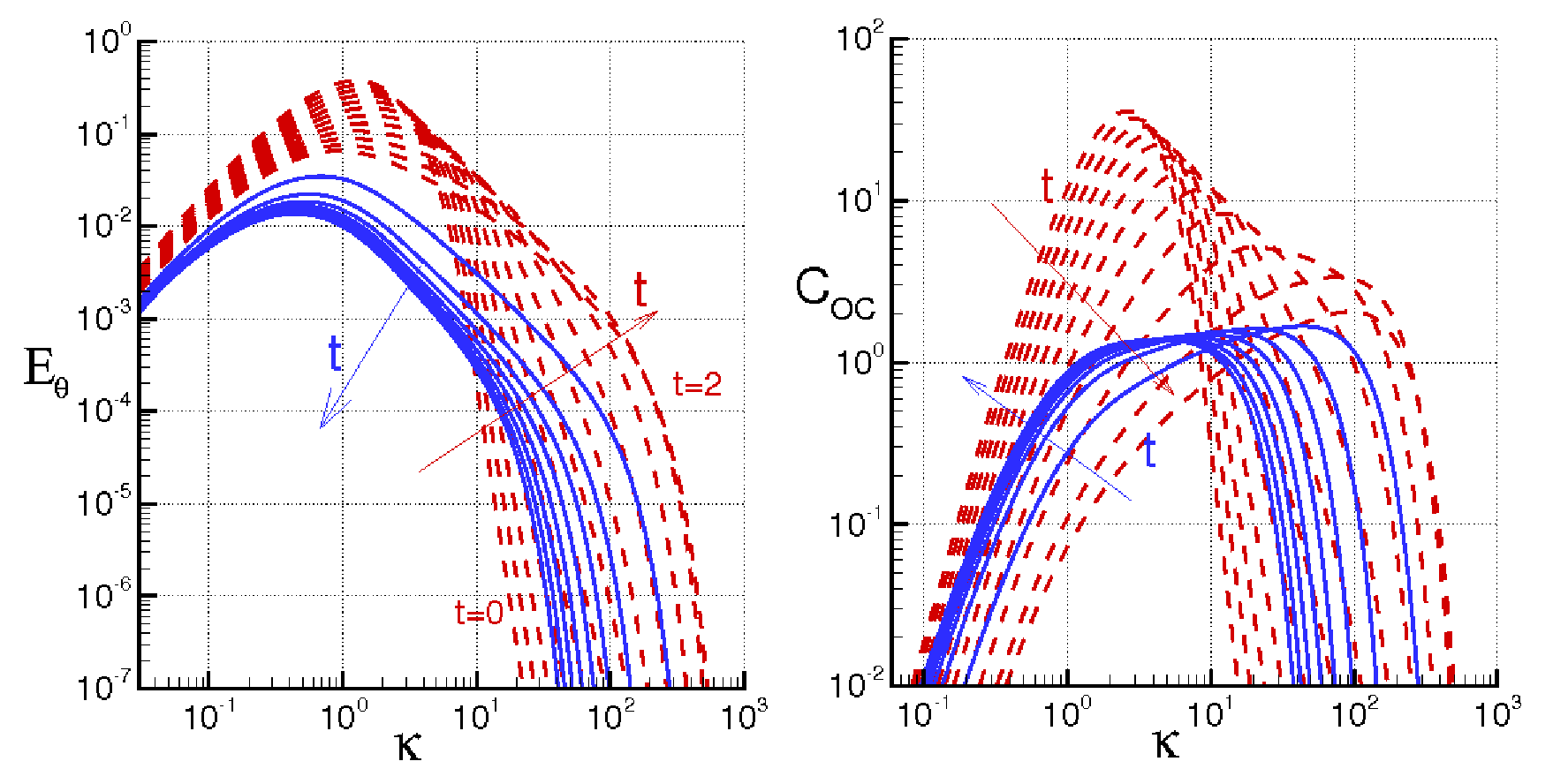}
	\caption{(a) Temperature spectra at different time instances. (b) Compensated Obukhov--Corrsin spectral functions at different time instances for the Gaussian initial condition \ref{G}. Red dashed lines represent the developing transient regime from $t=0$ to $t=2.2$ with a constant time increment of $\Delta t=0.2$ between successive curves, where the red arrow indicates the direction of increasing time. Blue  lines describe the diffusive regime at subsequent time instances $t > 2.2$, where the blue arrow denotes the direction of increasing time.  }
\label{f22}
\end{figure}

We now consider the spectra $E(k)$ and $E_\theta(k)$, alongside the corresponding compensated Kolmogorov and Obukhov--Corrsin spectral functions shown in Figs. \ref{f21} and \ref{f22}, respectively. It is found that $E(k)\approx k^4$ near the origin due to the Gaussian initial condition. Furthermore, the envelopes of both $E(k)$ and $E_\theta(k)$ spectra within the development regime closely follow the Kolmogorov $k^{-5/3}$ power law. { In this case as well, during the diffusive regime for $t < 33$, $\hat{C}_K$ and $\hat{C}_{O C}$ do not display a strict plateau but rather a relatively flat maximum, yielding $C_K \simeq 1.75$ and $C_{O C} \simeq 1.78$.  These values are in good agreement with the literature on decaying turbulence \cite{Wang1996, briard2016passive, sreenivasan1996passive}. }
For $t > 33$ (where $R_\lambda < 10$), HIT loses its physical conditions of existence, and thus the proposed closures (\ref{K closure}) are no longer applicable; within this regime, $E(k)$ no longer follows the Kolmogorov $k^{-5/3}$ scaling, and the compensated function $\hat{C}_K$ deviates from its canonical shape, tending to grow unboundedly as illustrated in Fig. \ref{f21}.

\begin{figure}[t]
	\centering
	\includegraphics[width=150mm, height=80mm,]{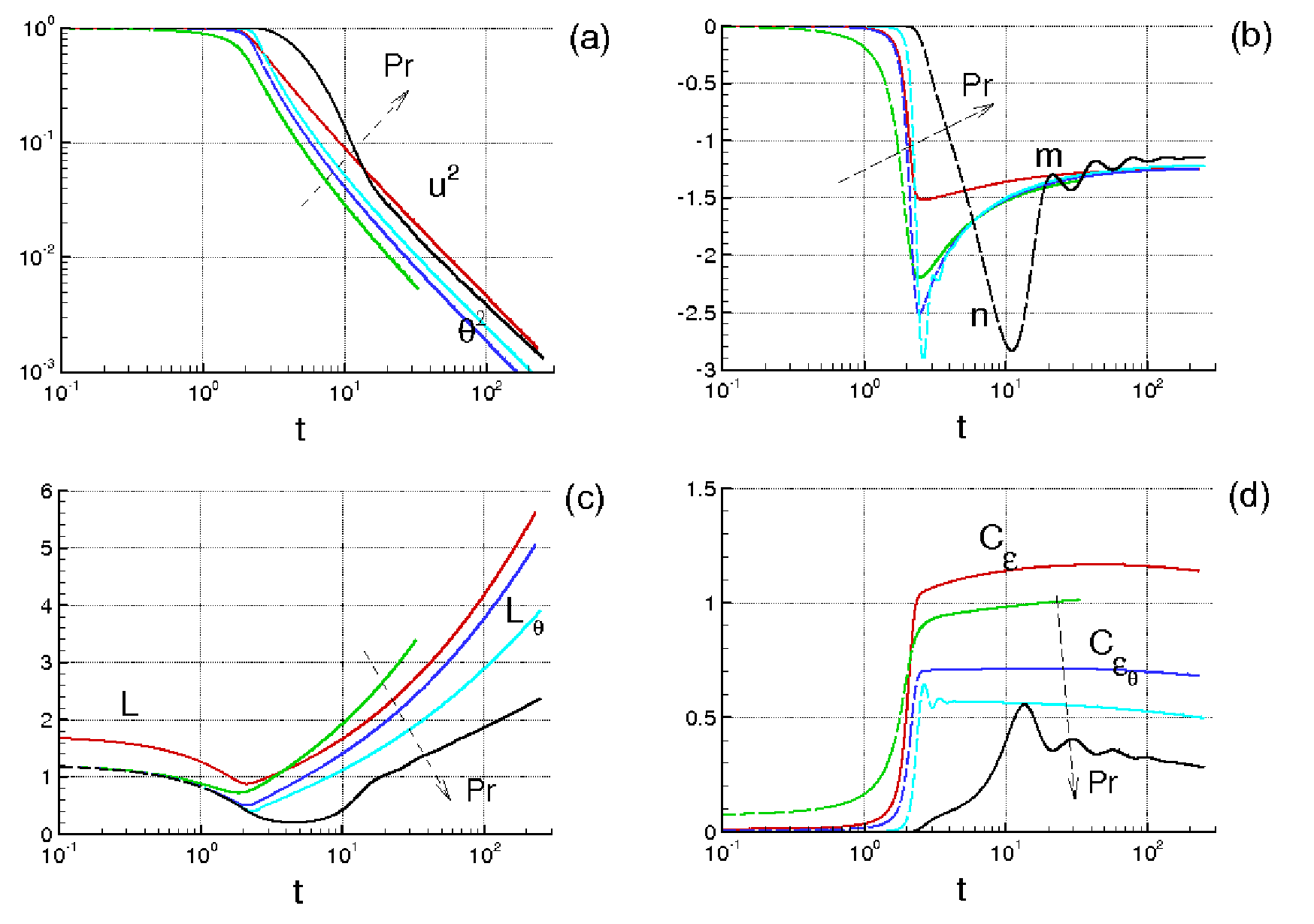}
	\caption{Evolution of physical quantities as a function of time starting from the Saffman--Birkhoff initial condition \ref{SB} at different Prandtl numbers ($Pr = 0.1, 1, 10,$ and $1000$). (a) Velocity and temperature standard deviations, (b) characteristic exponents of velocity and temperature, (c) integral scales, and (d) dissipation coefficients.}
\label{f23}
\end{figure}

To assess the effect of the Prandtl number on the temperature spectrum and its corresponding physical quantities, simulations were performed with different Prandtl numbers exclusively for the Saffman--Birkhoff initial condition (Eq. \ref{SB}). Figure \ref{f23} displays the velocity variation laws in red, as previously illustrated, while the remaining curves correspond to the temperature for $Pr = 0.1$, $1$, $10$, and $1000$, following the direction of the black dashed arrow. It is observed that the evolution of $\theta^2$ changes radically as a function of $Pr$. However, during the diffusive decay phase, the different curves tend to become mutually parallel, with asymptotic values of the characteristic exponents close to $-1.25$, except for the $Pr = 1000$ case, which exhibits a plateau at $n \simeq -1.15$. Regarding the thermal dissipation coefficients $C_{\epsilon_\theta}$, they demonstrate a strong dependence on $Pr$; specifically, these coefficients decrease as $Pr$ increases. For $Pr = 1000$, $C_{\epsilon_\theta}$ exhibits significant amplitude oscillations before stabilizing. This further confirms that $C_{\epsilon_\theta}$ is not a universal constant, but rather a quantity dependent on the specific flow field and initial conditions.

\begin{figure}[t]
	\centering
	\includegraphics[width=150mm, height=50mm,]{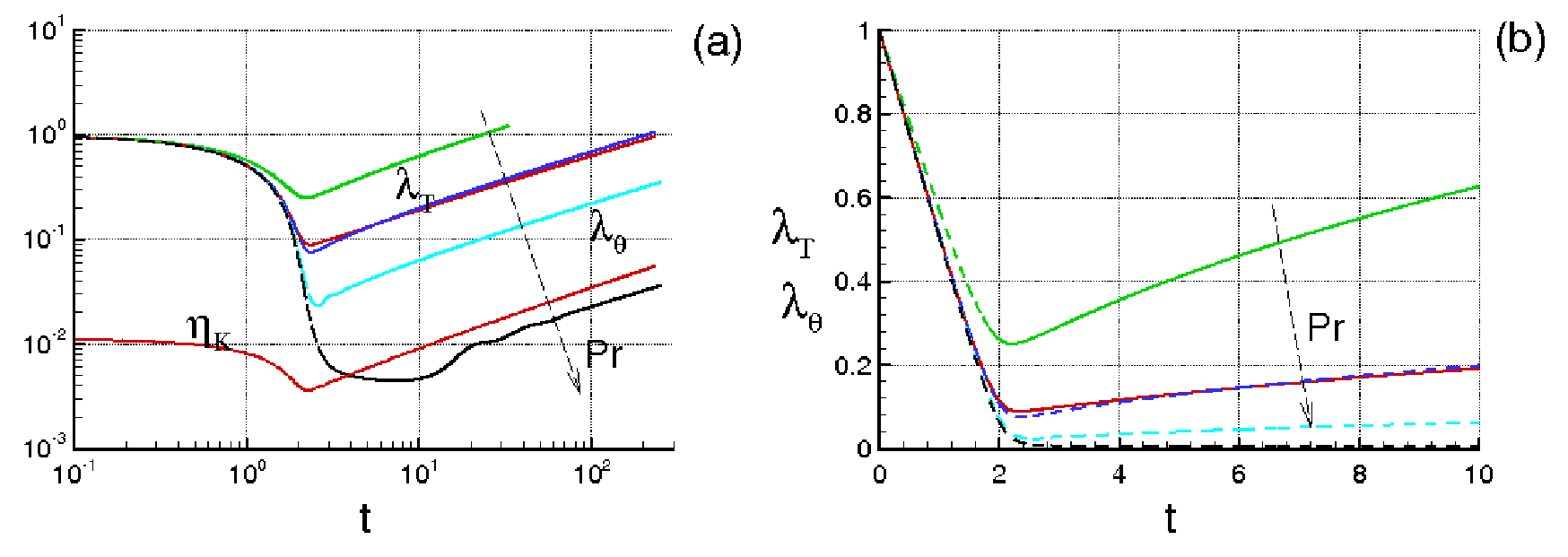}
\caption{Characteristic scales vs time at different Prandtl numbers: $Pr = 0.1, 1, 10,$ and $1000$.}
\label{f24}
\end{figure}

Figure \ref{f24} shows the correlation scales along with the Kolmogorov microscale $\eta_K$. It can be observed that as long as $Pr$ is relatively high, the Prandtl number does not significantly affect the development regime. Conversely, for $Pr = 0.1$, significant deviations of $\lambda_\theta$ are observed compared to the other cases, which otherwise exhibit a nearly linear trend with respect to $t$. When $Pr$ is very large (here, $Pr = 1000$), the correlation microscale $\lambda_\theta$ drops below the Kolmogorov scale. This behavior is in full agreement with Batchelor's theory for turbulent flows at high Prandtl numbers \cite{Batchelor_2}.

\begin{figure}[t]
	\centering
	\includegraphics[width=150mm, height=80mm,]{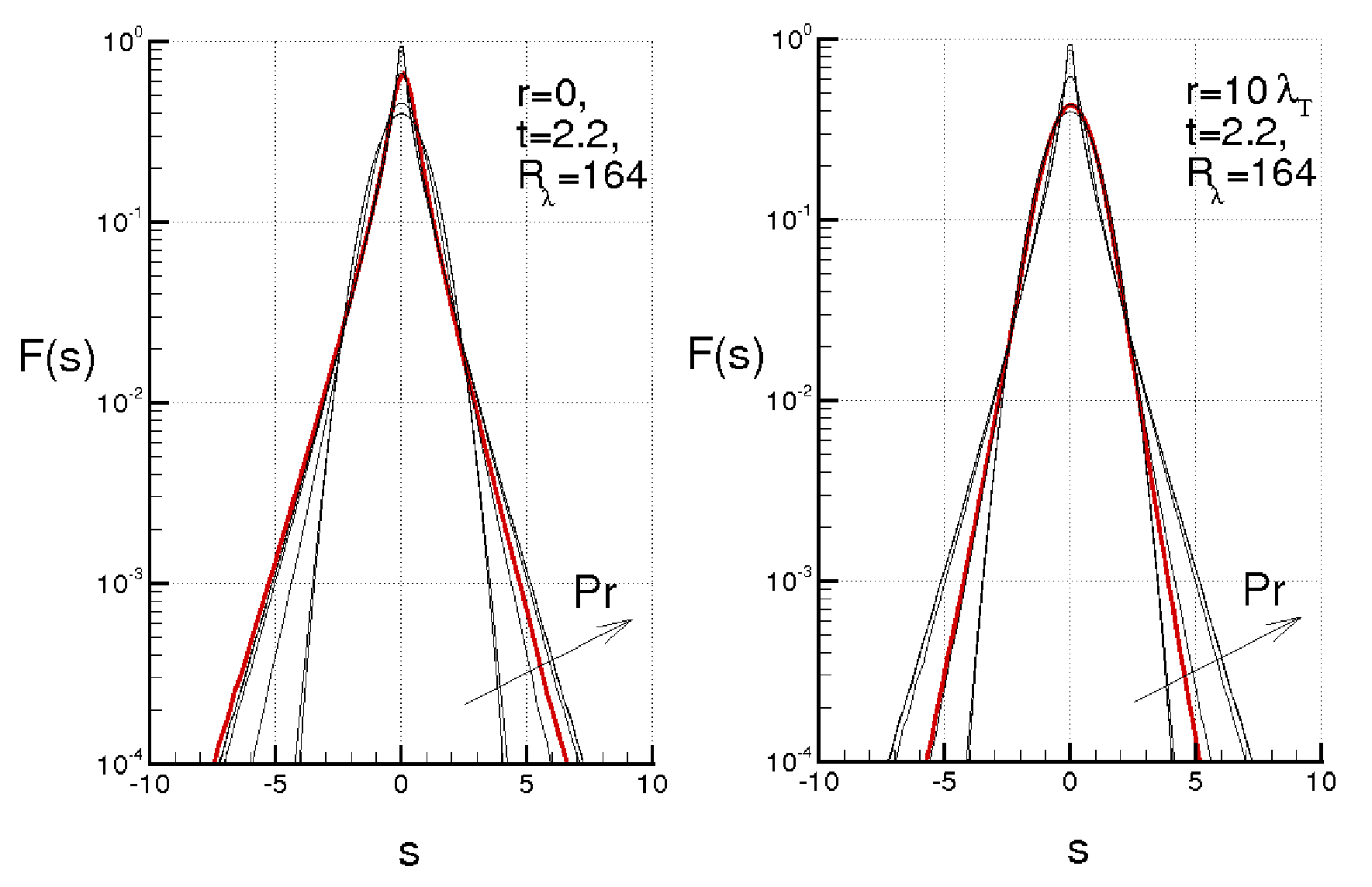}
\caption{PDF of the temperature increment $\Delta \vartheta$ at different Prandtl numbers: $Pr = 10^{-3}, 10^{-2}, 10^{-1}, 1, 10, 10^2,$ and $10^3$.  Red lines represent the PDF of longitudinal velocity difference. }
\label{f25}
\end{figure}

Subsequently, the probability density functions (PDFs) of the temperature increment are evaluated across different $Pr$ values with the proposed theory using  Eqs. (\ref{struct funs})-(\ref{Frobenius-Perron}). Figure~\ref{f25} displays these profiles in terms of the non-dimensional quantity $s$, defined as:
\begin{equation}
s = \frac{\Delta \vartheta}{\sqrt{\langle (\Delta \vartheta)^2 \rangle_E}}
\end{equation}
thereby yielding a family of curves that, despite representing different physical conditions, share a unit standard deviation. 
Specifically, the plots illustrate the various distributions as a function of $s$, computed at $r = 0$ and $r = 10 \lambda_T$ at the end of the development regime (i.e., at $t \simeq 2.2$), which corresponds to a Taylor-scale Reynolds number of approximately $164$. For comparison, the velocity increment PDF is also reported (indicated by the red line). The figure encompasses cases for $Pr = 10^{-3}$, $10^{-2}$, $10^{-1}$, $1$, $10$, $10^2$, and $10^3$. At $r = 0$ (see Fig.~\ref{f25}a), the velocity increment exhibits its maximum skewness value ($-3/7$) and a high kurtosis, computable via Eq.~(\ref{Tm1}).
The PDF of $\Delta \vartheta$ alters its shape, showing increased intermittency as $Pr$ increases. Specifically, for $Pr = 10^{-3}$ and $10^{-2}$, the PDF of $\Delta \vartheta$ is nearly Gaussian, whereas with increasing $Pr$, the intermittency (kurtosis) grows according to Eq. (\ref{Tm1}). In Fig. \ref{f25}b, where $r/\lambda_T = 10$, $\Delta u_r$ displays markedly lower asymmetry and intermittency compared to the $r = 0$ case. It can be seen that for $Pr = 10^{-3}$, $10^{-2}$, and $10^{-1}$, the PDF of $\Delta \vartheta$ is approximately Gaussian. Conversely, for $Pr = 1$, $\Delta \vartheta$ exhibits a kurtosis close to that of $\Delta u_r$, and for higher $Pr$ values, the kurtosis of $\Delta \vartheta$ increases accordingly following Eq. (\ref{Tm1}).
{ Even though spatial intermittency is markedly reduced at larger separations ($r/\lambda_T = 10$), the current framework consistently captures the systematic imprint of the Prandtl number on the PDF shapes. This persistence aligns with the established scaling behavior documented in both high-resolution DNS and laboratory experiments~\cite{Warhaft2000, Donzis}. Specifically, at small scales, our theory predicts that the kurtosis of the longitudinal temperature derivative varies within a well-defined range of approximately $5.5$ to $9$ as the Taylor-scale P\'eclet number increases from $100$ to $2000$. This quantitative behavior is in line with the classical grid-turbulence datasets by Warhaft~\cite{Warhaft2000}. Furthermore, while unconstrained high-resolution DNS often exhibits higher kurtosis values due to localized numerical forcing or subtle small-scale anisotropy, the bounds captured by the present framework reflect the statistical regularization inherent to a perfectly isotropic formulation of the von Kármán–Howarth and Corrsin equations, which naturally constrains higher-order dimensionless moments closer to their Gaussian limits.}
\begin{figure}[t]
	\centering
	\includegraphics[width=150mm, height=80mm,]{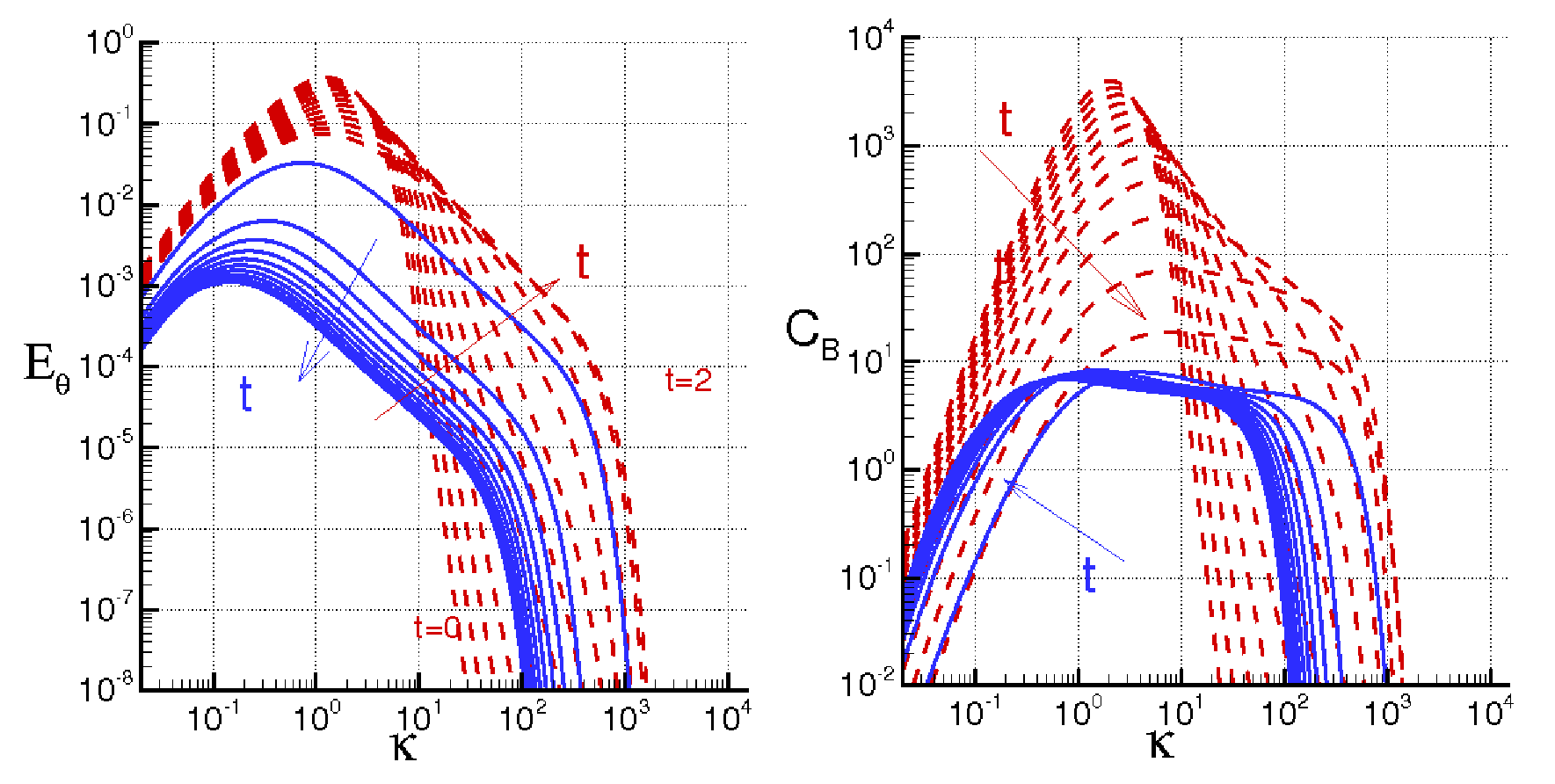}
\caption{Temperature spectrum evolution (left) and compensated Batchelor spectra (right) for $Pr = 10$.   Red dashed lines represent the developing transient regime from $t=0$ to $t=2.2$ with a constant time increment of $\Delta t=0.2$ between successive curves, where the red arrow indicates the direction of increasing time. Blue  lines describe the diffusive regime at subsequent time instances $t > 2.2$, where the blue arrow denotes the direction of increasing time.   }
\label{f26}
\end{figure}

\begin{figure}[t]
	\centering
	\includegraphics[width=150mm, height=80mm,]{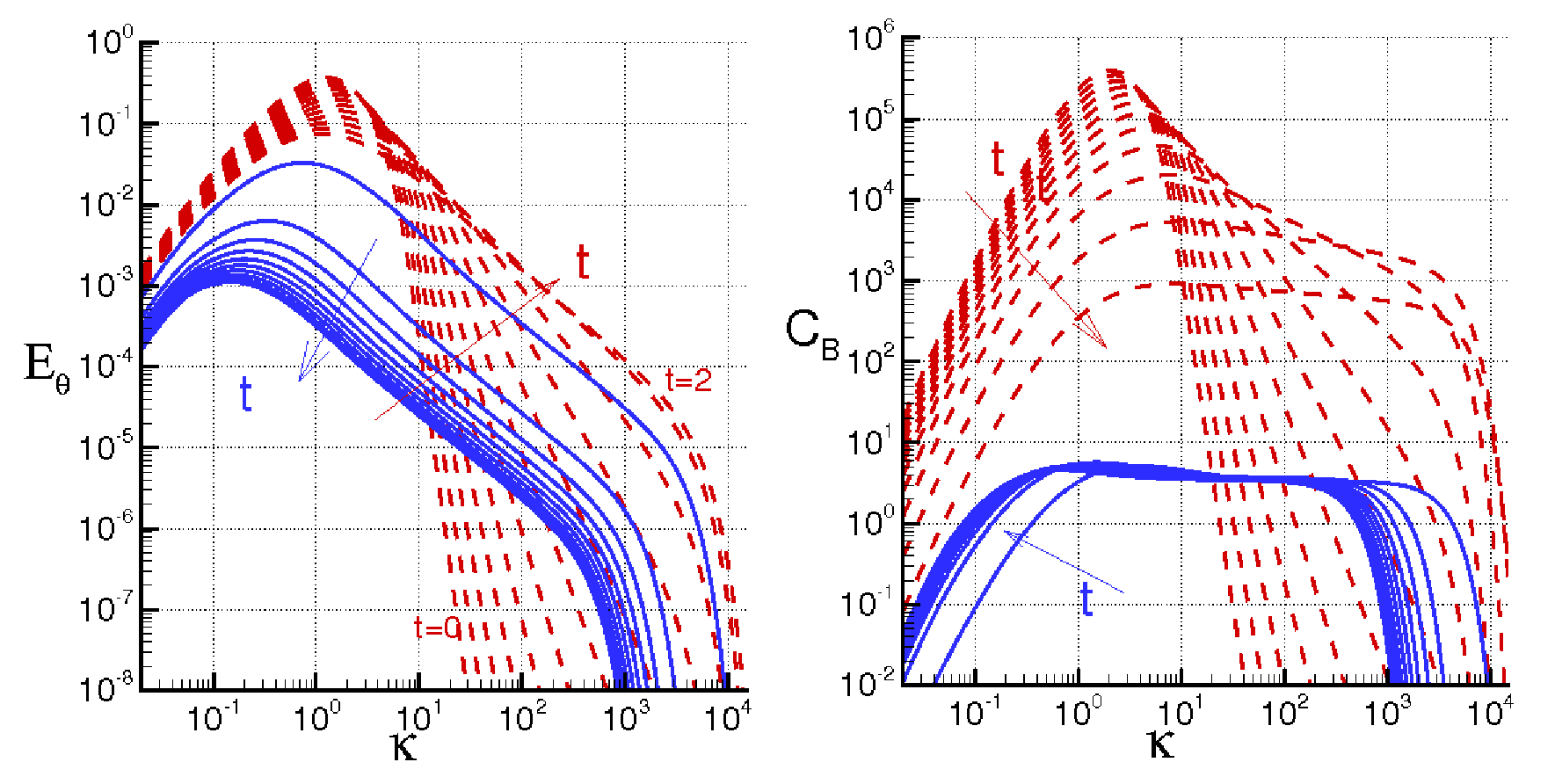}
\caption{Temperature spectrum evolution (left) and compensated Batchelor spectra (right) for $Pr = 1000$.   Red dashed lines represent the developing transient regime from $t=0$ to $t=2.2$ with a constant time increment of $\Delta t=0.2$ between successive curves, where the red arrow indicates the direction of increasing time. Blue lines describe the diffusive regime at subsequent time instances $t > 2.2$, where the blue arrow denotes the direction of increasing time.   }
\label{f27}
\end{figure}

\begin{figure}[t]
	\centering
	\includegraphics[width=150mm, height=80mm,]{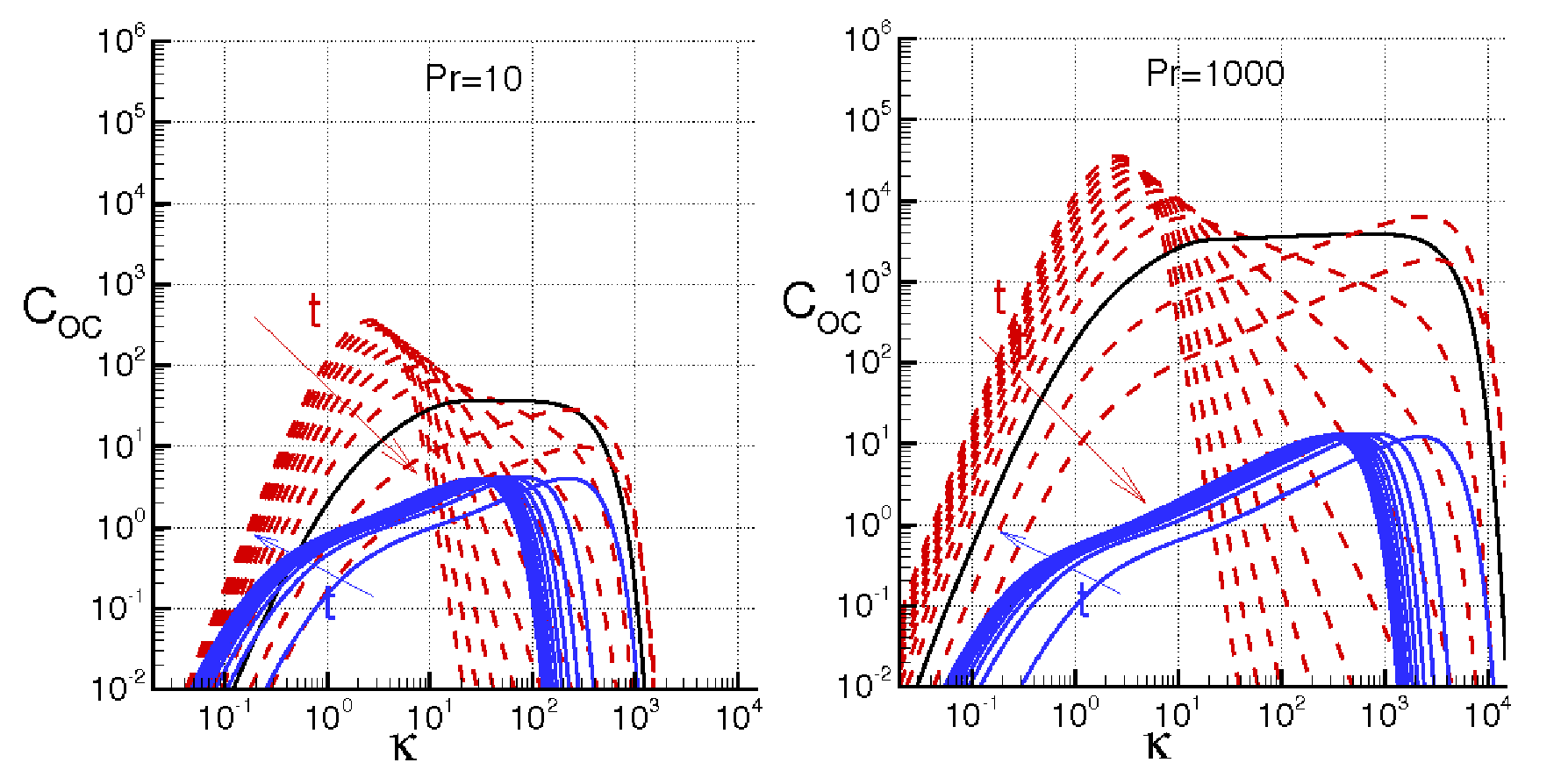}
\caption{Obukhov--Corrsin spectral function evolution for $Pr = 10$ (left) and $Pr = 1000$ (right).  Red dashed lines represent the developing transient regime from $t=0$ to $t=2.2$ with a constant time increment of $\Delta t=0.2$ between successive curves, where the red arrow indicates the direction of increasing time. Blue  lines describe the diffusive regime at subsequent time instances $t > 2.2$, where the blue arrow denotes the direction of increasing time.   }
\label{f28}
\end{figure}

Finally, we analyze the effect of the Prandtl number on the temperature spectra. Figures \ref{f26} and \ref{f27} display the results for $Pr = 10$ and $Pr = 1000$, respectively. In both cases, the spectra exhibit a $k^{-5/3}$ scaling law at approximately $t = 1.8$ (indicated by the black lines in Figs. \ref{f28}a and \ref{f28}b), whereas the envelopes of both spectra closely follow the $k^{-5/3}$ power law at the end of the development regime, i.e., at $t = 2.2$ (see Figs. \ref{f26} and \ref{f27}).
At subsequent time instances within the inertial range, both spectra exhibit a shallower slope than the $k^{-5/3}$ law due to the continuous decay of kinetic energy. Conversely, between the Kolmogorov microscale $\eta_K$ and the Batchelor microscale $\eta_B$, both temperature spectra display a $k^{-1}$ scaling behavior, in full agreement with Batchelor's theoretical arguments \cite{Batchelor_2}. We also introduce the compensated Batchelor spectral function, defined as:
\begin{equation}
C_B = \frac{E_\theta(k) k \sqrt{\epsilon/\nu}}{\epsilon_\theta}
\end{equation}
the behavior of which allows for the estimation of the Batchelor constant, $C_B$. 
The calculated values of $C_B$ are approximately $5$ for $Pr = 10$ and around $3.5$ for $Pr = 1000$, values in good agremment with the data of the literature. 

Regarding the Obukhov--Corrsin constant, its value is approximately $C_{OC} \simeq 1.8$ in both cases. This latter value is obtained from the compensated spectral functions $C_{O C}$ shown in Fig. \ref{f28}, corresponding to the oblique inflection point (as no strict plateau is present) for both curves.

{In conclusion, our analysis yields $C_B \in (3.4, 5.6)$---with $C_B \simeq 3.5$ via spectral fitting at high Prandtl numbers ($Pr = 1000$)---alongside a corresponding value of $C_{OC} \simeq 1.8$. While these values lie within the scatter documented in the literature, localized variations can be interpreted through two physical aspects:
\begin{enumerate}
    \item \textbf{Decaying vs. Forced Stationary Turbulence}: Many DNS data reporting canonical constants (such as the high-resolution simulations in \cite{Wang1996}) are obtained from statistically stationary, forced turbulence. Conversely, our work focuses on \textit{freely decaying} homogeneous isotropic turbulence (HIT). As discussed in \cite{briard2016passive}, the non-equilibrium evolution and the absence of a continuous spectral energy injection can modify the balance between dissipation and transfer scales, affecting the limits of both $C_B$ and $C_{OC}$.
    \item \textbf{Sensitivity to Initial Correlation Profiles}: The analysis illustrates that the Lyapunov--Liouville framework reproduces how different initial states (Saffman--Birkhoff, Loitsiansky, and Gaussian correlation profiles) lead to distinct multi-scale decay pathways. Since the non-equilibrium evolution of characteristic scales appears modulated by these initial configurations of the fields, the resulting spectral constants downstream retain a signature of the initial conditions---a feature that is documented in decaying passive scalar experiments and simulations \cite{sreenivasan1996passive}.
\end{enumerate} }

\bigskip

\section{Conclusions}

In this paper, we have presented a numerical investigation of the Lyapunov--Liouville theory applied to freely decaying homogeneous isotropic turbulence and passive scalar transport, assessing its performance against independent literature data. By integrating the closed von K\'arm\'an--Howarth and Corrsin equations through an automated, adaptive numerical solver, we have analyzed the statistical and structural evolution of fluid velocity and temperature increments starting from classical Saffman--Birkhoff, Loitsiansky, and Gaussian initial states under a wide range of Prandtl numbers ($Pr \in [10^{-1}, 10^3]$).

The main conclusions of this study can be summarized as follows:
\begin{itemize}
    \item The non-diffusive turbulence closures derived from the Lyapunov--Liouville analysis reproduce the dual-regime physics of decaying HIT, displaying a transition from a non-equilibrium transient developing phase to a diffusive decay regime.
    \item The solutions obtained are strongly influenced by the initial conditions, which is consistent with the non-universal nature of the decay exponents documented in the literature. While the Saffman--Birkhoff initial condition asymptotically yields the classical grid-turbulence values of $m \simeq -1.25$ and $n \simeq -1.25$, the Loitsiansky condition leads to an accelerated mechanical energy decay ($m \simeq -1.51$) alongside a higher thermal persistence ($n \simeq -0.89$).
    The Gaussian initial condition represents a highly dissipative regime where the mechanical energy decays extremely rapidly ($m \simeq -2.7$). This dissipation rate limits the duration of the self-preserving regime: at $t \simeq 33$, the Taylor-scale Reynolds number drops to $R_\lambda = 10$ and the dissipation rate exceeds the mean Lagrangian Lyapunov exponent. Beyond this threshold, our analysis indicates that the homogeneous isotropic turbulence (HIT) loses its physical condition of existence, marking the limit of the active turbulent state as both $R_\lambda$ and $u^2$ rapidly approach zero.
    \item The non-universal nature of the flow is also reflected by the mechanical and thermal dissipation coefficients, $C_\epsilon$ and $C_{\epsilon_\theta}$. In all examined cases, these coefficients maintain values of the order of unity but do not display, in general, extended universal plateaus. Instead, they exhibit either flat maxima or monotonic trends that appear dependent on the specific flow configurations and initial states.
    \item The inherent nonstationarity of the decaying flow modifies the scale-dependent transfer mechanics. Consequently, at the stationary points of the correlation scales ($d \lambda_T/dt \simeq 0, d \lambda_\theta/dt \simeq 0$), the spatial Kolmogorov and Yaglom functions display flat, localized maxima with peak values very close to $4/5$ and $2/3$, respectively, rather than the extended horizontal plateaus characteristic of stationary forced turbulence.
    \item For what concerns  the local scaling exponents, the curves corresponding to different time instances collapse onto a single master curve during the diffusive decay regime, confirming the onset of self-similarity despite the significant variations observed during the transient development phase.
\item { The analysis provides a description of multi-scale velocity statistics and energy transfer. Within the developing regime, the model reproduces the breakdown of the initial Gaussian symmetry at larger separations ($r/\lambda_T = 10$), where non-local interactions and direct energy transport expand the tails of the PDFs and alter their asymmetry. This physical behavior connects with the asymptotic trends evaluated at the smallest scales ($r \to 0$). In this limit, the present Lyapunov--Liouville theory yields a constant longitudinal velocity derivative skewness of $-3/7$, in compliance with foundational isotropic constraints. Under this condition, the calculated velocity derivative kurtosis remains bounded between approximately $4$ and $9$ as the Taylor-scale Reynolds number spans from $10$ to $2000$. This quantitative trend aligns with the landmark cryogenic turbulence experiments by Tabeling et al.~\cite{Tabeling1996} and Belin et al.~\cite{Belin1997}, suggesting that the proposed framework can replicate higher-order small-scale intermittency features under strictly isotropic conditions.}
\item The numerical results obtained with the proposed Lyapunov--Liouville framework yield predictions for the universal turbulence constants; specifically, the Kolmogorov constant settles within the range $C_K \simeq 1.72 - 1.75$, while the Obukhov--Corrsin constant matches $C_{OC} \simeq 1.8$, with both aligning with independent literature benchmarks.
\item The Prandtl number influences the passive scalar field, consistent with Batchelor's scaling theory for large-$Pr$ flows. Specifically, while the high Prandtl regimes investigated ($Pr = 10, 1000$) modify the scalar dynamics, it is for $Pr = 1000$ that the thermal correlation microscale $\lambda_\theta$ drops below the Kolmogorov scale $\eta_K$. Within these regimes, a best-fit procedure of the temperature spectra yields a Batchelor constant close to $C_B \simeq 5$ for $Pr = 10$, and $C_B \simeq 3.5$ for $Pr = 1000$, with $C_B$ estimated to vary between approximately $3.4$ and $5.7$ across the different cases investigated. In addition, the Obukhov--Corrsin constant stabilizes around $C_{OC} \simeq 1.8$ at the oblique inflection points of the compensated functions.
\item The reconstructed PDFs of temperature increments across different spatial separations describe how intermittency is modulated by the Prandtl number. At small scales, the PDF varies its shape from a nearly Gaussian profile at low $Pr$ ($10^{-3} - 10^{-2}$) to a distribution with broader tails and increased kurtosis at higher $Pr$. { Although intermittency dampens at larger scales ($r/\lambda_T = 10$), within the present framework, the systematic dependence of the PDF profiles on the Prandtl number is preserved, in line with the scaling behavior observed in literature DNS and experimental datasets~\cite{Warhaft2000, Donzis}. 
In particular, the small-scale statistics yielded by the model indicate that the kurtosis of the longitudinal temperature derivative bounds between approximately $5.5$ and $9$ over the analyzed Taylor-scale P\'eclet range ($100 \le Pe_\lambda \le 2000$). This outcome matches the quantitative trends observed in classical grid-turbulence experiments~\cite{Warhaft2000}. 
The kurtosis values yielded by the present framework can be lower than those from unconstrained DNS configurations; this outcome suggests how the assumption of perfect isotropy may act as a statistical regularizer, contributing to bound the growth of higher-order dimensionless moments under HIT conditions.
}
\end{itemize}
In conclusion, the satisfactory alignment between our numerical solutions and established direct numerical simulations (DNS) and experimental benchmarks supports the application of the Lyapunov--Liouville formulation to these configurations. This framework provides a structured approach to the closure problem, offering predictions for non-equilibrium features, spectral scaling behaviors, and Prandtl-dependent statistics within a self-contained formulation.

\bigskip

\section{Acknowledgments}

This work was partially supported by the Italian Ministry for the Universities 
and Scientific and Technological Research (MUR).

\bigskip

\end{document}